\documentclass[12pt,titlepage]{article}
\usepackage[footnotesize]{caption}
\pdfoutput =1
\usepackage{geometry,amsmath,amsfonts}
\usepackage{slashed}
\usepackage{epsfig}
\usepackage{hyperref}
\usepackage{latexsym}
\usepackage{graphicx}
\usepackage{amssymb}
\usepackage{verbatim}
\usepackage{bm}
\usepackage{dsfont}

\usepackage{cancel}
\usepackage{cite}

\usepackage{multirow}
\usepackage{tikz}

\usepackage{subfig}
\usepackage{color}

\usepackage{rotating}
\usepackage{ifthen}
\usepackage[utf8]{inputenc}

\numberwithin{equation}{section} 
\numberwithin{figure}{section}
\numberwithin{table}{section} 
\definecolor{cred}{RGB}{180,50,40} 
     
\newcommand{\Kahler}{\ensuremath{\text{K}\ddot{\text{a}}\text{hler}\,}}

\begin{document}

\begin{titlepage}

\thispagestyle{empty}
\setcounter{page}{0}

\vskip 1.5cm

\begin{center}
{\LARGE\bf PBH dark matter from axion inflation}

\vskip 2cm

{\large  Valerie~Domcke\footnote{valerie.domcke@apc.univ-paris7.fr},  Francesco Muia\footnote{francesco.muia@physics.ox.ac.uk}, Mauro Pieroni\footnote{mpieroni@apc.univ-paris7.fr}, \newline Lukas T.~Witkowski\footnote{lwitkow@apc.univ-paris7.fr}}\\[3mm]
{\it{
${}^{1,3,4}$ AstroParticule et Cosmologie (APC)/Paris Centre for Cosmological Physics, Universit\'e Paris Diderot, CNRS, CEA, Observatoire de Paris, Sorbonne Paris Cit\'e University.  \\
${}^2$ Rudolf Peierls Centre for Theoretical Physics, University of Oxford,\\
1 Keble Road, Oxford OX1 3NP, UK.
}}
\end{center}

\vskip 1cm
\centerline{ {\bf Abstract}}{
Protected by an approximate shift symmetry, axions are well motivated candidates for driving cosmic inflation. Their generic coupling to the Chern-Simons term of any gauge theory gives rise to a wide range of potentially observable signatures, including equilateral non-Gaussianites in the CMB, chiral gravitational waves in the range of direct gravitational wave detectors and primordial black holes (PBHs). In this paper we revisit these predictions for axion inflation models non-minimally coupled to gravity. Contrary to the case of minimally coupled models which typically predict scale-invariant mass distributions for the generated PBHs at small scales, we demonstrate how broadly peaked PBH spectra naturally arises in this setup. For specific parameter values, all of dark matter can be accounted for by PBHs.}

\vspace{1cm}

\center
\textit{Dedicated to the memory of Pierre Bin\'etruy}

\end{titlepage}
\tableofcontents

\section{Introduction\label{sec:introduction}}
The recent detection of gravitational waves (GWs) emitted by a black hole (BH) binary by the LIGO/VIRGO collaboration~\cite{Abbott:2016blz} marked not only the beginning of a new age of GW astronomy but can also be seen as the first direct evidence of BHs. With a lot more data from advanced LIGO/VIRGO and from the future space-mission LISA~\cite{LISA} on the horizon, these are exciting times for considering possible implications for fundamental physics and early universe cosmology. 

Prompted by the LIGO discovery and the lack of positive results in dedicated dark matter searches, the idea that primordial black holes (PBHs) could be (a fraction of) dark matter has recently received a lot of attention~\cite{Bird:2016dcv, Clesse:2016vqa,Sasaki:2016jop,Carr:2016drx,Clesse:2016ajp,Blinnikov:2016bxu,Garcia-Bellido:2017fdg,Georg:2017mqk}. PBHs are formed when local over densities collapse due to their gravitational instability. Here we will focus on the possibility that these local over densities are sourced by the primordial scalar power spectrum of inflation (see e.g.\ Refs.~\cite{Cotner:2016cvr,Davoudiasl:2016mwf } for other recently discussed possibilities). This requires an amplitude of the scalar power spectrum which is much larger than the amplitude observed in the cosmic microwave background (CMB), i.e.\ a highly blue spectrum. On the other hand, PBHs formed at very small scales are strongly constrained by the traces their decays leave in the CMB and in the extra-galactic gamma ray background~\cite{Carr:2009jm}. A viable model of PBH dark matter thus requires a non-trivial peaked structure of the primordial scalar power spectrum. It has been demonstrated that such spectra can be achieved in multifield models of inflation~\cite{GarciaBellido:1996qt,Lyth:2011kj,Bugaev:2011wy,Clesse:2015wea,Kohri:2012yw,Kawasaki:2012wr,Bugaev:2013vba,Inomata:2017okj,Inomata:2016rbd,Kawasaki:2016pql} or by arranging for the corresponding features in the scalar potential of inflation~\cite{Cheng:2016qzb,Garcia-Bellido:2017mdw,Garcia-Bellido:2016dkw}.

In this paper, we generate PBH dark matter in an inflation model which is based on two simple, generic features: an axion-like inflaton with a non-minimal coupling to gravity which increases towards the end of inflation. Axion-like particles are interesting inflaton candidates for several reasons: (i)~their approximate shift-symmetry protects them from large mass contributions, preserving the required slow-roll phase of inflation. (ii)~they appear abundantly in supergravity as part of complex scalars as well as in string theory~\cite{0605206}, where they typically arise upon compactification of form field gauge potentials. (iii)~they generically couple to the topological Chern-Simons term of any gauge group, leading to a wide range of potentially observable effects~\cite{Turner:1987bw, Garretson:1992vt, Anber:2006xt,Barnaby:2011qe}. These include a non-Gaussian and blue contribution to the scalar spectrum, potentially leading to PBH formation~\cite{Barnaby:2010vf,Linde:2012bt,Lin:2012gs,Meerburg:2012id,Bugaev:2013fya,Garcia-Bellido:2016dkw} as well as a blue, non-Gaussian and highly polarized stochastic GW spectrum, potentially observable by LIGO or LISA~\cite{Cook:2011hg,Anber:2012du,Bartolo:2016ami}.
The predictions for different universality classes of pseudoscalar inflation have recently been studied in Ref.~\cite{Domcke:2016bkh}. Here we extend this work by allowing for a non-minimal coupling to gravity, i.e.\ by considering simple Jordan frame inflation models. In recent years, such a setup was found to be appealing both for deriving the supergravity Lagrangian~\cite{Kallosh:2000ve} and from a phenomenological perspective, driven most notably by the discussion of Higgs inflation~\cite{Bezrukov:2007ep}. Such a non-minimal coupling to gravity is a characteristic feature of supergravity in the Jordan frame~\cite{Kallosh:2013hoa,Kallosh:2013tua,Ferrara:2013rsa,Galante:2014ifa,Broy:2015qna,Das:2016kwz}. The so-called attractor models at strong coupling~\cite{Kallosh:2013tua,Kallosh:2013yoa} are an example of how the non-minimal coupling to gravity can be used to construct classes of models which asymptote to the sweet spot of the Planck data, see also~\cite{Einhorn:2009bh,Ferrara:2010yw,Buchmuller:2013zfa,Giudice:2014toa,Pallis:2013yda,Pallis:2014dma,Pallis:2014boa,Ellis:2013xoa,Kallosh:2013xya,Nakayama:2010ga,Pieroni:2015cma} for other implementations.

Combining these two ingredients, the following picture emerges: at large scales (constrained by CMB observations~\cite{Ade:2015lrj}), i.e.\ early on in the inflationary epoch, both the coupling to the Chern-Simons term and the non-minimal coupling to gravity are suppressed. As inflation proceeds, the former generically induces a tachyonic instability in the equation of motion for the gauge fields, leading to a gauge-field background which in turn acts as an additional, classical source term for scalar and tensor perturbations. This results in a strong increase of both spectra. Once the gauge field production has reached a critical value, the back-reaction of the gauge fields on the inflaton dynamics slows the growth of both spectra~\cite{Anber:2009ua,Barnaby:2011qe,Barnaby:2011vw}. Finally, in the last stage of inflation the non-minimal coupling to gravity becomes relevant and both the scalar and tensor spectrum are suppressed. As a result, a significant fraction of dark matter (for specific parameter choices even all of dark matter) can be contained in PBHs. In part of the parameter space their mergers may be observable by advanced LIGO. We observe a remarkable complementarity between CMB observations, PBH constraints and GW searches in constraining the parameter space of this setup.

The remainder of this paper is organized as follows. After a review of axion inflation in Sec.~\ref{sec:review}, we extend the formalism to account for a non-minimal coupling to gravity in Sec.~\ref{sec:non-minimal}. This in particular includes the computation of the resulting scalar and tensor perturbations, sourced by both the quantum mechanical vacuum contribution and the classical gauge field contribution. In Sec.~\ref{sec:attractors} we apply these results to inflation models described by attractors at strong coupling. We dedicate Sec.~\ref{sec:PBHs} to the production of primordial black holes, discussing different regions of the parameter space based on a handful of benchmark models. We conclude in Sec.~\ref{sec:conclusions}. The main results of this paper are supported by three appendices. In App.~\ref{sec:axions} we discuss how the type of axions we consider here arise in the context of field theory, supergravity and string theory. Some of the details of the derivations of the scalar power spectrum are left to App.~\ref{sec:spectrum}. Finally, App.~\ref{app:PBH} analyses how uncertainties related to the PBH formation and evolution impact our results.

\section{A review of axion inflation\label{sec:review}}
In this section we give a review\footnote{For more details see for example~\cite{Barnaby:2011vw,Barnaby:2011qe,Linde:2012bt}.} of the case of a \emph{pseudoscalar} inflaton $\phi$ coupled to Abelian gauge fields. In particular we show that a generic higher-dimensional coupling between the inflaton and the gauge fields introduces an instability in the theory, which leads to an exponential production of the gauge fields~\cite{Turner:1987bw, Garretson:1992vt, Anber:2006xt}. Interestingly this effect induces a back-reaction both on the background dynamics~\cite{Anber:2009ua,Barnaby:2011qe,Barnaby:2011vw} and on the perturbations~\cite{Anber:2012du,Linde:2012bt}, leading to a wide set of observational consequences. Let us consider the action for a pseudo-scalar inflaton $\phi$, which is coupled to a certain number $\mathcal{N}$ of Abelian gauge fields $A_\mu^a$ (associated to $U(1)$ gauge symmetries,\footnote{Alternatively, these may be associated with a $SU({\cal N})$ symmetry in the weak coupling limit.} i.e.\ $a = \{1,2,...,\mathcal{N}\}$)~\cite{Turner:1987bw,Garretson:1992vt, Anber:2006xt,Anber:2012du,Anber:2009ua,Barnaby:2011qe,Barnaby:2011vw,Linde:2012bt,Domcke:2016bkh}:\footnote{The metric has signature $(-,+,+,+)$ and $m_p \simeq 2.4 \cdot 10^{18}\,$~GeV denotes the reduced Planck mass.} 
\begin{equation}
\label{review:action_pseudoscalar}
\mathcal{S}= \int \textrm{d}^4 x \sqrt{|g|} \left[m_p^2 \frac{R}{2} -\frac{1}{2} \, \partial_\mu \phi \, \partial^\mu \phi - V(\phi) - \frac{1}{4} F^a_{\mu \nu} F_a^{\mu \nu} - \frac{\alpha^a}{4 \Lambda} \phi F^a_{\mu \nu} \tilde{F}_a^{\mu \nu} \right ]\ ,
\end{equation}
where $V(\phi)$ is the inflaton potential, $F^a_{\mu \nu} $ ($\tilde{F}^{\mu \nu}_{a}$) are the (dual\footnote{The dual field-strength tensor is defined as $\tilde{F}_a^{\mu \nu} \equiv \tilde{F}_{\rho \sigma}^a \epsilon^{\mu \nu \rho \sigma}/(2 \sqrt{-g})$,
where $\epsilon^{\mu \nu \rho \sigma}$ is the Levi-Civita tensor.}) field-strength tensors for the gauge fields, $\Lambda$ is a mass scale (that suppresses the higher-dimensional scalar-vector coupling) and $\alpha^a$ are the dimensionless coupling constants of the Abelian gauge fields. In the following we consider $\alpha^a = \alpha$ for all $a$. Throughout this paper, the effective mass scale $\Lambda/\alpha$ will take sub-Planckian values, 
 as required for a reasonable effective field theory. Typical values found e.g.\ in Ref.~\cite{Domcke:2016bkh} for phenomenologically interesting scenarios lie in the range $\Lambda/\alpha \simeq 0.01 - 0.03 \, m_p$.
However, we stress that contrary to natural inflation models~\cite{Freese:1990rb}, we do not impose that this scale simultaneously sets the scale of the scalar potential or of the field excursion during inflation. A discussion on how Eq.~\eqref{review:action_pseudoscalar} may arise in the context of field theory, supergravity or string theory is provided in App.~\ref{sec:axions}.

\subsection{Background dynamics}
\label{review:background}
In order to describe the dynamics we can start by computing the background equations of motion for the inflaton $\phi(t)$ and for the gauge fields $A_\mu^a(t,x)$. Without loss of generality, we proceed by assuming $ \phi > 0, \ V_{,\phi}(\phi)>0, \ \dot{\phi} < 0$, and we choose to describe the problem in the Coulomb gauge ($A_0^a = 0$, $\partial^\mu A_\mu^a(t,x) = 0$). Under these assumptions the equations of motion can be expressed as:
\begin{align}
\label{review:eq_motion}
\ddot \phi + 3 H \dot{\phi} + \frac{\partial V}{\partial \phi} = \frac{\alpha}{2\Lambda} \frac{\varepsilon^{\mu\nu\rho\sigma}}{\sqrt{|g|}} \langle \partial_\mu A_\nu \partial_\rho A_\sigma \rangle & \equiv \frac{\alpha}{\Lambda} \langle \vec{E}^a \cdot \vec{B}^a \rangle \ ,\\
\frac{d^2}{d \tau^2}\vec{A}^a - \vec{\nabla}^2 \vec{A}^a - \frac{\alpha}{\Lambda} \frac{\textrm{d} \phi}{\textrm{d} \tau} \vec{\nabla} \times \vec{A}^a & = 0 \ , \label{review:eq_motionA}
\end{align}
where dots are used to denote derivatives with respect to cosmic time $t$, primes are used to denote derivatives with respect to $\tau$ (conformal time defined as $\textrm{d}t \equiv a \, \textrm{d}\tau$), $\vec{\nabla}$ is the $3$-dimensional flat space gradient operator, the brackets $\langle \cdot \rangle$ denote a spatial mean and the vectors $\vec{E}^a$ and $\vec{B}^a$ are the ``electric'' and ``magnetic'' fields associated with the fields $A_\mu^a(t,x)$ defined as:
\begin{equation}
	\label{review:electric_magnetic}
	\vec{E}^a \equiv -\frac{1 }{a^2} \frac{\textrm{d} \vec{A}^a}{\textrm{d} \tau} = -\frac{1 }{a} \frac{\textrm{d} \vec{A}^a}{\textrm{d} t} \ , \qquad \qquad  \vec{B}^a \equiv \frac{1}{a^2} \vec{\nabla} \times \vec{A}^a  \ .
\end{equation}
In order to completely specify the dynamics we also have to compute the Friedmann equation which can be expressed as:
\begin{equation}
\label{review:friedmann}
3 H^2 m_p^2 = \frac{1}{2} \dot{\phi}^2 + V(\phi) + \frac{1}{2} \langle \vec{E}^{a \, 2}  +  \vec{B}^{a \, 2}\rangle  \ .
\end{equation}
Clearly a general analytical solution for Eq.~\eqref{review:eq_motion}, Eq.~\eqref{review:eq_motionA}, and Eq.~\eqref{review:friedmann} does not exist. However, assuming $\dot{ \phi} $ to be slowly varying (which is a reasonable assumption during slow-roll inflation) we can find an analytical solution for Eq.~\eqref{review:eq_motionA}. By substituting this solution into Eq.~\eqref{review:eq_motion} and Eq.~\eqref{review:friedmann} we can then study the back-reaction on the system.

In order to solve Eq.~\eqref{review:eq_motionA} we start by performing a spatial Fourier transform and, taking $\vec{k}$ to be parallel to the x-axis $\hat{x}$, expressing the gauge fields in terms of the two helicity vectors\footnote{The two helicity vectors $\vec{e}_{\pm}$ are defined as $\vec{e}_{\pm} = (\hat{y} \pm i \hat{z})/\sqrt{2}$. It is crucial to notice that using the helicity vectors we get $\vec{k} \times \vec{A}^a =  A^{a}_{\pm} \vec{k} \times \vec{e}_{\pm} = \mp i A^{a}_{\pm} |\vec{k}| \vec{e}_{\pm}$.} as $\vec{A}^a = A_{+}^a \vec{e}_{+} + A_{-}^a \vec{e}_{-}$, the equation of motion for the gauge fields reads:
\begin{equation}
\label{review:eq_motionAfourier}
  \frac{\textrm{d}^2  A^{a}_{\pm}(\tau,\vec{k})}{\textrm{d} \tau^2}  + \left[ k^2 \pm 2\, k  \, \frac{\xi}{\tau} \right]A^{a}_{\pm}(\tau,\vec{k}) =  \ 0 \ , 
  \end{equation}
where we have introduced the parameter $\xi$ defined as:
\begin{equation}
\xi \equiv \frac{\alpha \, |\dot{\phi}|}{2 \, \Lambda \, H} \ .
\label{review:xi}
\end{equation}
Notice that Eq.~\eqref{review:eq_motionAfourier} describes a tachyonic instability for the $A_+$ mode. In particular for the modes with $(8 \, \xi)^{-1} \lesssim k/(aH) \lesssim 2 \, \xi$,  $A_+$ can be expressed as~\cite{Anber:2009ua,Barnaby:2010vf,Barnaby:2011vw}:
\begin{equation}
A_+^a \simeq \frac{1}{\sqrt{2\, k}} \left( \frac{k}{2 \,  \xi \, a \, H}\right)^{1/4} e^{ \pi \xi - 2 \sqrt{2 \xi k/(a H)}} \ ,
\end{equation}
so that the terms $\langle \vec{E}^a \cdot \vec{B}^a \rangle$ and $\langle \vec{E}^{a \, 2} + \vec{B}^{a \, 2} \rangle$ appearing in Eq.~\eqref{review:eq_motion} and in Eq.~\eqref{review:friedmann} can be expressed as:\footnote{These expressions only hold for $\xi \gtrsim 1$. For the correct expressions for small values of $\xi$ see~\cite{Anber:2009ua,Pieroni:2016gdg}.}
\begin{equation}
\langle \vec{E}^a \cdot \vec{B}^a \rangle \simeq \mathcal{N} \cdot \   2.4 \cdot 10^{-4} \frac{H^4}{\xi^4} e^{2 \pi \xi} \ , \quad \frac{1}{2} \langle \vec{E}^{a \, 2} + \vec{B}^{a \, 2} \rangle  \simeq \mathcal{N} \cdot \   1.4 \cdot 10^{-4} \frac{H^4}{\xi^3} e^{2 \pi \xi} \ .
\end{equation}
It is possible to show (see for example~\cite{Barnaby:2011vw}) that while the back-reaction on the Friedmann equation is fairly negligible throughout the whole evolution, the back-reaction on the equation of motion for the scalar field cannot be neglected in the last part of the evolution. In particular this back-reaction introduces an additional friction term that has an exponential dependence on $\xi$. As this parameter is proportional to $\dot{\phi}$, it typically increases towards the end of inflation in single-field inflation models. Correspondingly the new friction term may be negligible at CMB scales while significantly slowing down the last part of the evolution. As explained in~\cite{Domcke:2016bkh}, this effect induces a shift in the region of the potential that can be probed with CMB observations. It is crucial to stress that the gauge fields are not changing the total number $N \equiv - \int H \textrm{d}t$ of e-foldings (i.e.\ the CMB is still generated at $N_\text{CMB} \simeq 60$ in the complete evolution) but rather they slow down the increase of $\xi \sim |\dot{\phi}/H|$.

\subsection{Scalar and tensor power spectra}
\label{review:perturbations}
The gauge fields do not only affect the background dynamics but they also modify the equation of motion for the perturbations. In particular they induce a source term both for scalar and tensor perturbations leading to an exponential amplification of the spectra at small scales. 

Let us start by discussing the modified scalar power spectrum.\footnote{More details on the derivation of these formulas are given in Sec.~\ref{sec:non-minimal}, where we also discuss the generalization to a model where the inflaton is non-minimally coupled to gravity.} As a first step we express the inflaton field as $\phi(\vec{x},t)=\phi(t) + \delta \phi(\vec{x},t)$ and by solving the linearized equation of motion for $\delta \phi(\vec{x},t)$ (for details see~\cite{Anber:2006xt,Anber:2009ua,Barnaby:2010vf,Barnaby:2011vw,Barnaby:2011qe}) we can express the scalar power spectrum $\Delta^2_s(k)$ as~\cite{Linde:2012bt,Domcke:2016bkh}:
\begin{equation}
\Delta^2_s(k) = \Delta^2_s(k)_\text{vac} + \Delta^2_s(k)_\text{gauge} = \left(\frac{H^2}{2 \pi |\dot{\phi}|}\right)^2 + \left( \frac{\alpha \langle \vec{E}^a\cdot \vec{B}^a \rangle/ \sqrt\mathcal{N}}{3 \Lambda b H \dot{\phi}} \right)^2 \ ,
\label{review:scalar}
\end{equation}
where we have defined $b$ as:
\begin{equation}
b \equiv 1 - 2 \, \pi \, \xi \,  \frac{\alpha \langle \vec{E}^a \cdot \vec{B}^a \rangle}{3 \Lambda H \dot{\phi}}  \ .
\end{equation} 
As $\langle \vec{E}^a\cdot \vec{B}^a \rangle$ grows exponentially with $\xi$, the second term of Eq.~\eqref{review:scalar} (i.e.\ the gauge-field induced contribution) is typically negligible at CMB scales (where the stringent CMB constraints essentially require $\Delta^2_s(k) \simeq \Delta^2_s(k)_\text{vac}$) but may dominate over the first term at small scales. In particular it is possible to show that in the gauge-field dominated regime (i.e.\ typically in the last part of inflation) the spectrum is well approximated by~\cite{Linde:2012bt,Domcke:2016bkh}:
\begin{equation}
\Delta^2_s(k ) \simeq \frac{1}{\mathcal{N} (2 \pi \xi)^2} \ .
\label{review:scalar_strong}
\end{equation}
Notice that in presence of several $U(1)$s the power spectrum is suppressed at small scales. Moreover, the presence of the gauge fields leads to the generation of equilateral non-Gaussianities~\cite{Anber:2012du}. As a consequence, the non-observation of non-Gaussianities at CMB scales~\cite{Ade:2015lrj,Ade:2015ava} can be used to set constraints on these models. In particular, this implies $\left. \xi\right|_{\text{CMB}} \lesssim 2.5$, where $\left. \xi\right|_{\text{CMB}}$ is the value of $\xi$ at CMB scales~\cite{Barnaby:2011vw,Barnaby:2011qe,Anber:2012du,Barnaby:2010vf,Linde:2012bt}.

In order to compute the tensor spectrum $\Delta^2_{t}$ we start with the equation of motion for the traceless transverse part of the metric perturbations $h_{ij}(\vec{x},t)$, given by the linearized Einstein equation:\footnote{For details on the derivation of this equation see~\cite{Maggiore:1900zz}.}
 \begin{equation}
\frac{\textrm{d}^2 h_{ij}}{\textrm{d} \tau^2} + 2 \frac{\textrm{d} \ln a}{\textrm{d} \tau} \frac{\textrm{d} h_{ij}}{\textrm{d} \tau} - \vec{\nabla}^2 h_{ij} = \frac{2}{m_p^2} \Pi_{ij}^{\mu \nu} T_{\mu \nu} \ ,
\label{review:linearized_einstein}
 \end{equation}
where $\Pi^{ij}_{\mu \nu}$ is the transverse, traceless projector and $T_{\mu \nu}$ is the matter energy-momentum tensor (which acts as a source for GWs). By solving this equation we obtain $\Delta^2_{t,L}$ and $\Delta^2_{t,R}$, power spectra for the two polarizations\footnote{In order to decompose $h_{ij}(\vec{x},t)$ in terms of the two polarizations of the GW we can use the projector $\Pi_{ij,L/R} = e_{i, \, \mp} e_{j, \, \mp}$.} $(L,R)$ of the GWs. The tensor power spectrum $\Delta^2_{t}$ (given by the sum of the spectra for the two polarizations) can then be expressed as:
\begin{equation}
\Omega_{GW} \equiv \frac{\Omega_{R,0}}{24} \Delta^2_{t} \simeq \frac{1}{12} \Omega_{R,0} \left(  \frac{H}{ \pi \, m_p } \right)^2 \left(1 + 4.3 \cdot 10^{-7} \mathcal{N} \frac{H^2}{ m_p^2 \, \xi^6} e^{4 \pi \xi}\right)\ ,
\label{review:OmegaGW}
\end{equation} 
where $\Omega_{R,0} = 8.6 \cdot 10^{-5}$ is used to denote the radiation energy density today. Similarly to the case of the scalar power spectrum, the gauge-field sourced term (i.e.\ the second term in the parenthesis on the r.h.s. of Eq.~\eqref{review:OmegaGW}) is typically negligible at CMB scales and dominates over the first term at small scales (i.e.\ in the last part of inflation). It is also crucial to stress that the second term in Eq.~\eqref{review:OmegaGW} is only sourced by one of the two polarizations. As a consequence the GW signal that is generated during the last part of the evolution is expected to be strongly chiral. Chirality is a peculiar feature for a GW background and if detected\footnote{For details on methods to detect the chirality of GW background see~\cite{Crowder:2012ik}.} it would point to the existence of a parity violating source. In the framework of standard general relativity axion inflation is one of the simplest models which implement this parity violation.

As in the context of direct GW observations it is customary to express quantities in terms of the frequency $f = k / (2 \pi)$, it is useful to introduce the relation between $f$ and the number of e-foldings $N$~\cite{Barnaby:2011qe,Domcke:2016bkh}:
\begin{equation}
N = N_\text{CMB} + \ln \frac{k_\text{CMB}}{0.002 \text{ Mpc}^{-1}} - 44.9 - \ln\frac{f}{10^2 \text{ Hz}} \ ,
\label{review:Nf}
\end{equation}
where $k_\text{CMB} = 0.002 \text{ Mpc}^{-1}$ and $N_\text{CMB} \simeq 50 - 60$. Following the convention used throughout this work, the number of e-foldings $N$ decreases during inflation, reaching $N=0$ at the end of inflation and we will set $N_\text{CMB} = 60$.

\section{Non-minimal coupling to gravity\label{sec:non-minimal}}
In this section\footnote{To ease the notation, in Secs.~\ref{sec:non-minimal} and~\ref{sec:attractors} we will work in Planck units, $m_p = 1$, while reintroducing $m_p$ in Sec.~\ref{sec:PBHs}.} we generalize the analysis presented in Sec.~\ref{sec:review} to the case where the inflaton is non-minimally coupled to gravity. In general non-minimal couplings between the inflaton and gravity may naturally arise in the context of supergravity and string theory~\cite{Kallosh:2000ve} or from radiative corrections in the framework of QFT in curved space-time.\footnote{For an introduction to the topic see for example~\cite{Birrel:1984}.} Moreover, during the last years it was realized that the introduction of a non-minimal coupling between the inflaton and gravity (in the Jordan frame) can provide a mechanism to ``flatten'' the inflationary potential (in the Einstein frame)~\cite{Bezrukov:2007ep,Bezrukov:2009db,Kallosh:2013tua,Einhorn:2009bh,Ferrara:2010yw,Buchmuller:2013zfa,Giudice:2014toa,Pallis:2013yda,Pallis:2014dma,Pallis:2014boa,Ellis:2013xoa,Kallosh:2013xya,
Nakayama:2010ga} leading to predictions for the scalar spectral index $n_s$ and the  tensor-to-scalar-ratio $r$  that lie right in the sweet spot of the constraints set by CMB observations~\cite{Ade:2015lrj}. For these reasons it is interesting to consider a generalization of the mechanism discussed in Sec.~\ref{sec:review} in order to take these models into account.

We start by considering the action for a pseudoscalar field $\phi$ non-minimally coupled to gravity and coupled to an Abelian gauge field\footnote{The generalization to $\mathcal{N}$ Abelian gauge fields $A^a_{\mu}$ is straight-forward. To ease the notation, we will work with ${\cal N} = 1$ here and re-introduce the parameter ${\cal N}$ only in the final expressions.} through the topological Chern-Simons term,
\begin{equation}
	\label{non_minimal:jordan_frame}
	\mathcal{S} = \int \textrm{d}^4 x \sqrt{-g_J} \left[ \frac{\, \Omega(\phi ) \, R}{2} -X -V_J -\frac{g_J^{\mu\rho}g_J^{\nu\sigma}}{4} F_{\mu\nu}F_{\rho\sigma} -\frac{\alpha}{8 \Lambda} \frac{\varepsilon^{\mu\nu\rho\sigma}}{ \sqrt{-g_J}} \, \phi \, F_{\mu\nu}F_{\rho\sigma} \right] \ ,
\end{equation}
where $\varepsilon^{\mu\nu\rho\sigma}$ is the Levi-Civita symbol, $X \equiv g_J^{\mu \nu} \partial_{\mu} \phi \partial_{\nu} \phi /2$, $F_{\mu\nu} \equiv \partial_{\mu} A_{\nu} - \partial_{\nu} A_{\mu}$, $\alpha$ is the coupling between the pseudoscalar and the gauge field, $\Lambda $ is a mass scale that suppresses the higher dimensional operator $\phi F \tilde{F}$ and $\Omega(\phi)$ is function of $\phi$ that without loss of generality can be expressed as:
\begin{equation}
	\Omega(\phi) = 1 + \varsigma f(\phi) \ , 
\end{equation}
where $\varsigma$ is a the dimensionless coupling constant and $f(\phi)$ is a generic function of $\phi$. The case discussed in Sec.~\ref{sec:review} can be easily recovered by setting $\varsigma = 0$.

The action of Eq.~\eqref{non_minimal:jordan_frame} corresponds to a Jordan frame formulation of the theory (explaining the subscript J). The Einstein frame formulation (where gravity is described by a standard Einstein-Hilbert term) of the theory is found via a conformal transformation:
\begin{equation}
	g_{J\, \mu\nu} \rightarrow g_{\mu\nu} = \Omega(\phi) g_{J\, \mu\nu} \, ,
\end{equation}
so that the action reads:
\begin{equation}
	\label{non_minimal:einstein_frame}
	\mathcal{S} = \int \textrm{d}^4 x \sqrt{-g} \left[ \frac{R}{2} - K(\phi)X -V_E -\frac{g^{\mu\rho}g^{\nu\sigma}}{4} F_{\mu\nu}F_{\rho\sigma} -\frac{\alpha}{4 \Lambda} \frac{\varepsilon^{\mu\nu\rho\sigma}}{2 \sqrt{-g}} \, \phi \, F_{\mu\nu}F_{\rho\sigma} \right] \ ,
\end{equation}
where $V_E \equiv V_J/\Omega^2(\phi)$ and $K(\phi)$ is defined as:\footnote{In order to make expressions simpler sometimes we will write $K$ instead of $K(\phi)$.}
\begin{equation}
 	K(\phi) \equiv \Omega^{-1} +  \frac{3}{2} \left( \frac{\textrm{d} \ln \Omega }{\textrm{d} \phi} \right)^2 \, .
 \end{equation} 
Note that the action shown in Eq.~\eqref{non_minimal:einstein_frame} is the action for a pseudoscalar field that is minimally coupled to gravity but that has a non-standard kinetic term.\footnote{Pseudoscalar theories from string theory compactifications typically exhibit a canonically coupling to gravity, a coupling of the pseudoscalar to the topological term of a gauge theory as well as a non-canonical kinetic term for the pseudoscalar. It is hence in the formulation in Eq.~\eqref{non_minimal:einstein_frame} that our model may arise from string theory. For more details see appendix \ref{app:axions}.} As a consequence the treatment carried out in the rest of this section is valid for both these cases.

In the following we discuss the evolution of the system defined by the action of Eq.~\eqref{non_minimal:einstein_frame}. In particular we study both the evolution of the homogeneous background and of perturbations around this background. As a first step we start by computing the equation of motion for the gauge field. For this purpose we take the variation of the action (expressed in terms of conformal coordinates) with respect to $A_{\nu}$:
\begin{equation}
\frac{\delta \mathcal{S}}{\delta A_{\nu}}  = \partial_{\mu} \left[ \frac{\eta^{\mu\rho}\eta^{\nu\sigma}}{2}\left( \partial_{\rho} A_{\sigma} - \partial_{\sigma} A_{\rho} \right) \right] + \frac{\alpha}{4 \Lambda} \partial_{\mu} \left[ \phi \, \varepsilon^{\mu\nu\rho\sigma} \left( \partial_{\rho} A_{\sigma} - \partial_{\sigma} A_{\rho} \right) \right] = 0 \ ,
\end{equation}
where $\eta_{\mu \nu}$ is the Minkowski metric. At this point we can proceed by fixing the gauge. A convenient gauge to describe the problem is Coulomb gauge, i.e.\ $g^{\mu \nu }\partial_{\mu} A_{\nu} = 0 , A_0 = 0$ so that the equation of motion reads:
\begin{equation}
\eta^{\nu\sigma} \Box A_{\sigma} - \frac{\alpha}{\Lambda} \phi^{\prime} \varepsilon^{0\nu\rho\sigma} \partial_{\rho} A_{\sigma} = 0 \ , 
\end{equation}
where $\Box$ is the standard flat space d'Alembert operator $\Box \equiv \eta^{\mu\nu} \partial_{\mu} \partial_{\nu}$. As a consequence we can write:
\begin{equation}
\label{non_minimal:gauge_fields}
\Box \vec{A}  - \frac{\alpha}{\Lambda} \phi^{\prime} \, \vec{\nabla} \times \vec{A} = 0 \ , 
\end{equation}
where $\vec{\nabla} \times $ is the usual curl operator and a prime is used to denote a derivative with respect to $\tau$. This equation matches with the Eq.~\eqref{review:eq_motionA} and thus the solution for the gauge fields is exactly the same as in the case of a minimal coupling to gravity.

We can now consider the variation of the action (again in conformal time) with respect to $\phi$ to get the equation of motion for the inflaton:
\begin{equation}
	\partial_\mu \left( \eta^{\mu \nu} K(\phi) a^2 \, \partial_\nu \phi \right) -a^4 \left( \frac{\partial V_E}{\partial \phi}(\phi) + \frac{\alpha}{4 \Lambda} F^{\mu\nu}\tilde{F}^{\mu\nu}\right) = 0 \ ,
\end{equation}
which can be expressed as:
\begin{equation}
	\label{non_minimal:eom_phi_general}
	- K(\phi) \Box\phi -\frac{K_{,\phi}(\phi)}{2} \eta^{\mu\nu} \, \partial_\mu \, \phi \partial_\nu \phi+ 2a \mathcal{H} K(\phi)\phi^\prime + a^2 \, V_{E,\phi}(\phi) + a^2 \frac{\alpha}{\Lambda} \langle \vec{E}\cdot \vec{B}\rangle = 0  \, ,
\end{equation}
where ${}_{,\phi} $ is used to denote a derivative with respect to $\phi$, primes are used to denote a derivatives with respect to $\tau$ and $\mathcal{H}\equiv a^\prime/a$. At this point we can express $\phi$ as:
\begin{equation}
	\phi(\tau,\vec{x}) = \phi_0(\tau) + \delta \phi(\tau,\vec{x}) \ , 
\end{equation}
where $\phi_0(\tau)$ is the (homogeneous) background solution and $\delta \phi(\tau,\vec{x})$ is a small perturbation. By substituting this parametrization into Eq.~\eqref{non_minimal:eom_phi_general} it is easy to show that the equation of motion for the background field $\phi_0$ reads:
\begin{equation}
	\label{non_minimal:background_conformal}
	K(\phi_0) \phi_0^{\prime \prime} + \frac{K_{,\phi}(\phi_0)}{2} \phi_0^{\prime\, 2} + 2a \mathcal{H} K(\phi_0)\phi_0^\prime + a^2 \, V_{E,\phi}(\phi_0) + a^2 \frac{\alpha}{\Lambda} \langle \vec{E}\cdot \vec{B}\rangle = 0  \, .
\end{equation}
Conversely, the linearized equation of motion for the perturbation reads:
\begin{equation}
\begin{aligned}
	\label{non_minimal:perturbations}
	&\delta\phi^{\prime \prime} - \vec{\nabla}^2\delta\phi +\frac{1}{2}\frac{\textrm{d}}{\textrm{d} \phi} \left(\frac{\textrm{d} \ln K}{\textrm{d} \phi}\right) \phi_0^{\prime\, 2} \delta\phi + \frac{\textrm{d} \ln K}{\textrm{d} \phi} \phi_0^{\prime} \delta\phi^\prime + 2a \mathcal{H} \delta \phi^\prime +  \\ 
	&+ a^2 \frac{\textrm{d}}{\textrm{d} \phi} \left( \frac{V_{E,\phi}}{K}\right)\delta \phi  - \frac{a^2}{K^2} K_{,\phi} \frac{\alpha}{\Lambda} \langle \vec{E}\cdot \vec{B}\rangle \delta \phi + \frac{a^2}{K} \frac{\alpha}{\Lambda} \delta\left( \langle \vec{E}\cdot \vec{B}\rangle\right) = 0 \ .
\end{aligned}
\end{equation}
A detailed analysis on the evolution of the background and of the inflaton perturbation is presented in Sec.~\ref{sec:background} and in Sec.~\ref{sec:perturbations} respectively. 

\subsection{Background evolution}
\label{sec:background}
As a first step we express Eq.~\eqref{non_minimal:background_conformal} in terms of cosmic time ($\textrm{d}t = a \, \textrm{d}\tau$) and we divide by $K(\phi_0)$ to get:
\begin{equation}
	\label{non_minimal:background_cosmic}
	\ddot{\phi}_0 + \frac{1}{2} \frac{\textrm{d} \ln K}{\textrm{d} \phi_0} \dot{\phi}_0^2 + 3 H \dot{\phi}_0 + \frac{V_{E,\phi_0}}{K} + \frac{\alpha}{\Lambda K} \langle \vec{E}\cdot \vec{B}\rangle = 0 \ .
\end{equation}
We can also express Friedmann equation as:
\begin{equation}
	\label{non_minimal:background_friedmann}
	3 H^2 = K \frac{\dot{\phi}_0^2}{2} + V_E + \frac{1}{2} \langle \vec{E}^2 + \vec{B}^2 \rangle \ .
\end{equation}
At this point we can solve Eq.~\eqref{non_minimal:gauge_fields} and plug the solution into Eq.~\eqref{non_minimal:background_cosmic} and Eq.~\eqref{non_minimal:background_friedmann} to get:\footnote{Notice that these expressions for $\langle \vec{E}\cdot \vec{B}\rangle$ and $\langle \vec{E}^2 + \vec{B}^2 \rangle$ only hold for sizable values for $\xi$ (i.e.\ for $\xi \gtrsim 1$). Different expressions should be used for small values of $\xi$, see~\cite{Anber:2009ua}.}
\begin{eqnarray}
	\label{non_minimal:background_scalar}
	\ddot{\phi}_0 + \frac{1}{2} \frac{\textrm{d} \ln K}{\textrm{d} \phi_0} \dot{\phi}_0^2 + 3 H \dot{\phi}_0 + \frac{V_{E,\phi_0}}{K} - 2.4 \cdot 10^{-4} \, \frac{\alpha}{\Lambda K} \frac{H^4}{\xi^4} e^{2 \pi \xi} = 0 \ , \\
	\label{non_minimal:background_friedmann_2}
	3 H^2 = K \frac{\dot{\phi}_0^2}{2} + V_E + 1.4 \cdot 10^{-4} \frac{H^4}{\xi^3} e^{2 \pi \xi}	 \ ,
\end{eqnarray}
with the parameter $\xi$ defined as in Eq.~\eqref{review:xi},
\begin{equation}
	\label{non_minimal:xi_def}
	\xi \equiv \frac{\alpha}{2 \Lambda} \, \frac{|\dot{\phi}_0|}{H} \ .
\end{equation}
Note that here $\phi_0$ is not the canonically normalized field. If the field $\phi_0$ is slow-rolling these equations can be approximated as:
\begin{eqnarray}
	\label{non_minimal:slow_roll_gauge}
	 3 H \dot{\phi}_0 + \frac{V_{E,\phi_0}}{K} - 2.4 \cdot 10^{-4} \, \frac{\alpha}{\Lambda K} \frac{H^4}{\xi^4} e^{2 \pi \xi} \simeq 0 \ , \\
	 	\label{non_minimal:slow_roll_friedmann}
	3 H^2 \simeq V_E + 1.4 \cdot 10^{-4} \frac{H^4}{\xi^3} e^{2 \pi \xi} \ ,
\end{eqnarray}
so that in the first part of the evolution (where $\xi \simeq 1$) we recover the usual slow-roll equations:
\begin{equation}
	\label{non_minimal:slow_roll}
	\dot{\phi}_0 \simeq - \frac{V_{E,\phi_0}}{3 H K} \ , \qquad \qquad 3 H^2 \simeq V_E \ .
\end{equation}
As in Sec.~\ref{sec:review}, the back-reaction of the gauge fields on the Friedmann equation (i.e.\ Eq.~\eqref{non_minimal:slow_roll_friedmann}) is negligible throughout the whole evolution, but it may have a significant impact on Eq.~\eqref{non_minimal:slow_roll_gauge}. As above, the $\langle \vec{E}\cdot \vec{B} \rangle$ term plays the role of a friction term which grows exponentially with $\xi$. Notice that as long as the gauge field contribution is not too strong we find for the first slow-roll parameter $\epsilon_H$,\footnote{As long as the back-reaction of the gauge fields is negligible, this definition is equivalent to the more commonly used `potential' slow-roll parameter, $\epsilon_H \simeq \epsilon_V = m_p^2 (V'(\phi)/V)^2/(2 K)$.}
\begin{equation}
	\label{non_minimal:epsilon_H}
	\epsilon_H \equiv -\frac{\dot{H}}{H^2} \simeq \frac{1}{2} K(\phi_0) \left( \frac{\dot{\phi}_0}{H} \right)^2 \propto K(\phi_0) \xi^2 \ ,
\end{equation}
where we have used Eq.~\eqref{non_minimal:slow_roll} and Eq.~\eqref{non_minimal:xi_def}. As for a vast majority of slow-roll inflationary models $\epsilon_H$ grows during inflation,\footnote{In particular this happens for 
all the models considered in Sec.~\ref{sec:attractors} (see Fig.~\ref{fig:potentials_and_xi}).} for sufficiently large values of $\alpha/\Lambda$ we expect a strong gauge field production that strongly affects the dynamics towards the end of inflation. However, it is crucial to stress that in general $\xi$ is not \emph{monotonically} increasing for all the slow-roll inflationary models. For example, if both  $\epsilon_H$ and $K(\phi_0)$ monotonically increase during inflation, their interplay may lead to a local maximum in $\xi$, if the growth in $K(\phi_0)$ dominates the growth of $\epsilon_H$ at late times. Such a scenario will be particularly relevant for the discussion of PBHs in Sec.~\ref{sec:PBHs}. 

\subsection{Perturbations}
\label{sec:perturbations}
In order to study the evolution of the inflaton perturbations we start by expressing the $\delta\left( \langle \vec{E}\cdot \vec{B}\rangle\right)$ term appearing in Eq.~\eqref{non_minimal:perturbations} in a convenient form. In particular following the treatment of~\cite{Anber:2009ua} we proceed by using:
\begin{equation}
\label{non_minimal:gauge_field_contrib}
\delta\left( \langle \vec{E}\cdot \vec{B}\rangle\right) = [\vec{E} \cdot \vec{B} - \langle \vec{E} \cdot \vec{B} \rangle]_{\delta \phi = 0} + \frac{\partial \langle \vec{E} \cdot \vec{B} \rangle}{\partial \dot{\phi}_0}\frac{ \delta \phi^{\prime} }{a} \equiv \delta_{\vec{E}\cdot \vec{B}} \, -  \left( \frac{ \pi \alpha}{ \Lambda H } \right)  \langle \vec{E} \cdot \vec{B} \rangle \, \frac{\delta \phi^\prime}{a} \ .
\end{equation}
We can thus proceed by substituting this expression into Eq.~\eqref{non_minimal:perturbations} (expressed in terms of cosmic time) to get:
\begin{equation}
\label{non_minimal:scalar_fluctuations}
\begin{aligned}
	&\ddot{\delta\phi} - \frac{\vec{\nabla}^2}{a^2}\delta\phi +\frac{1}{2} \frac{\textrm{d}}{\textrm{d} \phi_0} \left(\frac{\textrm{d} \ln K}{\textrm{d} \phi_0}\right) \dot{\phi}_0^{ 2} \delta\phi + \frac{\textrm{d} \ln K}{\textrm{d} \phi_0} \dot{\phi}_0 \dot{\delta\phi} + 3H \dot{\delta \phi} + \frac{\textrm{d}}{\textrm{d} \phi_0} \left( \frac{V_{E,\phi_0}}{K}\right)\delta \phi \\ 
	& - \frac{1}{K} \frac{\textrm{d} \ln K}{\textrm{d} \phi_0}  \frac{\alpha}{\Lambda} \langle \vec{E}\cdot \vec{B}\rangle \delta \phi + \frac{1}{K} \frac{\alpha}{\Lambda} \left[ \delta_{\vec{E}\cdot \vec{B}} \, -  \left( \frac{ \pi \alpha}{ \Lambda H } \right)  \langle \vec{E} \cdot \vec{B} \rangle \dot{\delta \phi} \right] = 0 \ .
\end{aligned}
\end{equation}
Let us start by discussing the regime where the gauge field contribution is negligible before we proceed by considering the strong gauge field regime. If the gauge field contribution is negligible the scalar power spectrum $\Delta^2_s(k)$ is simply given by the standard vacuum amplitude:
\begin{equation}
	\left.   \Delta^2_s (k) \right|_\text{vac} \simeq   \frac{ 1 }{8 \, \pi^2 } \frac{H^2}{  c_s \ \epsilon_H }\ , 
\end{equation}
where $\epsilon_H$ is the first (Hubble) slow-roll parameter defined in Eq.~\eqref{non_minimal:epsilon_H} and $c_s$ is the speed of sound that for all the models considered in this work is equal to one. Using Eq.~\eqref{non_minimal:slow_roll} it is easy to show that the vacuum amplitude can be expressed as
\begin{equation}
	\label{non_minimal:vacuum}
	\left.   \Delta^2_s (k) \right|_\text{vac} \simeq   \frac{ K V }{12 \, \pi^2 }  \left( \frac{\textrm{d} \ln V_E}{\textrm{d} \phi_0}\right)^{-2} \ .
\end{equation}
Conversely, in the strong gauge field regime the gauge-field induced contributions (i.e.\ the terms proportional to $\langle \vec{E}\cdot\vec{B} \rangle $) are large. At this point we can neglect all the higher order terms in the slow-roll parameters as well as all the terms that are not exponentially enhanced\footnote{For more details on the derivation of this formula and for consistency checks see appendix~\ref{sec:spectrum}.} and the scalar power spectrum in the strong gauge field regime can be expressed as:
 \begin{equation}
 	\label{non_minimal:gauge_sourced}
 	\left.  \Delta^2_s (k) \right|_\text{gauge} \simeq  \left(\frac{\alpha \langle \vec{E}^a \cdot \vec{B}^a  \rangle / \sqrt{\mathcal{N}} }{3 \, b \, \Lambda \, \dot{\phi}_0 \, H \, K }\right)^2  \ ,
 \end{equation}
  where we have defined:
\begin{equation}
	b \equiv  1  -  \pi \left( \frac{\alpha}{\Lambda} \right)^2  \frac{\langle \vec{E}^a \cdot \vec{B}^a  \rangle}{3 \, H^2 K} \ ,
	\label{eq:b}
\end{equation}
and we have generalized the result to the case where $\mathcal{N}$ Abelian gauge fields are coupled to the inflaton. Finally we can use both Eq.~\eqref{non_minimal:vacuum} and Eq.~\eqref{non_minimal:gauge_sourced} to get the complete expression for the scalar power spectrum:
 \begin{equation}
 	\label{non_minimal:scalar_final}
 	\Delta^2_s (k) \simeq  \frac{ 1 }{K}  \left( \frac{H^2 }{2 \pi |\dot{\phi}_0|}\right)^{2} + \left(\frac{\alpha \langle \vec{E}^a \cdot \vec{B}^a  \rangle / \sqrt{\mathcal{N}} }{3 \, b \, \Lambda \, \dot{\phi}_0 \, H \, K }\right)^2 \ .
 \end{equation}
In the regime where the gauge fields dominate the evolution we can approximate $b$ by simply neglecting the constant term (i.e.\ the first term in Eq.~\eqref{eq:b}). One can then easily show that:
\begin{equation}
	\label{non_minimal:scalar_strong_approx}
 	\left.  \Delta^2_s (k) \right|_{\text{gauge}} \simeq  \frac{  1 }{ 4 \pi^2 \xi^2 \mathcal{N} }  \ ,
 \end{equation}
which corresponds (at least in form) to the expression given in Eq.~\eqref{review:scalar_strong}. There are some conceptual (and indeed physical) differences that must be pointed out:
\begin{itemize}
	\item While Eq.~\eqref{review:scalar_strong} is expressed in terms of a canonically normalized field, Eq.~\eqref{non_minimal:scalar_strong_approx} is expressed in terms of the non-canonically normalized field. As a consequence, in order to match the two equations we should express the dynamics in terms of a canonically normalized field $\varphi$ defined as $(\textrm{d} \varphi / \textrm{d} \phi_0)^2 \equiv K(\phi_0)$. Using this definition Eq.~\eqref{non_minimal:scalar_strong_approx} can be turned into:
	\begin{equation}
	\label{non_minimal:scalar_strong_approx_varphi}
 	\left.  \Delta^2_s (k) \right|_\text{gauge} \simeq  \frac{  K(\phi_0) }{ 4 \pi^2 \mathcal{N}   \tilde{\xi}^2 }  \propto \frac{K}{\epsilon_H} ,
 \end{equation}
	where $\tilde{\xi} \equiv \alpha |\dot{\varphi}| / (2 \Lambda H )$. The difference between Eq.~\eqref{non_minimal:scalar_strong_approx_varphi} and Eq.~\eqref{review:scalar_strong} is now obvious: in terms of the canonical field $\varphi$ in the Einstein frame, the introduction of a non-minimal coupling to gravity implies that the coupling to the gauge fields is altered as $\dot \varphi \mapsto \dot \varphi/\sqrt{K}$ and hence all effects sourced by the gauge fields are suppressed as $K$ is increased. This can be traced back to the choice of coupling the Chern-Simons term to the canonically normalized field $\phi_0$ in Jordan frame in Eq.~\eqref{non_minimal:jordan_frame}.
	
	\item In the minimally coupled case with monotonically increasing $\epsilon_H$, Eq.~\eqref{review:scalar_strong} holds from some critical value until the end of inflation. Now, if $K(\phi_0)$ strongly increases towards the end of inflation, a monotonic growth of $\epsilon_H$ no longer implies a monotonic growth of $\xi \propto \sqrt{\epsilon_H/K(\phi_0)}$, see Eq.~\eqref{non_minimal:epsilon_H}. Consequently, $\xi$ can be strongly suppressed towards the end of inflation, shutting off the tachyonic instability for the gauge fields. In particular after the usual nearly scale invariant power spectrum of slow-roll inflation at CMB scales and the following gauge-field induced increase, we can achieve another regime with $ b \simeq 1$ and thus $\Delta^2_s (k)$ is much smaller than the prediction of Eq.~\eqref{non_minimal:scalar_strong_approx}.\footnote{Note that Eq.~\eqref{non_minimal:scalar_strong_approx} only holds for $b \gg 1$. For $b \simeq 1$, a reduction in $\xi$ leads to an (exponential) reduction in the scalar power spectrum.} In this setup, the scalar power spectrum thus features a bump. As discussed in~\cite{Clesse:2015wea} such a particular shape for the spectrum can be extremely interesting for the generation of PBHs. A detailed discussion of this topic is carried out in Sec.~\ref{sec:PBHs}.

	\item The analysis performed here is complementary to the proposal of Ref.~\cite{Garcia-Bellido:2016dkw}, where a bump in the scalar spectrum (and hence in the PBH spectrum) is achieved based on the action~\eqref{review:action_pseudoscalar}. Instead of introducing a non-minimal coupling to gravity as we do here, they instead modify the scalar potential so that the velocity of the inflaton (and hence the parameter $\xi$) is reduced in the last stages of inflation. On the contrary, after a conformal transformation to the Einstein frame and a canonical normalization of the inflaton field, our scalar potential is essentially featureless. The non-trivial evolution of $\xi$ in this frame is sourced by the coupling between the inflaton and gauge fields.
\end{itemize}

Before concluding this section briefly discuss the spectrum $\Delta_t^2$ of GWs that are generated in these generalized models of inflation. In order to compute the shape of the spectrum we  start once again by considering the linearized Einstein equation. In particular, it can be expressed as in Eq.~\eqref{review:linearized_einstein}, where $T_{\mu\nu}$ only depends on the gauge fields.\footnote{Notice that $\sqrt{-g} \phi F_{\mu \nu} \tilde{F}^{\mu \nu} $ does not depend on $g_{\mu \nu}$ and $h^{ij}\partial_i \phi \partial_j \phi = 0$ at first order.} Moreover, as \begin{equation}
 \sqrt{-g_J} \, g_J^{\mu \rho} g_J^{\nu \sigma} F_{\mu \nu} F_{\rho \sigma} = \sqrt{-g} \, g^{\mu \rho} g^{\nu \sigma} F_{\mu \nu} F_{\rho \sigma}  \ , 
\end{equation}
the source term for GWs is not affected by the presence of a non-minimal coupling between the inflaton and gravity. As a consequence the spectrum of GWs is once again given by Eq.~\eqref{review:OmegaGW}. However, it is crucial to stress that while the expression of the spectrum has exactly the dependence on $\xi$ as in Sec.~\ref{sec:review}, the evolution of $\xi$ is now expected to be different and thus (consistently with the examples shown in Sec.~\ref{sec:attractors}) we expect the spectrum to be different.

\section{An example: attractor models\label{sec:attractors}}
In general all the single field models of inflation can be described by an Einstein frame Lagrangian that is similar to the one shown in Eq.~\eqref{non_minimal:einstein_frame} i.e.\ by:
\begin{equation}
	\label{attractors:einstein_frame}
	\mathcal{S} = \int \textrm{d}^4 x \sqrt{-g} \left[ \frac{R}{2} - K(\phi)X -V_E(\phi)  \right] \ .
\end{equation}
A particular case are the so-called \emph{T-models}~\cite{Kallosh:2013hoa,Kaiser:2013sna,Kallosh:2013daa,Kallosh:2013tua,Kallosh:2013yoa,Kallosh:2014rga,Kallosh:2015lwa} which lead to the class of $\alpha$-attractors. \emph{T-models} are described by the action of Eq.~\eqref{attractors:einstein_frame} with:
\begin{equation}
\label{eq:Tmodels}
	K(\phi) = \left(1 - \frac{\phi^2}{6 \alpha} \right)^{-2} \ , \qquad \qquad V_E(\phi) = \frac{m^2 \phi^2}{ 2} \ .
\end{equation}
Other interesting examples are the case of Higgs inflation introduce by Bezrukov and Shaposhnikov~\cite{Bezrukov:2007ep,Bezrukov:2009db} and its generalization. in terms of the so-called attractor at strong coupling~\cite{Kallosh:2013tua}: 
\begin{equation}
	\label{attractors:model_def}
	K(\phi) = \frac{1+\varsigma h(\phi) + 3/2 \, \varsigma^2 h_{,\phi}^2(\phi)}{(1+\varsigma h(\phi))^2} \ , \qquad \qquad V_E(\phi) = \lambda^4 \frac{ h^2(\phi)}{ \left( 1 + \varsigma h(\phi) \right)^2} \ ,
\end{equation}
where $\lambda$ is a mass scale (that fixes the normalization of the inflationary potential) and $h(\phi)$ is a generic function\footnote{Higgs inflation corresponds to $h(\phi)=\phi^2$.} of $\phi$.

Over the last years these models have received a lot of attention because they predict values of $n_s$ and $r$ that are in good agreement with the values that are favored by Planck data~\cite{Ade:2015xua,Ade:2015lrj}. In the case of T-models we have:
\begin{equation}
	n_s -1 = - \frac{2}{N} \ , \qquad \qquad r = \frac{12 \alpha}{N^2} \ ,
\end{equation}
and in the case of the attractor at strong coupling (i.e.~$\varsigma \gg 1$):
\begin{equation}
	n_s -1 = - \frac{2}{N} \ , \qquad \qquad r = \frac{12 }{N^2} \ .
\end{equation}
We can immediately notice that, for both these models:\footnote{However, it is crucial to stress that these expressions for $n_s -1$ and $r$ only hold for large values of $N$. This will be particularly relevant for the possibility of observing the GW spectrum generated by these models.}
\begin{equation}
\epsilon_H \simeq \frac{\mathcal{O}(1)}{N^2} \ .
\end{equation}
Interestingly, using the classification of inflation models~\cite{Mukhanov:2013tua,Roest:2013fha,Binetruy:2014zya} based on first slow-roll parameter $\epsilon_H$:
\begin{equation}
	\epsilon_H \simeq \frac{\beta_p}{N^p} + \mathcal{O}(1/N^{p+1}) \ ,
\end{equation}
all these models fall in the same universality class (i.e.\ the Starobinsky-like class\footnote{The wording `Starobinsky-like' refers here to a class of models characterized by a scalar potential which asymptotes to an exponentially flat plateau. The original Starobinsky model~\cite{Starobinsky:1980te} is constructed from the Ricci-scalar which is not a pseudoscalar quantity.} with $p=2$). As shown in~\cite{Domcke:2016bkh}, in the context of higher-dimensional couplings with Abelian gauge fields this happens to be the most promising scenario from the point of view of possible observational signatures. As a consequence it seems interesting apply the framework developed in Sec.~\ref{sec:non-minimal} to this class of models.

\begin{figure}
\centering
\includegraphics[width=0.555\textwidth]{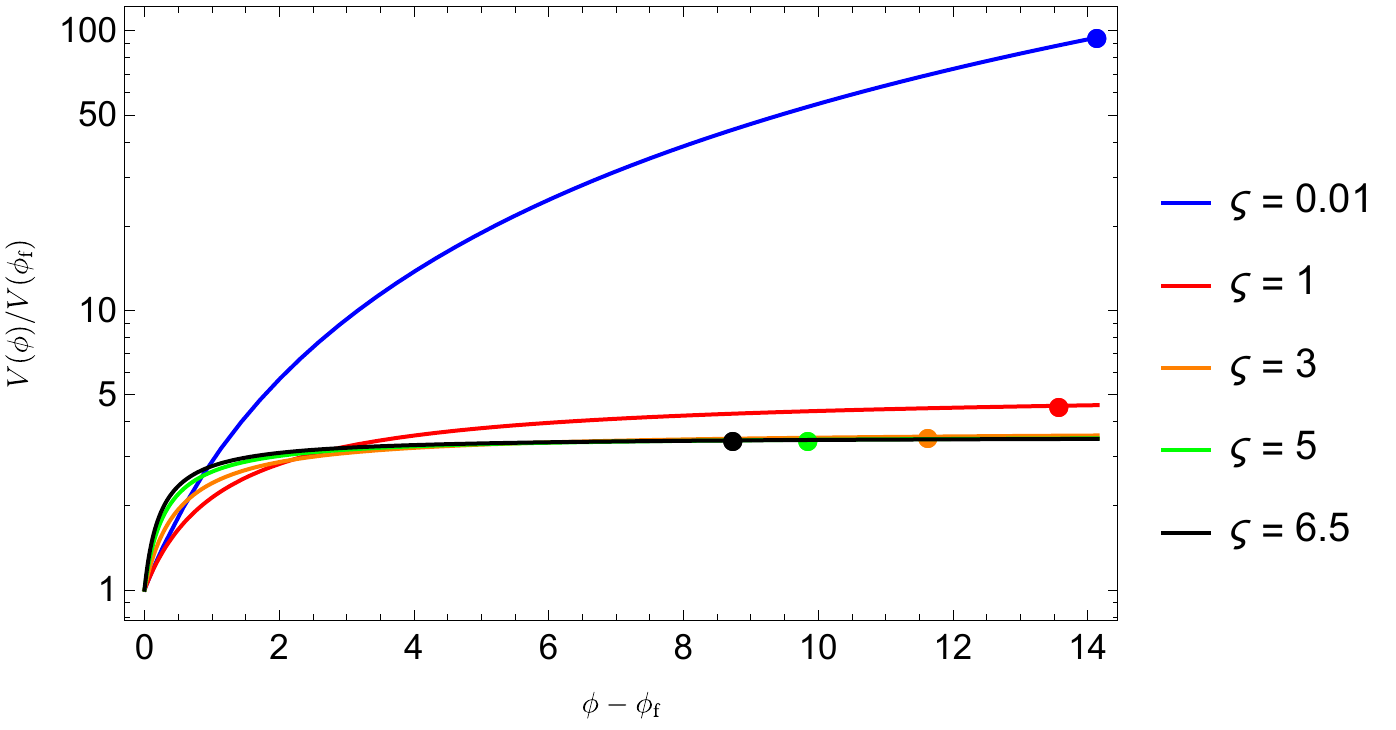}
\includegraphics[width=0.435\textwidth]{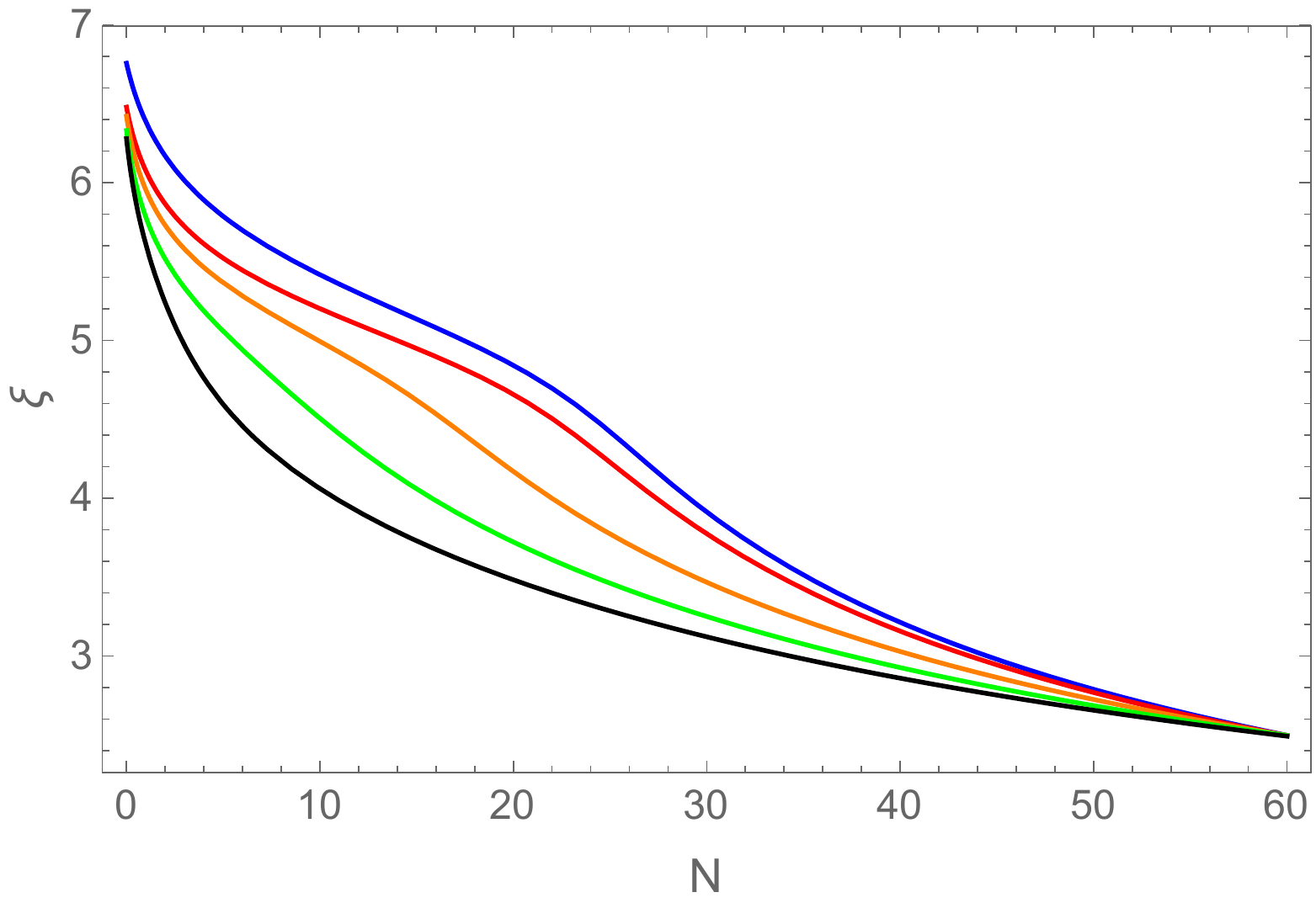}
\caption{\textbf{Left panel:} Profile of the inflationary potential~\eqref{attractors:model_example} normalized to its value at the end of inflation, $V(\phi_\textrm{f})$, as a function of the field excursion $\phi - \phi_\textrm{f}$ for different values of the non-minimal coupling $\varsigma$ to gravity. Dots are used to denote the points that corresponds to $N = 60$ in the case with $\alpha/\Lambda = 0$. \textbf{Right panel:} Evolution of the parameter $\xi$ governing the effects of the gauge fields as a function of the number of e-folds $N$ (right plot).}
\label{fig:potentials_and_xi}
\end{figure}

\begin{figure}
\centering
\includegraphics[width=0.9\textwidth]{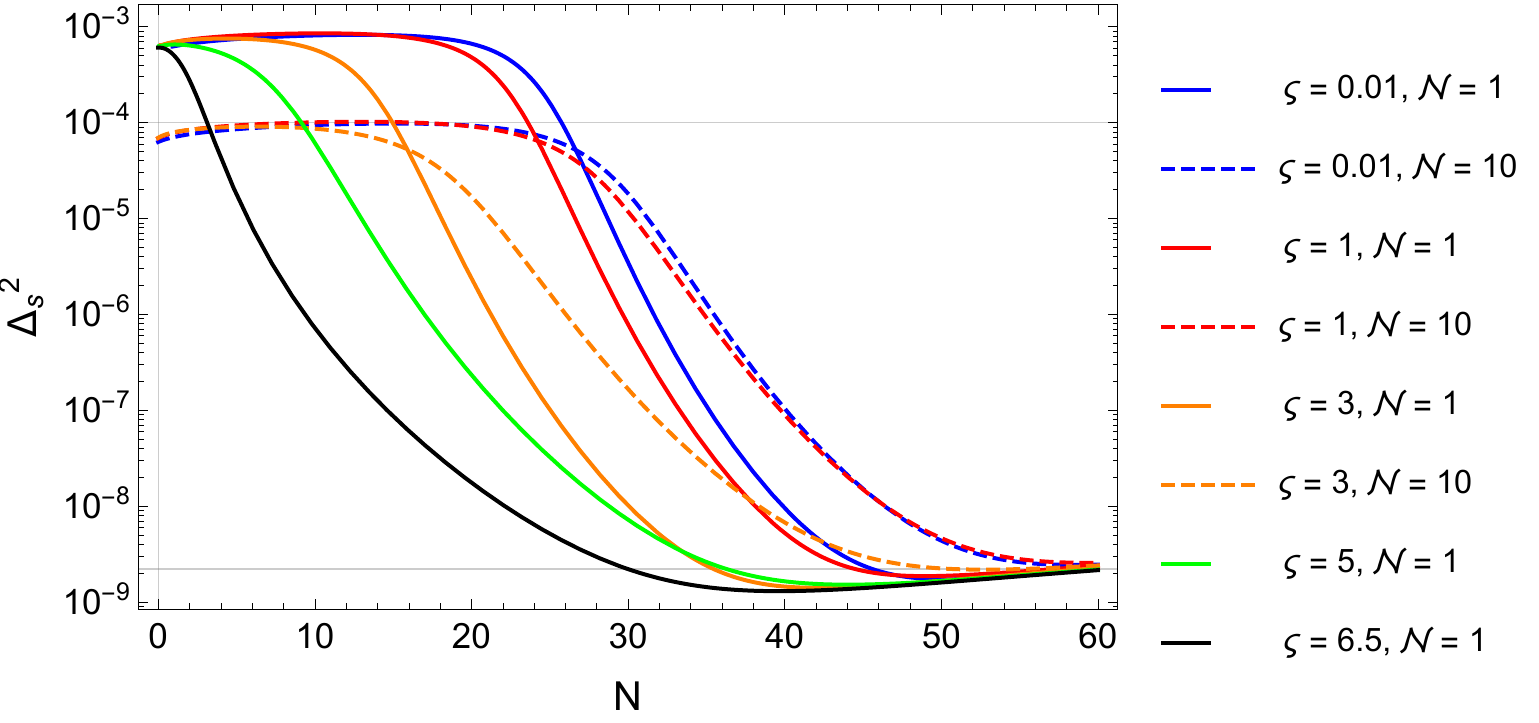}
\caption{Scalar power spectrum for all the models of Fig.~\ref{fig:potentials_and_xi} (dashed lines corresponds to models with $\mathcal{N}=10$).
 The lower horizontal line denotes the observed normalization at CMB scales (according to~\cite{Ade:2015lrj}) and the upper horizontal line is the estimate of~\cite{Linde:2012bt} for the PBH bound. More details on this bound on the generation of PBHs are reported in Sec.~\ref{sec:PBHs}.}
\label{fig:scalar}
\end{figure}

In order to study the effect of non-minimal couplings on the dynamics and on the scalar and tensor power spectra, we focus on the family of models defined by Eq.~\eqref{attractors:model_def}. As a first example, let us consider models with $h(\phi) = \phi$ so that:
\begin{equation}
	\label{attractors:model_example}
	K(\phi) = \frac{1+\varsigma \phi + 3/2 \, \varsigma^2 }{(1+\varsigma \phi)^2} \ , \qquad \qquad V_E(\phi) = \lambda^4 \frac{ \phi^2}{ \left( 1 + \varsigma \phi \right)^2} \ .
\end{equation}
Notice that for $\varsigma = 0 $ this model simply reduces to the case of chaotic inflation with $V(\phi) = \frac{1}{2} m^2 \phi^2$ i.e.\ it corresponds to the $p=1$ class of~\cite{Domcke:2016bkh}. 
By increasing the value of $\varsigma$ we progressively flatten the potential, see left panel of Fig.~\ref{fig:potentials_and_xi}.\footnote{For all the models shown in this plot the constant $\lambda$ is fixed in order to respect the observed normalization of the scalar spectrum (according to~\cite{Ade:2015lrj}) at CMB scales. In particular this can be noticed by looking at the plot of Fig.~\ref{fig:scalar}, where we show the scalar spectra associated with all the different models discussed in this section.} This clearly induces a strong modification on the evolution of the inflaton field. In particular, the field excursion during inflation is decreased as $\varsigma$ increases and the evolution of $\xi$ is altered compared to the minimally coupled case (see right panel of Fig.~\ref{fig:potentials_and_xi}).
 Notice that for increasing values of $\varsigma$ the growth of $\xi$ (as a function of $N$) becomes slower. In particular the suppression of $\xi$ over a wide range of e-folds leads to a strong suppression in both the scalar and tensor power spectra.

\begin{figure}
\centering
\includegraphics[width=0.9\textwidth]{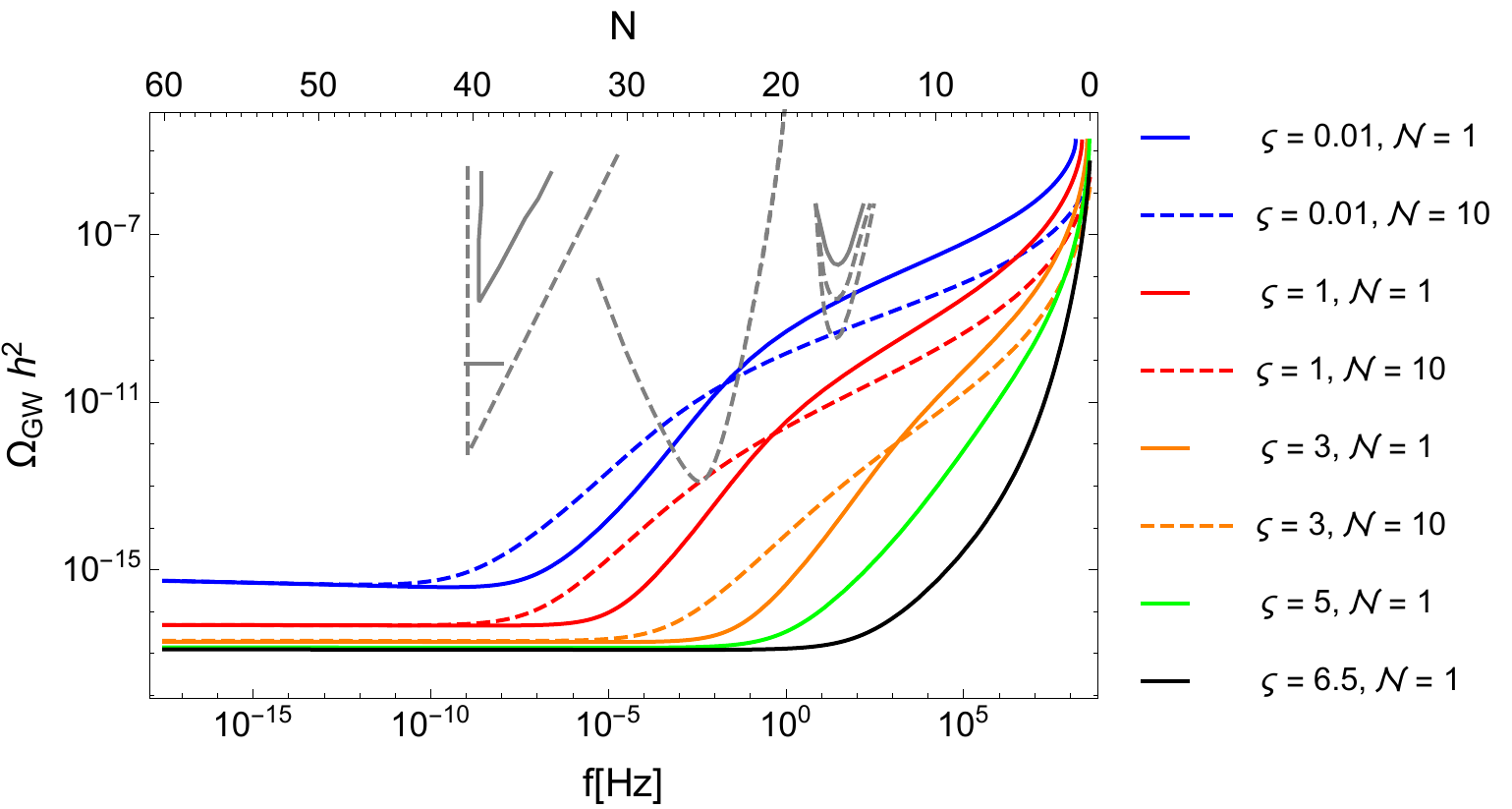}
\caption{Power spectrum of tensor perturbations for all the models of Fig.~\ref{fig:potentials_and_xi} (dashed lines corresponds to models with $\mathcal{N}=10$). We are also showing the sensitivity curves for (from left to right): milli-second pulsar timing, LISA, advanced LIGO. Current bounds are denoted by solid lines, expected sensitivities of upcoming experiments by dashed lines. See main text for details on these curves.}
\label{fig:GW}
\end{figure}

As described in Sec.~\ref{sec:non-minimal}, the changes in the tensor spectrum are simply due to the different evolution of $\xi$. In particular, as for increasing values of $\varsigma$ the growth of $\xi$ becomes slower over a broad range of $N$-values, we expect the GW spectrum (which is exponentially sensitive to $\xi$) to be strongly suppressed. This effect is clearly visible in Fig.~\ref{fig:GW} where we compare the GW spectra for the models of Fig~\ref{fig:potentials_and_xi} with the sensitivity curves of some current (solid lines) and upcoming (dashed lines) direct GW detectors. For the millisecond pulsar timing arrays covering frequencies around $10^{-10}$~Hz, we show the constraint depicted in Ref.~\cite{Smith:2005mm}, the update from EPTA~\cite{vanHaasteren:2011ni} and the expected sensitivity of SKA~\cite{Kramer:2004rwa}. The laser interferometer space antenna (LISA)~\cite{LISA} is a future mission designed to probe the mHz range~\cite{Petiteau, Caprini:2015zlo,Bartolo:2016ami}, where as ground-based detectors are sensitive at a few 10 Hz (LIGO/VIRGO~\cite{TheLIGOScientific:2016wyq}). It is interesting to notice that for $\varsigma \sim 1$ the spectrum is marginally detectable (for $\mathcal{N}=10$) by LISA but not by advanced LIGO. For larger values of $\varsigma$ the suppression of the spectrum is so strong that the signal is well below the expected sensitivities of upcoming GW detectors.

In Fig.~\ref{fig:scalar} we show the scalar power spectrum (corresponding to Eq.~\eqref{non_minimal:scalar_final}) for all the models shown in the previous plots. Notice that, in order to respect the COBE normalization (that sets the amplitude of the spectrum  at CMB scales) all the spectra meet at the same value for $N\simeq 60$. Moreover, it is worth mentioning that all the models shown in this section are in agreement\footnote{Except for the model with $\varsigma = 0.01$, for which the predicted value of $r$ is marginally excluded by Planck~\cite{Ade:2015lrj} at $95\%$ CL.} with the constraints set by Planck on $n_s$ and $r$ and on the generation of primordial non-Gaussianities.\footnote{ This can directly be checked from Fig.~\ref{fig:potentials_and_xi}, $\xi|_\text{CMB} \lesssim 2.5$.} Interestingly, for none of the models shown in this plot the spectrum features a peak. This can be explained by considering the expression of $K(\phi)$ in terms of $N$. While for $\varsigma \ll 1 $ $K(\phi)$ is essentially constant, for sizable values of $\varsigma$ we have $K(\phi) \propto 1/(\varsigma N)^2$ to leading order in $N$. By numerically solving the evolution it is possible to show that $K(\phi)$ becomes large only for $N\simeq 2 \div 4 $ depending on the model, and the suppression of $\xi$ is not strong enough to disrupt its monotonic growth before the end of inflation. As a consequence, in order for the spectrum to feature a peak, we require a strong growth of $K(\phi)$ at sufficiently large values of $N$. Such a behavior can be achieved by considering different parameterizations for $h(\phi)$.

As discussed in Ref.~\cite{Domcke:2016bkh}, the scalar and tensor spectra in axion inflation can be well understood in terms of universality classes of inflation~\cite{Mukhanov:2013tua,Roest:2013fha,Binetruy:2014zya}, categorized by the scaling of $\epsilon_H(N)$ in the regime where the back-reaction of the gauge fields is weak. This is immediately understood (for a minimal coupling to gravity) since the growth of both the scalar and tensor spectrum is driven by $\xi \propto \sqrt{\epsilon_H}$. However, when a non-minimal coupling is considered, the instability is controlled by $\xi \propto \sqrt{\epsilon_H/K}$ (see Eq.~\eqref{non_minimal:epsilon_H}). Whereas the examples depicted so far correspond to $\epsilon_H/K \propto 1/N^p$ with $p \lesssim 1$~\cite{Binetruy:2014zya},\footnote{While it is easy to see that $p = 1$ for $\varsigma = 0$ (see~\cite{Binetruy:2014zya}), this statement may seem surprising for $\varsigma \neq 1$, when the scalar potential asymptotically approaches the exponentially flat Starobinsky-like potential (that for a minimal coupling is characterized by $\epsilon_H \propto 1/N^2$~\cite{Binetruy:2014zya}). This subtlety arises due to the coupling of the gauge fields to the canonically normalized inflaton field of the Jordan frame. By considering the expression of $K(\phi)$ given in Eq.~\eqref{attractors:model_def} and using the definition $N \equiv \int \sqrt{K/2\epsilon_H} \, \textrm{d}\phi$, one finds that $\xi \propto \sqrt{\epsilon_H/K}$ is in fact independent of $N$ at small values of $N$ (ignoring here a possible black-reaction of the gauge-fields).} 
different behaviors can be achieved. In particular here, in order to obtain a scalar spectrum which rises at larger values of $N$ (i.e.\ at an earlier stage during inflation) we require a quicker increase of  $\epsilon_H(N)$, with decreasing $N$. At the same time, in order to obtain a suppression of $\xi$ which leads to the appearance of a bump in the spectra, we also require the kinetic function $K(N)$ to become sizable at sufficiently large values of $N$. A simple example of this type can obtained by choosing:
\begin{equation}
	\label{attractors:inv_definition}
 	h(\phi) = (1 - 1/\phi) \ , \qquad \qquad  V_E(\phi) = \lambda^4 \frac{ (\phi-1)^2}{ \left( \phi + \varsigma (\phi-1) \right)^2} \ ,
 \end{equation} 
 which for a vanishing non-minimal coupling and a vanishing coupling to gauge fields ($\varsigma = 0$, $\alpha = 0$) leads to $\epsilon_H \propto 1/N^{4/3}$ while the kinetic function $K(N)$ interpolates from an approximately constant behaviorw at large $N$ to $K(N) \propto e^{-N}$ at small $N$. We will also return to models of this type when discussing PBH dark matter.

\begin{figure}[t]
\centering
\includegraphics[width=0.9\textwidth]{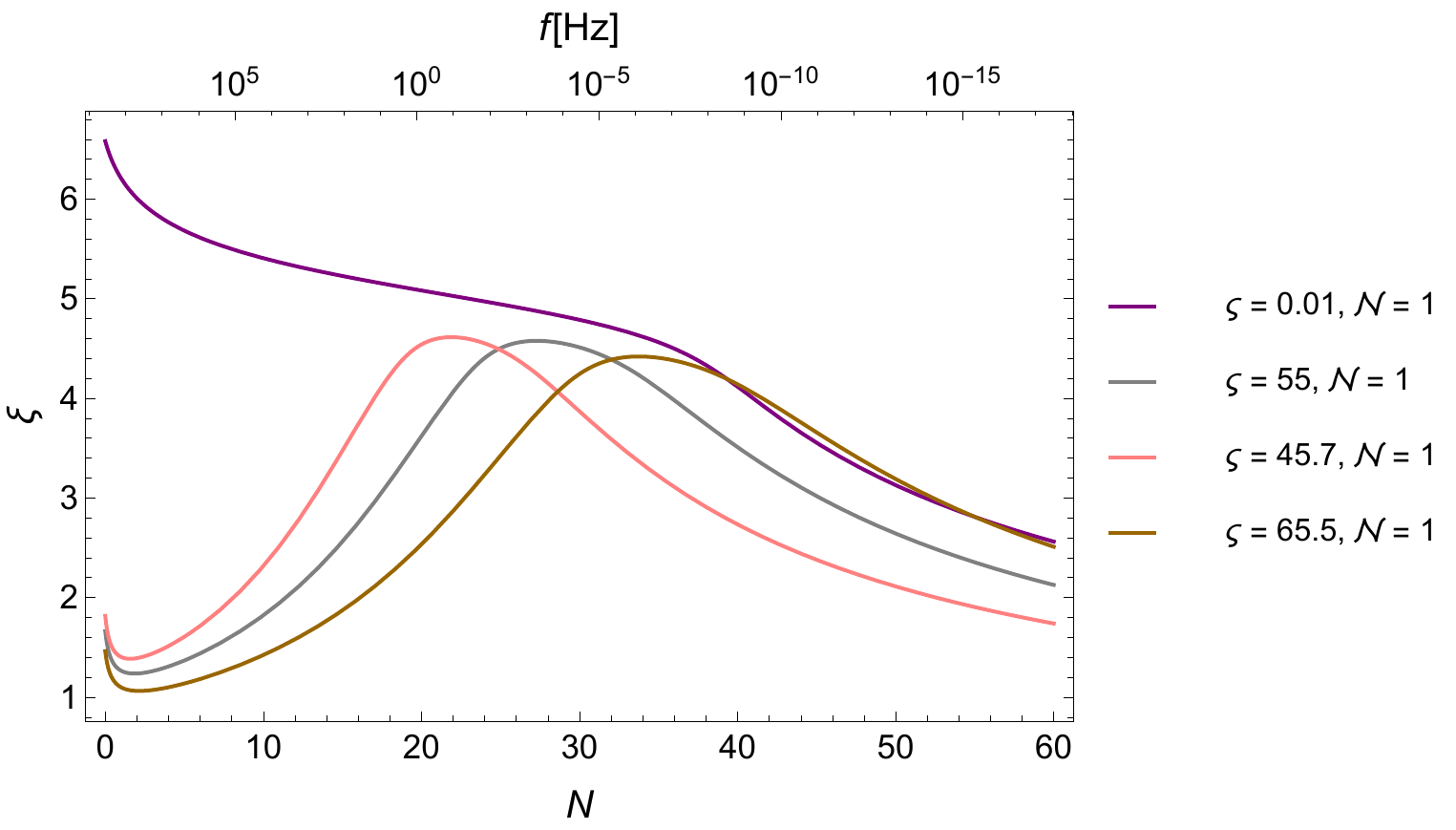}

\caption{Evolution of $\xi$ as a function of $N$ for models based on Eq.~\eqref{attractors:inv_definition}. The model parameters are chosen to obey CMB and PBH constraints, see main text for details. In particular,  the depicted curves are obtained for couplings to the gauge-fields of $\alpha/\Lambda \simeq \{ 97, 63, 71, 74 \}$ (for increasing value of the non-minimal coupling to gravity $\varsigma$, respectively).}
\label{fig:xi_inv}
\end{figure}

\begin{figure}[htb!]
\centering
\includegraphics[width=0.97\textwidth]{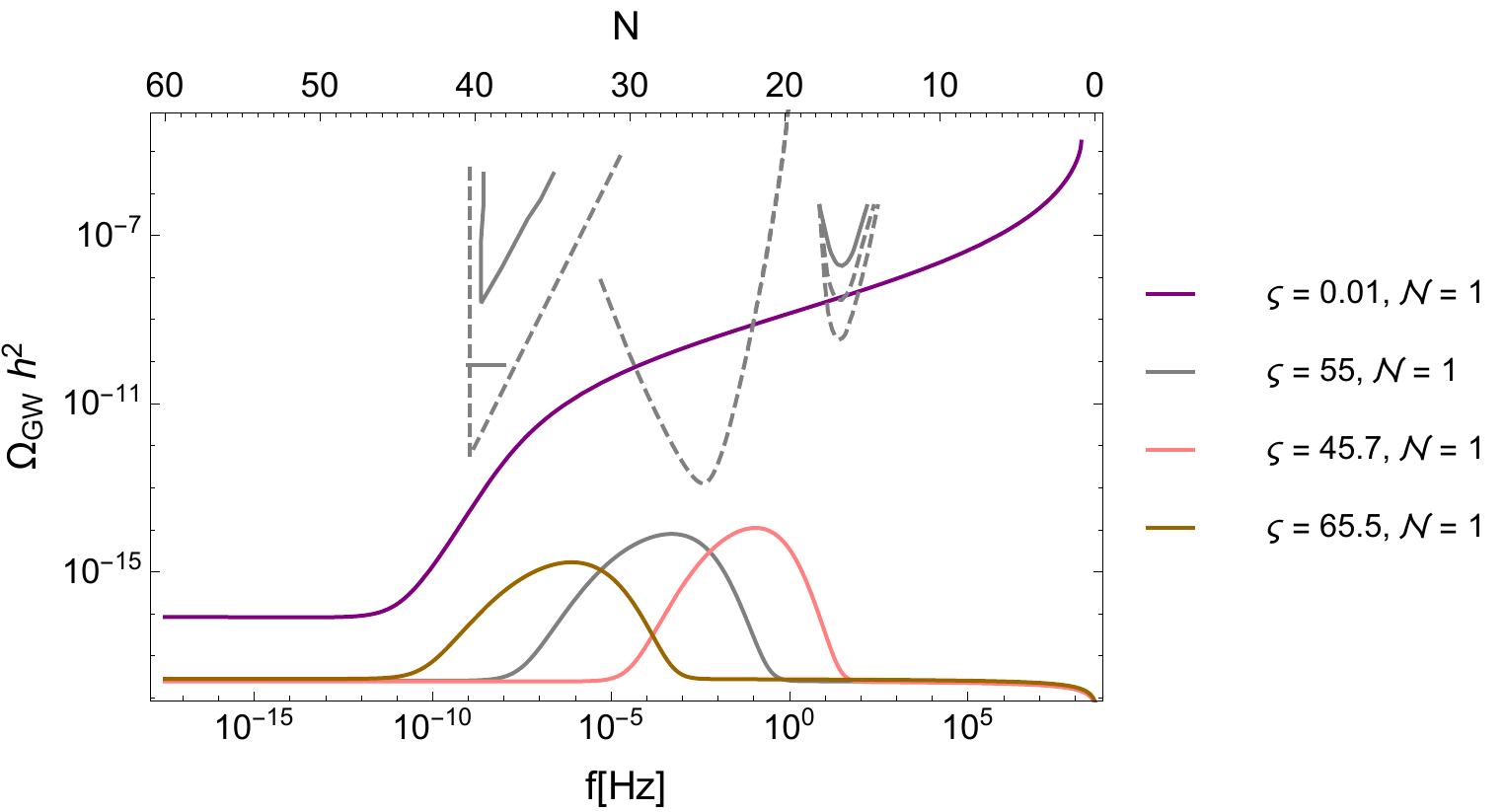}\vspace{5mm}
\includegraphics[width=0.97\textwidth]{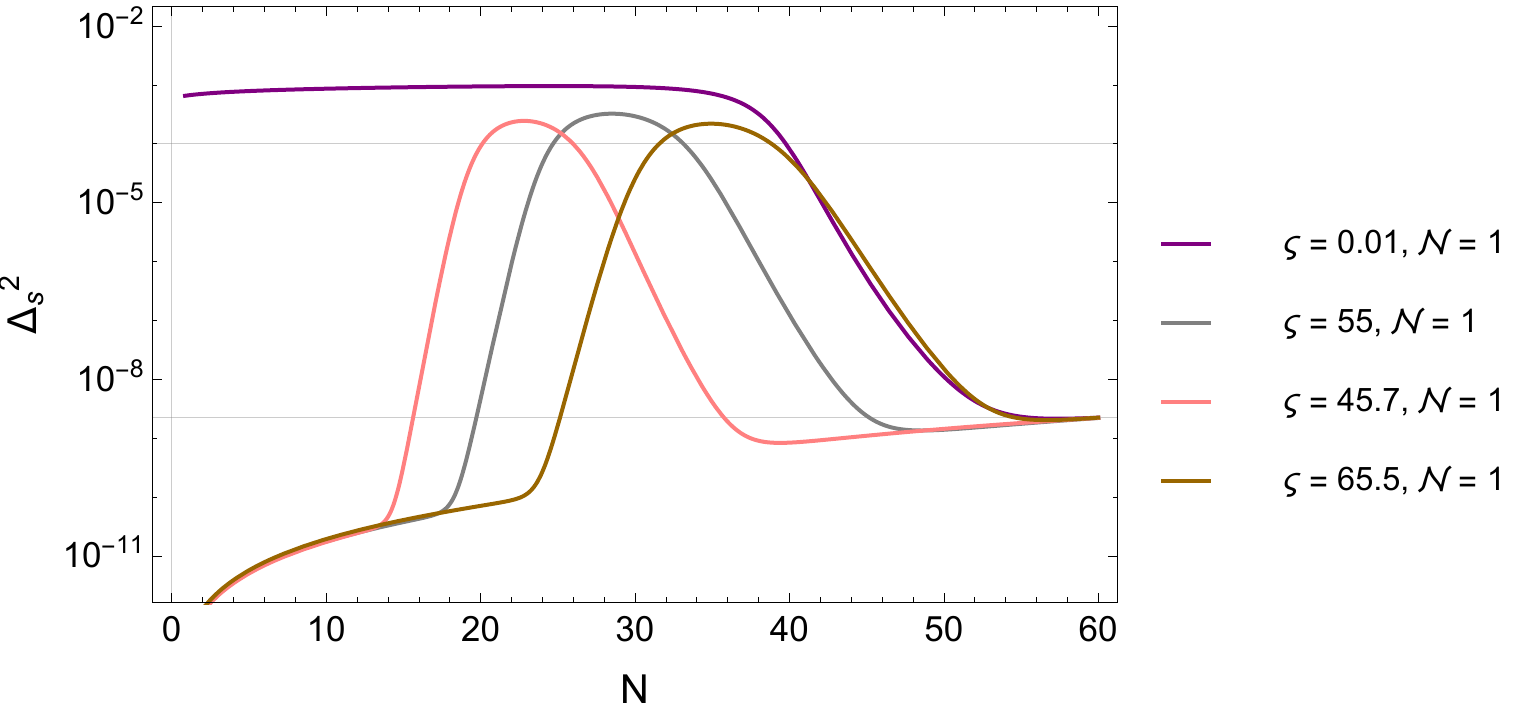}
\caption{Scalar and tensor spectrum for models based on Eq.~\eqref{attractors:inv_definition} with model parameters and color-coding as in Fig.~\ref{fig:xi_inv}.
 The experimental constraints (in grey) are as in Fig.~\ref{fig:GW} and Fig.~\ref{fig:scalar}, respectively.}
\label{fig:spectra}
\end{figure}

In Fig.~\ref{fig:xi_inv} we depict the evolution of the parameter $\xi$  for some representative models of the class defined by Eq.~\eqref{attractors:inv_definition}. As before, the normalization of the scalar potential is fixed by the COBE normalization that fixes the amplitude of the scalar power spectrum at CMB scales and we ensure agreement with all CMB constraints. The remaining parameters are chosen to obtain sizable gauge-field induced effects while conveying an idea of the possible parameter space. 
 We clearly see the peaked features around $N \simeq 20 - 40$ in all spectra with a significant value of the non-minimal coupling $\varsigma$, which translates into a broad peak in the scalar and tensor spectra, depicted in Fig.~\ref{fig:spectra}.
For increasing values of $\alpha/\Lambda$ and $\varsigma$ the bump in the spectra shifts to larger values of $N$ i.e.\ to larger scales. This is consistent with the physical intuition: For larger values of $\alpha/\Lambda$ the increase in the spectrum arises earlier (see also~\cite{Domcke:2016bkh}), for larger values of $\varsigma$ the subsequent suppression due to the non-canonically normalized kinetic term takes effect at earlier times.
Moreover, increasing $\varsigma$ we reduce the field excursion during inflation, leading to a suppression of $\xi$ over a broad range of $N$-values. As a consequence (and indeed consistently with Fig.~\ref{fig:GW} and Fig.~\ref{fig:scalar}), increasing the non-minimal coupling $\varsigma$ we note a suppression both in the scalar and tensor power spectra. For $\varsigma \simeq 63$ and $71$ the signal is below the predicted sensitivity of LISA by roughly an order of magnitude. Finally, we point out that the peaks in the scalar spectrum of Fig.~\ref{fig:spectra} correspond to values of $\xi \lesssim 4.6$. Contrary to the case of minimal coupling, we do hence not enter into the regime $\xi \gtrsim 4.8$ where perturbativity breaks down~\cite{Peloso:2016gqs} and where significant uncertainties become attached to the predictions for the scalar spectrum.

Let us conclude this section by summarizing the key feature of $K(\phi)$ and $V(\phi)$ which allow for the appearance of a peak in the spectra within the last 60 e-folds. A first requirement is that the instability occurs at sufficiently large values of $N$. This can be rephrased as the requirement that $\xi \propto \sqrt{\epsilon_H/K}$ grows sufficiently fast with decreasing $N$ for large values of $N$. After this first regime, we need the effects of the non-canonical kinetic term to become important. Concretely, in order to observe a bump in the spectra we require a rapid increase in $K(\phi)$ at a sufficiently large value of $N$ (for example at $N \simeq 20$). As the growth of $K(\phi)$ turns into a suppression of $\xi$, this mechanism shuts off the instability, leading to the appearance of a bump both in the scalar and tensor power spectra. To illustrate this we chose to study two example models as defined by Eq.~\eqref{attractors:model_def} and Eq.~\eqref{attractors:inv_definition}. However, one can also perform a more model-independent analysis. In particular, we can make predictions for the scalar and tensor spectra by directly parameterizing $K$ and $\epsilon_H$ as functions of $N$:
\begin{equation}
 	\epsilon_H \simeq \frac{\beta}{(1+N)^\alpha} \ , \qquad K \simeq 1 + \frac{\gamma}{N^\delta} \ .
 \end{equation} 
Note that one can always reconstruct the expressions for $V_E(\phi)$ and $K(\phi)$ by expressing $N$ as a function of $\phi$. Hence, in case of a detection of peaks in the scalar and tensor spectra one could determine the corresponding expressions for $V_E(\phi)$ and $K(\phi)$.

\section{Primordial black holes \label{sec:PBHs}}
\subsection{A quick review of PBH formation}

The models discussed in the previous sections generically lead to a strong enhancement of the scalar power spectrum at small scales. If the scalar fluctuations are sufficiently large, this leads to the formation of PBHs in the subsequent evolution of the universe. Here we briefly review the resulting PBH distribution following Ref.~\cite{Carr:2009jm}, collecting the relevant tools to apply this formalism to our discussion.

Primordial scalar fluctuations with a power spectrum $P(\zeta)$ form PBHs upon re-entry into the horizon if  $\zeta > \zeta_c$. The mass $M$ of the resulting black hole can be estimated to be determined by the mass $M_H$ contained in the horizon at the  time of horizon re-entry $t_N$,
\begin{equation}
M(N) = \gamma M_H
 \simeq  \gamma \, \frac{4 \pi \, m_p^2}{H_\text{inf}}\,  e^{j N} \simeq 55 \, g \, \gamma \, \left( \frac{10^{-6} \, m_p}{H_\text{inf}}\right) \, e^{j N} \,,
 \label{eq:MH}
\end{equation}
with $N$ counting the number of e-folds from the end of inflation when the fluctuations in question exited the horizon, $H_\text{inf}$ denoting the Hubble parameter at this time and $j$ parameterizing if the equation of state between the end of inflation and the re-entry of the fluctuation was mainly matter dominated ($j = 3$) or radiation dominated ($j = 2$). In the setup of Sec.~\ref{sec:review} we expect reheating to be efficient due to the inflaton gauge field coupling. In the following we will thus work with $j = 2$. Finally, $\gamma$ is a numerical factor depending on the details of the gravitational collapse. Following~\cite{Green:2004wb, Carr:2016drx} we will use $\gamma = 0.4$. 

The fraction of the energy density of the universe which collapses into PBHs at any given time $t_N$ is given by 
\begin{equation}
\beta(N) = \int_{\zeta_c}^\infty  \frac{ M(N)}{M_H(N)} P_N(\zeta) \, \textrm{d}\zeta =  \int_{\zeta_c}^\infty  \gamma \, P_N(\zeta) \, \textrm{d}\zeta\,.
\label{eq:beta0}
\end{equation}
Here $P_N(\zeta)$ denotes the probability distribution of fluctuations sourced at $N$ e-folds before the end of inflation. In our case, the fluctuations are characterized by a strong equilateral non-Gaussianity~\cite{Anber:2012du}, and can in particular be expressed as $\zeta = g^2 - \langle g^2 \rangle$ with $g \propto (\vec E \cdot \vec B)^{1/2}$ following a Gaussian distribution (see Eq.~\ref{eq:phiEB})~\cite{Linde:2012bt}. For such a positive $\chi$-squared form of non-Gaussianities, the probability distribution is given by (see e.g.~\cite{0510052, 1201.4312, Byrnes:2012yx}):
\begin{equation}
P_N(\zeta) = \frac{\exp \left( - \frac{\zeta + \sigma_N^2}{2 \sigma_N^2}\right)}{\sqrt{2 \pi \sigma_N^2 (\zeta + \sigma_N^2)}} \;,
\label{eq:Pnongauss}
\end{equation}
with $\sigma_N^2 = (\Delta_s^2(N))^{1/2}$ and $\int_{- \sigma_N^2}^\infty P_N(\zeta) \, \textrm{d}\zeta = 1$.  
Since Eq.~\eqref{eq:MH} provides a unique relation between the time of formation $t_N$ and the mass of the PBH $M$, we can equivalently express the fraction $\beta(N) \, \textrm{d}N$ of the universe collapsing in the time interval $[t_N, t_{N + dN}[$ as a function of the PBH mass $M$:
\begin{equation}
\beta(N) \, \textrm{d}N = \beta(M) \frac{dN}{dM}\, \textrm{d}M = \frac{\beta(M)}{2 \, M} \, \textrm{d}M \,.
\label{eq:betaM}
\end{equation}
We stress that $\beta$ denotes the fraction of the universe collapsing into PBHs at the respective time of formation. Within the $\Lambda$CDM model, we can translate this to the fraction of the energy density in the Universe at some later time. Ignoring the decay of PBHs for the moment (we will return to this point below) and assuming an adiabatically expanding universe, the ratio of the number density of PBHs $n_{\text{PBH}}(t)$ and the entropy $s(t)$ is constant. We can hence relate the function $\beta(M)$ to the number density of PBHs at any later time $t$ as~\cite{Carr:2009jm}:
\begin{equation}
\beta(M) = \frac{M n_\text{PBH}(t_N)}{\rho(t_N)} = \frac{4\,  M \, n_\text{PBH}(t)}{3 \, T(t_N) \, s(t)} \,,
\label{eq:nPBH}
\end{equation}
with $\rho(t_N) = 3/4 \, T(t_N) \, s(t_N)$ denoting the energy density of the Universe, assuming that the universe is dominated by radiation at $t_N$. The temperature $T(t_N)$ is related to $H(t_N) = H_\text{inf} \exp(- 2 N)$ through the Friedmann equation. Substituting $H_\text{inf} \exp(- 2 N)$ into Eq.~\eqref{eq:MH}, we find that the fraction of dark matter today formed by PBHs is
\begin{align}
f(M)  & = \frac{M n_\text{PBH}(t_0) }{\Omega_\text{CDM} \, \rho_c} \simeq \nonumber \\ 
& \simeq 4.1 \cdot 10^8 \, \gamma^{1/2} \left( \frac{g_{*}(t_N)}{106.75}\right)^{-1/4} \left( \frac{h}{0.68}\right)^{-2} \left( \frac{M}{M_\odot} \right)^{-1/2} \beta(M)\,,
\label{eq:f}
\end{align}
where we have inserted $\rho_c = 3 H_0^2 m_p^2$ with $H_0 = h \, 100 \,$~(km/s)/Mpc for the critical density and $\Omega_\text{CDM} \simeq 0.21$ for the dark matter fraction today. In the following, we will set $h = 0.68$ and $g_*(t_N) = 106.75$. PBHs with masses below $M_\text{PBH} < M_* \simeq 5 \cdot 10^{14}$~g have a life time shorter than the current age of the universe and would have evaporated through Hawking radiation by now.\footnote{The numerical value $M_*$ might be enhanced when taking into account accretion and merger processes in the matter dominated regime which stabilize light PBHs, see App.~\ref{app:PBH}.} The interpretation of $f(M)$ as the fraction of dark matter hence only applies for $M > M_*$, for smaller values of $M$ the ratio $f(M)$ may simply be seen as a convenient way of parameterizing the initial PBH abundance $\beta(M)$.

The production of PBHs is subject to various constraints, see Ref.~\cite{Carr:2009jm} for an overview and Ref.~\cite{Carr:2016drx} for updated bounds on the mass range $M > M_*$, which will be the range most relevant for us here. As pointed out in Ref.~\cite{Carr:2016drx}, applying these bounds to an extended mass function requires some care. Essentially, a flat constraint $f_\text{max}$ in a certain mass range implies that the total amount of PBHs in this mass range may not exceed the value $f_\text{max}$. A general constraint can then be treated as a sequence of approximately flat constraints. Here we follow the procedure suggested there, requiring
\begin{equation}
\int_{M_1}^{M_2} \frac{\textrm{d}M}{2 M} f(M) < f_\text{max}^{[M_1,M_2]} \,,
\label{eq:fbound}
\end{equation}
with the constraints $f_\text{max}$ taken from Tab.~1 of Ref.~\cite{Carr:2016drx} and $f_\text{max}^{[M_1,M_2]}$ denoting the weakest constraint in the mass range $[M_1,M_2]$. Here we vary the integration boundaries $M_{1,2}$ so that all relevant mass intervals are covered. We have used Eq.~\eqref{eq:betaM} to obtain a dimensionless integrand in Eq.~\eqref{eq:fbound}. 

We note that the PBH formation as sketched above is subject to theoretical uncertainties, in particular concerning the effects of critical collapse, non-spherical properties of the primordial fluctuations and late time accretion and mergers. We discuss these issues in more detail in App.~\ref{app:PBH}. The upshot is that these uncertainties are to large degree degenerate with our model parameters, and hence will not change the overall picture presented here.

\subsection{PBHs in models with non-minimal coupling to gravity}

\begin{figure}
\centering
\includegraphics[width=0.6\textwidth]{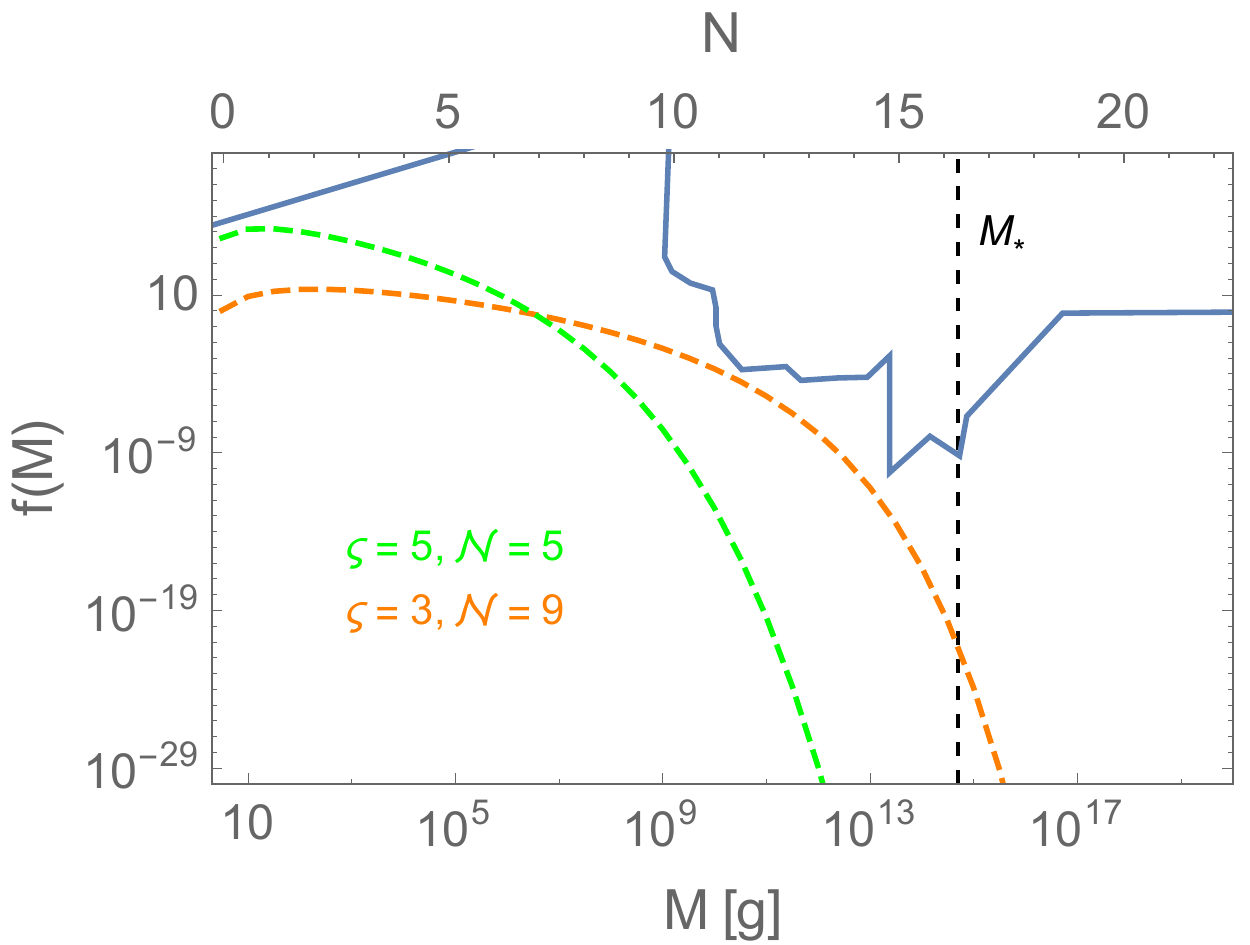}\hfill
\caption{Would-be fraction of PBH dark matter, compared to the constraints in Ref.~\cite{Carr:2009jm,Carr:2016drx} for models based on Eq.~\eqref{attractors:model_example}. The parameters of these curves are chosen to maximize the PBH contribution while obeying the depicted PBH bounds as well as the CMB constraints. 
 Note that the quantity $f(M)$ corresponds to the fraction of dark matter today  only for $M > M_* \simeq 5 \cdot 10^{14}$~g.}
\label{fig:BH1}
\end{figure}

In Fig.~\ref{fig:BH1} we depict two typical results for $f(M)$ for the first model discussed in Sec.~\ref{sec:attractors}, characterized by a non-minimal coupling to gravity through $h(\phi) = \phi$. The parameters\footnote{Here $m^2 = 2 \lambda^4$ in Eq.~\eqref{attractors:model_def}.}  $\alpha/\Lambda$, $m$, $\varsigma$ and ${\cal N}$ are chosen to meet the following conditions: (i) the amplitude and spectral tilt of the scalar spectrum at the CMB scales match the observed values, (ii) the effect of the gauge field contribution is maximized while respecting the CMB non-Gaussianity bounds and (iii) the PBH abundance is maximized while respecting the depicted PBH bounds. Moreover, the orange curve corresponds to the situation which roughly maximizes the contribution to PBH dark matter in this setup. The parameter values for the depicted curves are $m = 7.6 \cdot 10^{-5}\, m_p$, $\alpha/\Lambda = 43.2$ for the $\varsigma = 5$ example and $m = 5.4 \cdot 10^{-5}\, m_p$, $\alpha/\Lambda = 37$ for the $\varsigma = 3$ example. The resulting mass functions increase towards low PBH masses at $M \ll M_*$, hence only the high-mass tails of these distributions can contribute to PBH dark matter today. The strongest constraints close to the dark matter threshold come from evaporating PBH around $M \sim M_*$, which leave traces in the anisotropies of the CMB~ and in the (extra-) galactic photon background (see \cite{Carr:2009jm} for details), which imply $\beta(M) \lesssim 10^{-28}$ in this mass range. For the depicted examples, we find for the scalar spectral index $n_s$, the tensor-to scalar ratio $r$ and the fraction of PBH dark matter $f_\text{tot}$,
\begin{align}
\varsigma &= 5: \quad n_s = 0.98 \,, \quad r = 1 \cdot 10^{-3} \,, \quad f_\text{tot} \approx 0  \,,\\
\varsigma &= 3: \quad n_s = 0.97 \,, \quad r = 5 \cdot 10^{-3} \,, \quad f_\text{tot} = 3 \cdot 10^{-23} \,.
\end{align}

To obtain a mass distribution peaked at heavier PBHs, we need to modify the functional dependence of the non-minimal coupling, $h(\phi)$. In particular we require a scalar spectrum which rises at larger values of $N$, i.e.\ at an earlier stage during inflation. As discussed in Sec.~\ref{sec:attractors}, this can be realized by considering for example $h(\phi) = 1 - 1/\phi$. In this case, for sufficiently large values of $\varsigma$ the scalar spectrum features a peak at large scales (see Fig.~\ref{fig:spectra}) that can suitably induce the generation of a distribution of PBHs. The exact position and the shape of the peak are controlled by the parameters $\alpha/\Lambda$ and $\varsigma$.
In particular, increasing the value of $\varsigma$ moves the peak towards larger values of $N$ (see also Fig.~\ref{fig:spectra}) and suppresses the amplitude, whereas increasing the value of $\alpha/\Lambda$ enhances the spectrum while also shifting the peak towards larger values of $N$ (see e.g.\ \cite{Domcke:2016bkh}). 
Since both parameters also impact the CMB observables, the depicted curves roughly demonstrate the range of possibilities for this choice of $h(\phi)$.

\begin{figure}
\centering
\includegraphics[width=0.85\textwidth]{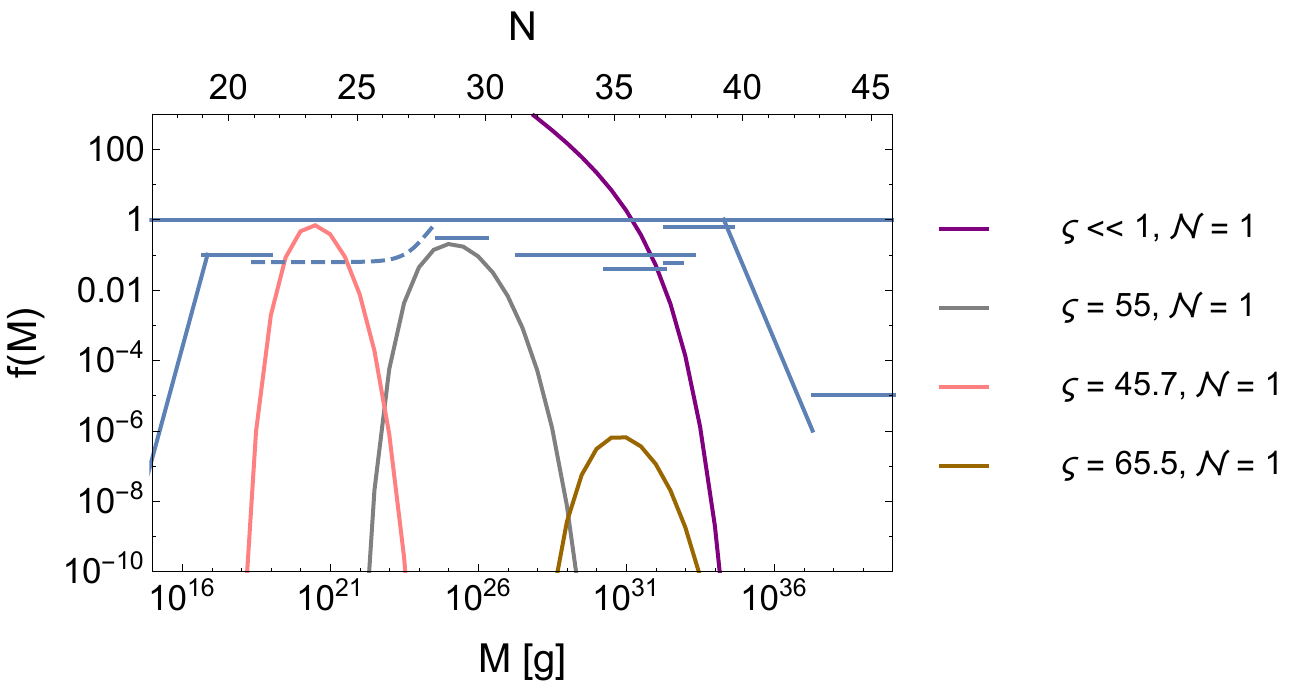}
\caption{Fraction of energy in PBH at time of formation, compared to the constraints in Ref.~\cite{Carr:2016drx}, for models based on Eq.~\eqref{attractors:inv_definition}. Model parameters and color coding as in Fig.~\ref{fig:xi_inv}. Compatibility with the depicted constraints is confirmed using Eq.~\eqref{eq:fbound}.}
\label{fig:BH2}
\end{figure}

 For the examples shown in the plot of Fig.~\ref{fig:BH2} we have: 
\begin{align}
\varsigma &= 45.7, \quad \alpha/\Lambda \simeq 63: \quad n_s = 0.963 \,, \quad r = 3.2 \cdot 10^{-4} \,,\\
\varsigma &= 55, \quad \ \ \alpha/\Lambda \simeq 71: \quad n_s = 0.957 \,, \quad r = 3.3 \cdot 10^{-4}  \,,
\\\varsigma &= 65.5, \quad \alpha/\Lambda \simeq 74: \quad n_s = 0.958 \,, \quad r = 7.4 \cdot 10^{-4} \,.
\end{align}
For reference, we also show the result for a negligible non-minimal coupling (purple curve). The bumps in the spectra for the three models shown in Fig.~\ref{fig:BH2} lead to the generation of PBH distributions peaked around $M \sim 10^{20}, \, 10^{25}$ and $10^{31}$~g, respectively, thus in all three cases the entire PBH population contributes to dark matter today. In all cases we restrict ourselves to a single gauge field, ${\cal N} = 1$. As in Fig.~\ref{fig:BH1}, the model parameters for these benchmark points have been chosen in order to maximize the PBH contribution while obeying all relevant bounds. The reduced amplitude for the $\varsigma = 65.5$ example is due CMB constraints which become more relevant as $\alpha/\Lambda$ is increased. 

In this mass range, the strongest constraints come from neutron star capture~\cite{Capela:2013yf} as well as from micro lensing constraints from the Kepler, MACHO, EROS and OGLE experiments~\cite{Griest:2013aaa,Tisserand:2006zx,Novati:2013fxa}. 
In particular, the micro lensing constraints restrict the total amount of PBHs to be less than $4\%$ to $30\%$ in their respective mass ranges, whereas the neutron star capture constraint requires the total amount of PBHs to be less than $6\%$ for $10^{18}~\text{g} \lesssim M \lesssim 10^{24}$~g. 
This last constraint (shown as a dashed curve in Fig.~\ref{fig:BH2}) has been disputed, as it relies on assumptions about the dark matter content in globular clusters~\cite{Carr:2016drx}. For the pink curve in Fig.~\ref{fig:BH2} we thus choose to omit this constraint, demonstrating that in this case PBHs can account for \textit{all} of dark matter.\footnote{Upon finalizing this paper, we became aware of the very recent microlensing constraint from the Subaru Hyper Suprime-Cam~\cite{Niikura:2017zjd}, which sets strong constraints on most of this window, leaving only a narrow slice around $M \simeq 10^{20}$~g for a significant PBH contribution to dark matter in this mass range.} Of course, if new results confirm this bound, the fraction of PBH dark matter in this mass range must be less than the $6\%$ mentioned above.

The total fraction of PBH dark matter today is given by
\begin{equation}
f_\text{tot} = \frac{\Omega_{\text{PBH}}}{\Omega_{\text{CDM}}} = \int_{M_*}^\infty \frac{\textrm{d} M}{2 M}  \, f(M) \,,
\end{equation}
where in practice we need to integrate only over the strongly enhanced part of the scalar spectrum, as it is well-known that the contribution from the spectrum at CMB scales (where the amplitude is fixed by CMB observations) is completely negligible.\footnote{Strictly speaking, our analysis here applies only to PBH formed in the radiation dominated regime, i.e.\ for $M < M(N_\text{eq})$ with $N_\text{eq} \simeq N_\text{CMB} - 1.2$ labeling the primordial fluctuations entering the horizon at matter-radiation equality. However as these scales are strongly constrained by the CMB, we do not need to worry about this subtlety.} For example, for the curves in Fig.~\ref{fig:BH2} we find 
\begin{equation}
f_\text{tot}^{\varsigma = 45.7} = 98.6 \%,  \, \quad f_\text{tot}^{\varsigma = 55} = 39.4 \% , \, \quad f_\text{tot}^{\varsigma = 65.5} = 1.2 \cdot 10^{-4} \, \% \,.
\end{equation}
In summary, while for the $h(\phi) = \phi$ case the contribution to dark matter is completely negligible, PBHs can contribute a very significant fraction of dark matter for $h(\phi) = 1 - 1/\phi$ in this setup.

\subsection{Searching for PBHs with GW interferometers \label{app:GWs}}

PBHs can form binary objects which source GWs similar to the event observed by LIGO~\cite{Abbott:2016blz}. In this subsection, we address the question if these GWs may be observable in future GW interferometers, see also~\cite{Garcia-Bellido:2017fdg}.  We will normalize the results of this subsection to $M = M_\odot = 1.99 \cdot 10^{33}$~g, which roughly corresponds to the brown curve in Fig.~\ref{fig:BH2} after taking into account the mass increase of the PBH during the matter dominated phase due to accretion and mergers (see App.~\ref{app:PBH}).

Let us first consider GW signal from inspiraling PBHs (see e.g.\ \cite{Maggiore:1900zz}). Comparing the orbital distance of two (point-like) objects rotating with frequency $\omega$,
\begin{equation}
R_\text{orb} = \left( G M_\text{sys} / \omega^2 \right)^{1/3} \,,
\end{equation}
where $G = 6.67 \cdot 10^{-11} \text{m}^3/(\text{kg} \, \text{s}^2)$ and $M_\text{sys}$ is the sum of the two masses, to the sum of the Schwarzschild radii,
\begin{equation}
R_\text{schw} = G M_\text{sys} /c^2 \,,
\end{equation}
we find for the peak frequency of the GW signal of a black hole binary, $f =  \omega/\pi$,
\begin{equation}
f_\text{max} \sim 6.4 \cdot 10^4 \text{ Hz } \left( \frac{2 \cdot M_\odot}{M_\text{sys}}\right) \,.
\end{equation}
For comparison, the  frequency band of LIGO extends to about few times $ 10^3$~Hz. In the early inspiral phase, the system will emit GWs at lower frequencies, thus potentially `crossing' the LIGO frequency band as the system approaches coalescence. 
The time $\tau$ between when signal is at the LIGO peak sensitivity of $f \sim 100$~Hz and coalescence is given by
\begin{equation}
\tau = 3 \, \text{s} \, \left( \frac{M_\odot}{M_c}\right)^{5/3} \left( \frac{100~\text{Hz}}{f} \right)^{8/3} \,,
\end{equation} 
where $M_c$ is the so-called chirp mass, $M_c^5 = (M_1 M_2)^3/(M_1 + M_2) $, constructed from the masses of the two individual BHs.
The amplitude of the emitted GWs, detected at a distance $r$ from the source is given by
\begin{equation}
h \simeq \frac{4}{r} \left( \frac{G M_c}{c^2} \right)^{5/3} \left(\frac{\pi f}{c} \right)^{2/3} \,.
\end{equation}
Comparing this to the strain sensitivity of advanced LIGO, $h \simeq 10^{-24}/\sqrt{\text{Hz}}$~\cite{TheLIGOScientific:2016agk}, we find that assuming a bandwidth of 10~Hz, a signal originating from a binary BH system can be detected if a source is within 
\begin{equation}
r = 8.0 \cdot 10^2 \, \text{Mpc} \, \left( \frac{M_c}{M_\odot} \right)^{5/3} \left( \frac{f}{100 \text{ Hz}} \right)^{2/3} 
\label{eq:detectorreach}
\end{equation}
of the detector. For comparison, the size of the Milky Way is about 30 - 55 kpc and the size of the Virgo super cluster is about 33~Mpc. 
This has to be compared to the expected rate of BH mergers, which has been estimated to be~\cite{Bird:2016dcv,Clesse:2016vqa,Sasaki:2016jop}
\begin{equation}
\Gamma =  10^{-8} \, f_\text{tot} \, \delta_\text{PBH}^\text{loc}\, \text{yr}^{-1}\,  \text{Gpc}^{-3} \,.
\label{eq:rate}
\end{equation}
Here $\delta_\text{PBH}^\text{loc}$ denotes the local PBH density contrast, where $\delta_\text{PBH}^\text{loc} = 1$ yields the expected merger rate if PBH are distributed uniformly throughout the universe. Clustering of PBHs into sub-halos may however increase the value to $\delta_\text{PBH}^\text{loc} \sim 10^7 - 10^9$~\cite{Clesse:2016vqa}, dramatically enhancing the prospects for detection. Note that Eq.~\eqref{eq:rate} is independent of the PBH mass, which drops out when combining the dark matter number density and the BH - BH capture cross section (see e.g.~\cite{Bird:2016dcv}). Hence for $M \sim M_\odot$ and $f_\text{tot} = {\cal O}(0.1)$ one might indeed expect a handful of these events per year at $f = 10^2$~Hz, baring in mind the huge uncertainty in $\delta_\text{PBH}^\text{loc}$.  Note however that since the reach of the detector is sensitive to the PBH mass (see Eq.~\eqref{eq:detectorreach}), the total rate drops significantly for lighter PBHs.

In addition to the transcendent GWs signal from PBH binaries discussed above, there will also be a diffuse stochastic background arising from binaries which are too faint to be detected as individual sources. For a recent analysis see Ref.~\cite{Clesse:2016ajp}, where it was concluded that depending on the PBH merger rate (see above) and the shape of the PBH distribution, such a stochastic background might indeed be detectable by LISA.\footnote{In addition, the formation process of BPHs
may lead to a stochastic GW background similar to first order phase transitions~\cite{Garcia-Bellido:2017fdg}. }

\subsection{Discussion}

The peaks of the mass distributions in Fig.~\ref{fig:BH2} are in an interesting mass region. In the lower mass region (pink and grey curves) the bounds on $f(M)$ are weaker (in particular if one omits the neutron star capture bound), and it becomes possible to account for a significant fraction, or maybe even all of dark matter, in PBHs. Moreover, the mass ranges found in these benchmark models lie below the lower bound expected from BHs formed in stellar evolution, $M \gtrsim 3 \, M_\odot$~\cite{Misner:1974qy}. If detected, such light BHs thus immediately point towards a possible primordial origin. On the other hand, the higher mass region (brown curve) provides PBHs which might be in the correct ballpark to be detected through their gravitational wave signatures in advanced LIGO. However, we emphasize that for the specific, simple choice of $h(\phi)$ discussed here, the abundance of PBHs in this mass range is strongly constrained by CMB observations, rendering a detection with LIGO very challenging even for large density contrasts~$\delta_\text{PBH}^\text{loc}$.

We stress that the spectra depicted in Fig.~\ref{fig:BH2} are not a unique prediction but only three representative benchmark models. Clearly, varying the parameters (as discussed in Sec.~\ref{sec:attractors}) the amplitude of this signal can be adjusted. Moreover, as can be seen by comparing Figs.~\ref{fig:BH1} and \ref{fig:BH2}, the position of the peak is very sensitive to the choice of the function $h(\phi)$. In this context, we point out that Ref.~\cite{Carr:2016drx} recently identified two interesting `windows' for PBH dark matter, in which dark matter could still consist of PBH to $100\%$. The lunar BH window $M_\text{PBH} \sim 10^{20} - 10^{24}~g$ opens when omitting the neutron star capture constraint, as mentioned above. The second window is located at $M_\text{PBH} \sim 1\,  M_\odot - 10^3 \, M_\odot$ (LIGO-like BHs) and may be reached by either considering more complicated functions $h(\phi)$ or by a better understanding of the accretion and merger processes in the matter dominated regime.

Given the different spectra depicted in Fig.~\ref{fig:scalar}, one might wonder if a sizable PBH population might be achieved without invoking a non-minimal coupling to gravity, i.e.\ for $\varsigma = 0$. In this case the characteristic feature of the scalar spectrum is a plateau extending to small scales. The function $\beta(M)$ will hence also be constant, and the strongest constraint come from searches for anisotropies in the CMB and for anomalies in the (extra-) galactic photon background, restricting $\beta \lesssim 10^{-28}$ at $M \sim 10^{13}$~g~\cite{Carr:2009jm}. This implies $f \sim 10^{-20} \, (M_\odot/M)^{1/2}$. Integrating over the entire range which could contribute to PBHs, the lower bound $M > 5 \cdot 10^{14}~g $ dominates the integral leading to $f_\text{tot} \lesssim 10^{-11}$, i.e.\ a completely negligible fraction of dark matter. In the setup discussed here, the non-minimal coupling $\varsigma$ is hence essential to obtain a significant fraction of dark matter in PBHs.

\section{Conclusion and Outlook \label{sec:conclusions}}
Axion-like particles are one of the most natural candidates to drive cosmic inflation since their perturbative shift-symmetry protects the flatness of their potential against quantum corrections. These pseudoscalar fields typically couple to the gauge fields present in the theory through a Chern-Simons term in the Lagrangian. As reviewed in Sec.~\ref{sec:review}, in models of axion-inflation, such a coupling can lead to several interesting and possibly observable phenomenological features. In particular, the production of chiral gravitational waves is a peculiar signature of these models that could be observed in the near future in experiments like the ground-based interferometer LIGO or the space-mission LISA. Hence, it is extremely interesting to continue exploring the rich phenomenology provided by these inflationary models. In this paper we showed that if the axion-inflaton is non-minimally coupled to gravity, these models of axion-inflation represent a natural framework for the production of PBH dark matter. In particular, in Sec.~\ref{sec:non-minimal} we worked out the main generic features of an axion-inflation model with non-minimal coupling to gravity, deriving both the background dynamics and the perturbation power spectra. As explained in Sec.~\ref{sec:attractors}, the interplay between the two main features of this model, i.e. the coupling to gauge fields and the non-minimal coupling to gravity, makes the generation of a broad peak in the scalar spectrum possible. In particular, the increase of the spectrum is sourced by the instability caused by the coupling between the inflaton and the gauge fields. If the kinetic function $K(\phi)$ (see Eq.~\ref{non_minimal:einstein_frame}), corresponding to a non-minimal coupling to gravity in the Jordan frame, features a rapid increase after the advent of this instability, a subsequent suppression of the spectrum arises, generating a peak as shown in Fig.~\ref{fig:BH2}. The position of the peak is determined by the choice of $K(\phi)$ and by the ratio $\alpha/\Lambda$ (see Eq.~\ref{non_minimal:einstein_frame}), which parametrizes the strength of the inflaton coupling to gauge fields. In this context, an interesting class of models is given by the \textit{attractors at strong coupling}, defined in Eq.~\eqref{attractors:model_def}. As shown in Sec.~\ref{sec:PBHs}, by choosing the kinetic function $K(\phi)$ in such a way that the previously mentioned condition is met (see Eq.~\eqref{attractors:inv_definition}), PBHs produced with this mechanism can account for a sizeable fraction of the dark matter present in the universe in a wide region of the parameter space. As explained in the main text, if we ignore the disputable bounds based on neutron star capture, PBHs can even account for the totality of dark matter, in the mass range $10^{18}$~g $\, - \, 10^{24}$~g. Heavier PBHs with masses around a solar mass, although accounting only for a fraction of dark matter, may in turn lead to observable GW signals in LIGO.

The inflationary model employed and the PBH production mechanism proposed in this paper can be further developed following several complementary directions. For instance, they should be embedded in a consistent cosmological scenario, which includes the reheating of the Standard Model degrees of freedom. Since the inflaton is coupled to gauge fields, (p)reheating would primarily produce these massless vector fields~\cite{Adshead:2015pva}. If the number of gauge fields $\mathcal{N}$ is larger than one\footnote{Or if $\mathcal{N}=1$ but the related gauge field is not the Standard Model hypercharge.}, this could lead to a tension with the constraints on the amount of dark radiation present in the universe at the time of Big Bang Nucleosynthesis and of recombination~\cite{Ade:2015xua} if the gauge fields remain massless. Hence, the inclusion of the Standard Model degrees of freedom and of the (p)reheating should be consistent with these bounds. A detailed study of the (p)reheating process could also provide interesting clues for the generation of the baryon asymmetry of the universe, possibly sourced by the CP-violating axion-gauge field coupling~\cite{Kusenko:2014uta,Anber:2015yca,Adshead:2015jza}.

The high sensitivity of inflation to quantum corrections makes it crucial to embed any effective model into an UV complete theory. As remarked in App.~\ref{app:axions}, string theory contains all the ingredients necessary for the realization of an explicit embedding of the PBH production mechanism described in this paper: inflaton-dependent non-canonical kinetic terms for the axion-inflaton, a potential that supports axion-inflation and a coupling between the axion-inflaton and topological terms of Abelian gauge theories. The realization of an explicit string theory construction is beyond the scope of this paper, and we leave it for future work.

In Sec.~\ref{sec:PBHs} we have only reported a few benchmark models in order to show that the mechanism is well-suited for the production of PBHs. It would be interesting to perform a systematic analysis in order to understand which choices of the scalar potential and of the non-minimal coupling to gravity lead to the production of PBHs that compose a sizable fraction of the dark matter present in the universe - or vice versa how to reconstruct these fundamental ingredients of the theory for a given PBH distribution.

Finally, it would be interesting to extend this inflationary model to the case in which the axion-inflaton is coupled to non-Abelian gauge fields, and to study how the PBH production would be modified in such a setup.

\vspace{1cm}
\subsubsection*{Acknowledgements}

We thank D.~Figueroa, E.~Kiritsis, A.~Kusenko, and L.~Sorbo for valuable comments and discussions at various stages of this work. We are especially indebted to Pierre Bin\'etruy who, to our great sorrow, unexpectedly passed away during the completion of this work. Pierre sparked our interest in this topic, and his advice and guidance was invaluable in the early stages of this work.

We acknowledge the financial support of the UnivEarthS Labex program at Sorbonne Paris Cit\'e (ANR-10-LABX-0023 and ANR-11-IDEX-0005-02) and the Paris Centre for Cosmological Physics. L.\,W.\ is partially supported by the Advanced ERC grant SM-grav, No 669288. V.\,D.\ would like to thank the Berkeley Center for Theoretical Physics for kind hospitality during the final stages of this work. F.\,M.\ is supported by the European Research Council starting grant ‘Supersymmetry Breaking in String Theory’ (307605).

\appendix

\section{Axions in field theory, supergravity and string theory \label{sec:axions}}
\label{app:axions}
\subsubsection*{A brief review of the QCD axion}
An axion-like particle (axion, for short) is a pseudoscalar particle $\phi$ that enjoys a continuous perturbative shift-symmetry of the form
\begin{equation}
\phi \rightarrow \phi + a	\,,
\end{equation}
where $a$ is a real constant. Historically, axions arose in field theory to solve the long-standing strong-CP problem of QCD~\cite{Crewther:1979pi}. The symmetries of QCD allow for a topological term in the Lagrangian
\begin{equation}
\mathcal{L}_{\text{QCD}} \supset \frac{\theta}{32 \pi^2} \text{tr} F_{\mu \nu} \tilde{F}^{\mu \nu} \,,
\end{equation}
where $F_{\mu \nu}$ is the QCD field strength. Such a term violates the CP symmetry and would give rise to an electric dipole moment for the neutron in contrast with observations, unless the parameter $\theta$ is tuned very small $\lesssim 10^{-10}$\cite{Baker:2006ts}.\footnote{The physical quantity is $\theta_{\text{phys}} = \theta + \text{arg} \left(\text{det} \left(M_u M_d\right)\right)$, where $M_u$ and $M_d$ are the quark mass matrices. The shift comes from a chiral rotation of quark fields in the Standard Model Lagrangian.} The introduction of an axion $\phi$, proposed by Peccei and Quinn~\cite{Peccei:1977hh, Wilczek:1977pj, Weinberg:1977ma}, solves this naturalness problem of QCD. The field $\phi$ couples with the topological term $\left(\phi/f\right) \, F \tilde F$, giving rise to an effective $\theta$ parameter
\begin{equation}
\theta_{\text{eff}} = \theta_{\text{phys}} + \frac{\phi}{f}\,,
\end{equation}
where $f$ is the axion decay constant. Due to the shift-symmetry of the theory, $\theta_{\text{phys}}$ can be absorbed into the axion field. The shift-symmetry is broken by non-perturbative effects~\cite{Vafa:1984xg} (instantons in the case of QCD) to a discrete shift symmetry of the form
\begin{equation}
\phi \rightarrow \phi + 2 \pi f\,,
\end{equation}
and a cosine-like potential for the axion is generated
\begin{equation}
\label{eq:axionpotential}
V(\phi) = m_u \Lambda_{\text{QCD}}^3 \left[1 - \cos \left(\frac{\phi}{f}\right)\right]\,,
\end{equation}
where $\Lambda_{\text{QCD}}$ is the QCD scale and $m_u$ is the mass of the \textit{up} quark. Such a potential dynamically sets the coefficient of the topological term to zero. This procedure is very generic: non-perturbative effects always induce cosine-like potentials for axions. A coupling $\phi \, F \tilde F$ between the axion and the topological term is also allowed for Abelian gauge groups, i.e.~such a term is generically not forbidden by any symmetries. This is the basis for the coupling between the axion and Maxwell theory considered in this paper.

An axionic field with a coupling of the form $\phi \, F \tilde F$ is often associated with theories exhibiting global chiral or gauge anomalies. In the case of a gauge anomaly the axion is eaten by the gauge boson, giving rise to a consistent effective theory with a massive gauge boson. As a result the axion is removed from the low energy spectrum. This can be avoided in more complicated constructions with multiple U(1) factors and a carefully chosen fermion spectrum \cite{0605225}: In particular, one can construct theories with massless gauge bosons that couple to axions as $\phi \, F \tilde F$. 

\subsubsection*{Axions as inflaton fields}

Due to their perturbative shift-symmetry, axions are one of the most natural candidates to drive inflation: such a symmetry protects the flatness of their scalar potential against quantum corrections. The arguably simplest axion inflation model employs the cosine potential~\eqref{eq:axionpotential}.\footnote{The inflationary potential is obtained from eq. \eqref{eq:axionpotential} substituting $m_u \Lambda_{\text{QCD}}^3 \rightarrow \Lambda^4$, where $\Lambda$ is the scale of non-perturbative effects that generate the potential for the inflaton, i.e.~$V = \Lambda^4 \left[1 - \cos\left(\phi/f\right)\right]$.} Note that inflation models based on such an oscillatory potential are models of large-field inflation: the slow-roll conditions are satisfied for $\phi \gtrsim m_p$. As the potential is periodic, this implies that the period needs to take super-Planckian values, i.e.~natural inflation requires $f \gtrsim m_p$. 

More general inflationary potentials can only be generated if the shift-symmetry is broken completely. If the breaking is explicit, the protection from quantum corrections would generically be lost. However, we can retain a sufficient level of control by only breaking the shift-symmetry weakly: note that it is technically natural to take the symmetry-breaking terms  to be small. Then the quantum corrections resulting from the breaking are also suppressed. Another way of preserving the protection offered by the shift-symmetry is to only break it spontaneously. This can be realized by coupling the axion to a 3-form theory~\cite{0507215, 0811.1989, 1101.0026}. To be explicit, consider the theory of an axion $\phi$ and a 3-form $C_{\nu \rho \sigma}$ with corresponding field strength $F_{\mu \nu \rho \sigma} = 4 \partial_{[ \mu} C_{\nu \rho \sigma]}$ described by the following Lagrangian density:
\begin{equation}
\mathcal{L} = \frac{1}{2} (\partial \phi)^2 - \frac{1}{48} F_{\mu \nu \rho \sigma}^2 + \frac{\mu}{24} \phi \, \epsilon^{\mu \nu \rho \sigma} F_{\mu \nu \rho \sigma} \, .
\end{equation}
Here $\mu$ is a parameter with the dimensions of mass. As a 3-form does not have any propagating degrees of freedom in 4 dimensions, it can be integrated out, leading to
\begin{equation}
\mathcal{L} = \frac{1}{2} (\partial \phi)^2 - \frac{1}{2} (q+\mu \phi)^2 \, ,
\end{equation}
where the parameter $q$ corresponds to a VEV for $F_4$~\cite{0811.1989}. Thus one arrives at a non-periodic potential for the axion of the form $V= \tfrac{1}{2}(q+\mu \phi)^2$. The discrete shift symmetry is still intact as a shift in $\phi$ is accompanied by a shift in $q$:
\begin{equation}
\phi \rightarrow \phi + 2 \pi f \, , \qquad q \rightarrow q - 2 \pi \mu f \, .
\end{equation}
However, the symmetry is broken spontaneously once a minimum $\phi = - q / \mu$ is chosen. As $q$ is quantized and takes discrete values one finds a family of potentials of the form $V= \tfrac{1}{2}(q+\mu \phi)^2$, one for each possible value of $q$.\footnote{The existence of a family of potentials introduces the danger that the axion may tunnel to another branch of the family rather than perform slow-roll on one branch.} The model still profits from a protection from large quantum corrections: the symmetry implies that corrections to $V$ can only come in the form of powers of $V/\Lambda^4$ with $\Lambda$ the UV cutoff.   

Above we saw how this approach can give rise to a quadratic axion potential. This method can also be used to realize more complicated axion potentials~\cite{0507215}. In particular, by modifying the 3-form kinetic term as
\begin{equation}
- \frac{1}{48} F_{\mu \nu \rho \sigma}^2 \rightarrow -\frac{M^4}{24} \mathcal{G} \left(\frac{\epsilon^{\mu \nu \rho \sigma}F_{\mu \nu \rho \sigma}}{M^2} \right) \, ,
\end{equation}
for some function $\mathcal{G}(x)$ and a scale $M$, one arrives at an axion potential
\begin{equation}
V = M^2 \int \textrm{d} \phi \, (\mathcal{G}')^{-1}\left(\frac{q-\mu \phi}{M^2} \right) \, ,
\end{equation}
where $(\mathcal{G}')^{-1}$ is the inverse function of $\mathcal{G}'$.

So far we have discussed how theories of axions with various potentials can be constructed in field theory. For our purposes we also want to consider axion theories that -- in Einstein frame -- exhibit a non-canonical kinetic term. As we had not included gravity in our discussions of axions yet, there was no need to consider non-canonical kinetic terms, as one could always normalize. However, we will return to the question of non-canonical kinetic terms in the next sections, when we consider axions in supergravity and string theory. 

While considering axions as inflatons helps overcome many UV problems of inflation in effective field theory, one cannot remain completely agnostic about UV physics. For example, natural inflation needs $f \gtrsim m_p$, which seems to be an unnatural feature in the context of quantum gravity~\cite{ArkaniHamed:2006dz, 0605206, Conlon:2012tz}. Hence, in the following we will describe how axions and their properties can arise when the UV physics is given by string theory. As string theory models of axion inflation are typically formulated in the effective supergravity theory arising from string theory compactifications, we turn to axions in supergravity next.

\subsubsection*{Axions in supergravity}
Axions can be very naturally embedded in supergravity. They can be described by either the phase or the real or imaginary parts of complex scalar fields in supermultiplets. Consider for example a theory of a chiral superfield $\Phi$ described by:
\begin{equation}
\mathcal{K} \equiv \mathcal{K}\left(\Phi + \overline{\Phi}\right)\,, \quad W \equiv W_0 = \textrm{const} \, .
\end{equation}
As $\mathcal{K}$ only depends on the combination $(\Phi + \overline{\Phi})$ and $W$ is independent of $\Phi$, the theory is symmetric under continuous shifts of $\textrm{Im}(\Phi)$, which we identify with the axion: $\phi \equiv \textrm{Im}(\Phi)$. 

To generate a potential for the axion, the shift symmetry has to be broken. This can occur due to non-perturbative effects as in the non-supersymmetric case. To this end we introduce a non-Abelian gauge sector with its corresponding field strength superfield $\mathcal{W}^{\alpha}$. If we choose the corresponding gauge kinetic function to be $f=\Phi$, the axion $\phi$ will couple to the topological term of the non-Abelian gauge theory:
\begin{equation}
\label{gaugekin}
{\left. \left( \Phi \ \text{tr} \, \mathcal{W}_{\alpha} \mathcal{W}^{\alpha} + \textrm{h.c.} \right) \right|}_{F} \supset  -\phi \ \text{tr} \, F_{\mu \nu} \tilde{F}^{\mu \nu}  \, .
\end{equation}
Then non-perturbative effects will introduce a periodic (Einstein-frame) potential for $\phi$ while at the same time breaking the continuous shift symmetry down to a discrete one. 

In this paper we are also interested in the coupling between the axion and the topological term of an Abelian gauge theory. This will arise from a coupling of type~\eqref{gaugekin} where $\mathcal{W}^{\alpha}$ is now the field strength superfield of the Abelian gauge theory.  

One can generate more general (i.e.~non-peridiodic) potentials for $\phi$ if the chiral superfield $\Phi$ appears in the superpotential explicitly, e.g.~$W = m h (\Phi)$ where $h$ is a holomorphic function. Then the shift symmetry of the axion is broken explicitly by $W$ (see however~\cite{0811.1989, 1101.0026}) and one may generate a polynomial Einstein-frame potential $V_E$. As $\mathcal{K}$ is still shift-symmetric, some of the axionic protections are still active. In particular, the breaking of the shift-symmetry is completely controlled by the parameter $m$. This is the ansatz for inflation models realizing $F$-term axion monodromy inflation~\cite{1404.3040, 1404.3542, 1404.3711}.

For our purposes we are particularly interested in theories where the axion kinetic term depends non-trivially on the axion itself. Given a K\"ahler potential $\mathcal{K}$ the function $K(\phi)$ appearing in~\eqref{non_minimal:einstein_frame} is given by $K \equiv \partial_{\Phi} \partial_{\overline{\Phi}} \mathcal{K}$. We can then make the following observation. If $\mathcal{K}$ preserves the axionic shift-symmetry, it only depends on the combination $(~\Phi~+~\overline{\Phi}~)$ and the function $K$ will only depend on the saxion $\textrm{Re}(\Phi)$, but not on the axion $\phi = \textrm{Im}(\Phi)$. Hence it seems that in supergravity we cannot get any non-trivial axion-dependence in $K$, without breaking the axionic shift-symmetry in $\mathcal{K}$.\footnote{This finding holds for all axions, i.e.~also for axions arising as phases or the real parts of complex scalars in chiral superfields. As long as $\mathcal{K}$ does not depend on the axion as required by shift-symmetry, there cannot be any axion-dependence in $K$.} 

It is possible to break the shift-symmetry in the K\"ahler potential in different ways. One possibility is to introduce an explicit breaking term as in~\cite{Ferrara:2010yw,Ferrara:2010in,Buchmuller:2012ex}. In such a case, in order not to lose all the advantages deriving from the shift-symmetry, the amount of breaking has to be small. In particular, we can always include further fields in the theory. Hence, a possibility is to break the shift-symmetry through loop effects, as for example for theories whose superpotential $W \supset \kappa \Phi S_+ S_-$~\cite{Gaillard:1993es,Stewart:1996ey,Stewart:1997wg}, where $S_\pm$ are heavy fields. Upon integrating out $S_{\pm}$, axion-dependent loop corrections modify the \Kahler potential. The amount of shift-symmetry breaking is naturally small because loop-suppressed. 

Furthermore, it is important to observe that in supergravity we can never get a theory of just an axion. As the axion arises from a complex scalar, there will at least always be the saxion partner as an additional field. If we consider effective supergravity theories from string theory, the existence of further fields is rather generic. Note that these generically introduce an explicit axion-dependence in the axion kinetic term through backreaction. 

Finally, note that in the Peccei-Quinn mechanism the axion arises from a phase of a complex field, whereas here we focused on axions corresponding to the imaginary part of a complex scalar: $\phi= \text{Im}(\Phi)$. However, both descriptions are related. If the real part of a complex scalar is a scalar, then calling the imaginary part or the phase a pseudoscalar is equivalent (in both cases, a complex conjugation corresponds to a change in sign in the pseudoscalar part). An explicit connection between the two cases can be made as follows. In particular, let
\begin{equation}
f= \Phi \equiv \sigma + i \phi \, ,
\end{equation}
with $\phi$ an axion and $\sigma$ denoting the scalar component (saxion). Then we can define
\begin{equation}
\Phi_{\textrm{PQ}} = e^{f} = e^{i \phi} (\cosh \sigma + \sinh \sigma) \, ,
\end{equation}
such that the axion now appears as a phase.
Hence the descriptions of an axion as a phase vs.~the imaginary part are equivalent descriptions related by a non-linear map.

\subsubsection*{Axions from String Theory and Quantum Gravity Constraints}

Axions are also a very natural prediction of string theory. They arise from the dimensional reduction of form field gauge potentials on sub-manifolds of the string theory compactification space. The shift-symmetry in the four-dimensional effective field theory is a remnant of the gauge symmetry of gauge fields in the ten-dimensional theory. In addition there are also universal axions (the axionic part of the axio-dilaton and the axions from dualizing to 2-form gauge potentials (see e.g.~\cite{1404.2601} for a brief review). It is even expected that string theory compactifications may give rise to a \textit{string axiverse}~\cite{Arvanitaki:2009fg, Cicoli:2012sz}. 

There has been a sustained effort in the string theory community to embed inflation in string theory. Due to issues of theoretical control, models of inflation are typically constructed in a regime where an effective field theory description is applicable, rather than working in string theory directly. In particular, models are typically realized in the effective supergravity theory. There exists a multitude of proposed models of inflation in string theory (for a review until 2014 see~\cite{1404.2601}, for subsequent progress and references until September 2014 see~\cite{1409.5350}). However, most models come with open questions regarding control of quantum corrections, such that it is difficult to make robust predictions.

However, there are certain properties which appear to be universal as far as embedding axion inflation in string theory is concerned. In particular, axions with a super-Planckian period ($f > m_p$) seem to be inconsistent with string theory compactifications~\cite{0605206}. This implies that the simplest models of natural inflation cannot arise from string theory. This is also consistent with general quantum gravity arguments~\cite{ArkaniHamed:2006dz, Conlon:2012tz}.

This does not imply that axion inflation in general is forbidden in string theory. However, to evade the constraints on the axion field range, models have to become more involved. The following mechanisms have been proposed to allow for inflation despite the constraint on the axion period:
\begin{itemize}
\item \textit{Alignment mechanisms}~\cite{Kim:2004rp}. The idea is to create a long super-Planckian inflaton trajectory in the field space of two or more sub-Planckian axions. It is necessary to tune the potential~\cite{Kim:2004rp} or to rely on kinetic terms~\cite{1404.7496, Shiu:2015xda} to create this long trajectory within a compact sub-Planckian field space. 
\item \textit{N-flation}~\cite{Dimopoulos:2005ac, Cicoli:2014sva, Das:2014gua}: This approach exploits the fact that a diagonal direction in the field space of $N$ axions can be super-Planckian for sufficiently large $N$, even if every single axion has a sub-Planckian field range.
\item \textit{Axion monodromy}~\cite{Silverstein:2008sg, 0808.0706}: These models make use of an explicit breaking of the shift-symmetry (e.g.~due the presence of branes) in the spirit of~\cite{0507215, 0811.1989, 1101.0026} to generate a perturbative non-periodic potential for the axion. The originally periodic axion field space is effectively unfolded. 
\end{itemize}
String theory compactifications contain all necessary `ingredients' for a successful embedding of any one of the above approaches: the existence of multiple axions is rather generic and D-branes and fluxes give rise to monodromies. Consequently, there exist many proposed models based on these ideas (see e.g.\cite{1409.5350} for a review until Sep.~2014). Yet, there are many open questions regarding the viability of any one model of axion inflation from string theory. Generalizations of the Weak Gravity Conjecture pose serious constraints on models based on alignment or N-flation~\cite{1503.00795, 1506.03447}. At the same time there are persistent problems with control as far as the stabilization of additional scalar fields (moduli) is concerned. Overall, at the time of writing there is no final verdict regarding axion inflation from string theory: While string theory exhibits many properties for successful axion inflation, there does not yet exist a model that withstands deeper scrutiny.\footnote{Models of axion inflation are typically models of large-field inflation which will give rise to measurable tensor modes. However, measurable tensor modes can also be generated in stringy inflation models that do not employ axions~\cite{Cicoli:2008gp, Burgess:2016owb, Cicoli:2016chb}. For such inflationary models explicit string theory embeddings have been built, including a controlled moduli stabilization procedure and the presence of a chiral visible sector~\cite{Cicoli:2016xae, ChiralGlobalFibre}.}

What does this imply for the models of axion inflation considered in this work? In this paper we explore models that -- in Einstein frame -- exhibit a range of axion potentials, but also have non-canonical kinetic terms for the axion that depend on the axion itself. In addition, the axion should couple to the topological term associated with Abelian gauge theories. Overall, we find that string theory exhibits all the properties to accommodate the axion inflation models discussed in this work -- at least in principle. 
\begin{itemize}
\item As far as the potential is concerned, models of axion inflation in string theory were shown to give rise to a wide range of possible potentials. The possibilities increase further once back-reaction of other scalar fields (moduli) is taken into account~\cite{1011.4521,1405.3652}. 
\item As the existence of moduli is a generic feature of string theory compactifications, back-reaction effects will typically play a role. As described in the section on supergravity, back-reaction effects can also induce axion-dependence of the kinetic terms. 
\item Furthermore, couplings between axions and the topological terms of (non-)Abelian gauge theories is generic in string theory. These couplings are a crucial ingredient for the cancellation of gauge anomalies in consistent string vacuums via the generalized Green-Schwarz mechanism.\footnote{In this case the gauge bosons receive string scale masses.} Interestingly, couplings of the form $\phi F \tilde{F}$ can also arise if the gauge theory is non-anomalous (see e.g.~\cite{0906.1920}).
\end{itemize}
However, to make any more precise statements would require the construction of an explicit string embedding, which is beyond the scope of this paper. We hence leave this for future work.

\section{Scalar power spectrum and consistency checks \label{sec:spectrum}}
\begin{figure}
\centering
\includegraphics[width=0.565\textwidth]{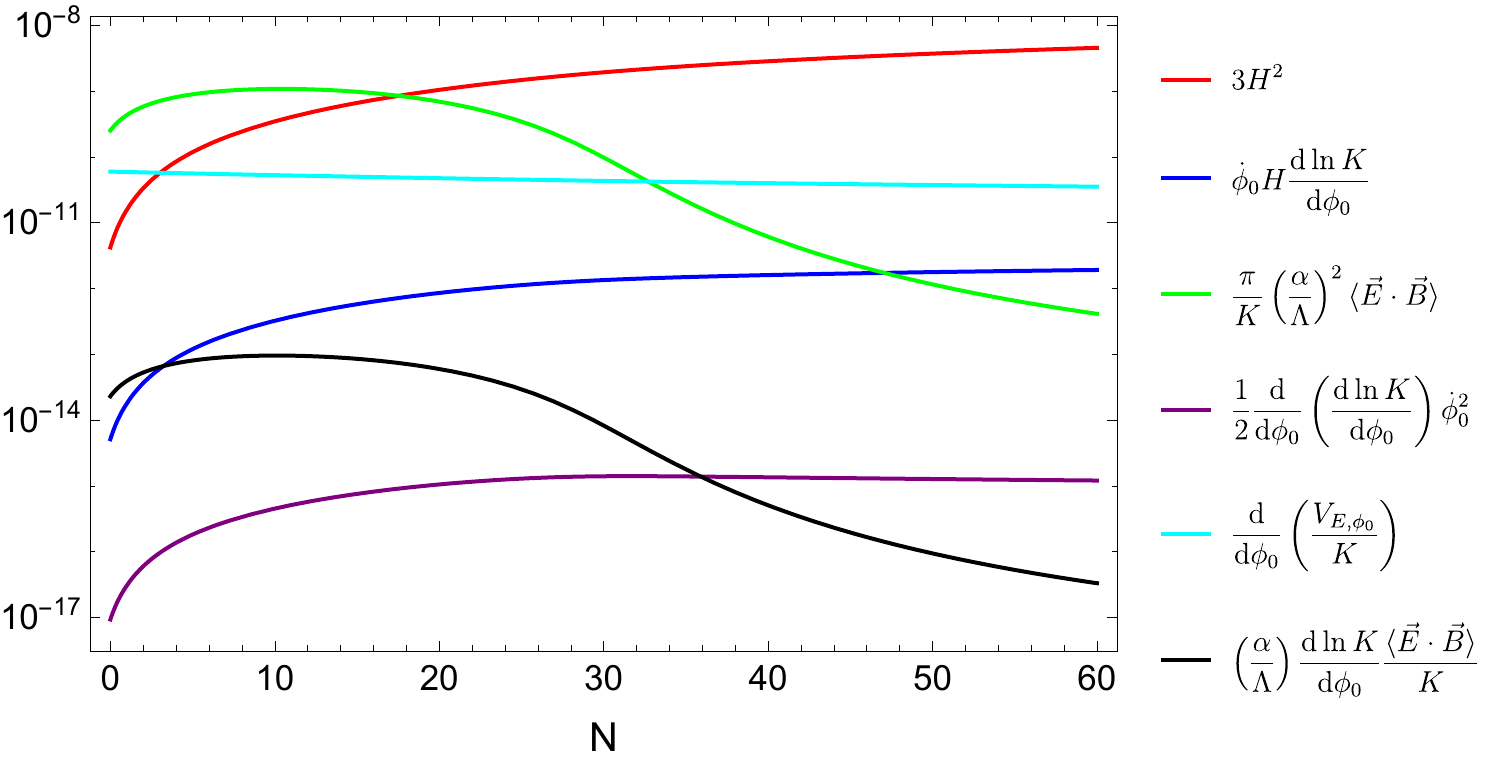}
\includegraphics[width=0.425\textwidth]{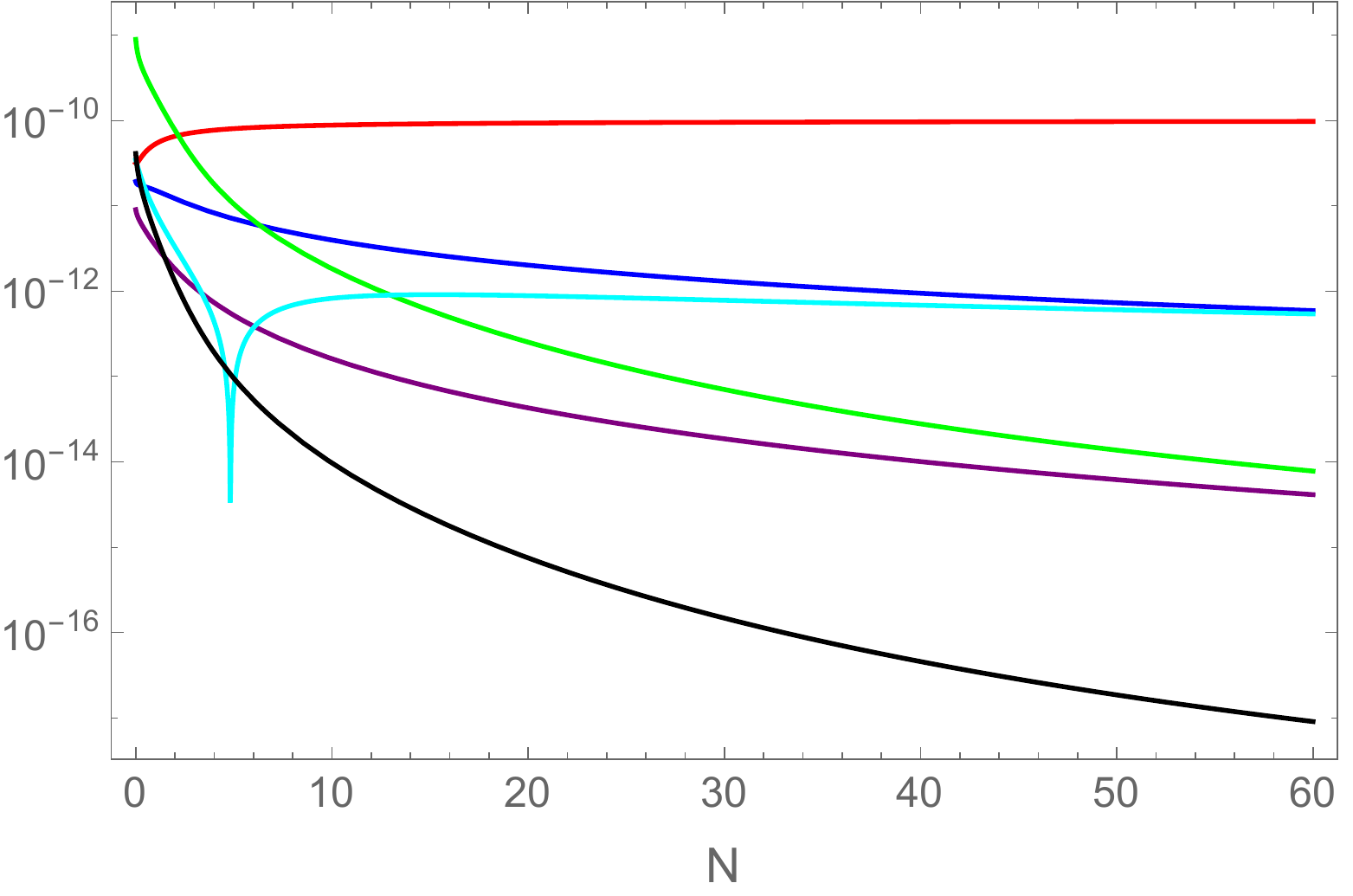}
\caption{Evolution of the terms in Eq.~\eqref{spectrum:perturbations_approx} for the case $h(\phi) = \phi$ with $\varsigma = 0.01$, $\mathcal{N}=10$ (left plot) and for the case $\varsigma = 7$, $\mathcal{N}=1$ (right plot) expressed in natural units. In particular the different colors correspond to the six terms on the right hand side of Eq.~\eqref{spectrum:perturbations_approx} respectively.}
\label{fig:checks}
\end{figure}

In this appendix we present some details of the derivation of Eq.~\eqref{non_minimal:gauge_sourced} and we also show some plots that ensure the validity of the approximations performed during this procedure. Let us start by expressing Eq.~\eqref{non_minimal:scalar_fluctuations} as:
\begin{equation}
\begin{aligned}
	\label{spectrum:perturbations_final}
	&-\frac{\alpha}{\Lambda} \frac{\delta_{\vec{E}\cdot \vec{B}}}{K}  = \ddot{\delta\phi} + \left[ 3H + \dot{\phi}_0 \frac{\textrm{d} \ln K}{\textrm{d} \phi_0}   - \frac{\pi}{ H K} \left( \frac{\alpha}{\Lambda} \right)^2  \langle \vec{E} \cdot \vec{B} \rangle \right] \dot{\delta \phi} \, +  \\ 
	&\qquad \qquad + \left[ \frac{1}{2} \frac{\textrm{d}}{\textrm{d} \phi_0} \left(\frac{\textrm{d} \ln K}{\textrm{d} \phi_0}\right) \dot{\phi}_0^{ 2}  + \frac{\textrm{d}}{\textrm{d} \phi_0} \left( \frac{V_{E,\phi_0}}{K}\right) - \left(\frac{\alpha}{\Lambda}\right) \frac{\textrm{d} \ln K}{\textrm{d} \phi_0}   \frac{\langle \vec{E}\cdot \vec{B}\rangle}{K} - \frac{\vec{\nabla}^2}{a^2} \right] \delta \phi \  .
\end{aligned}
\end{equation}
By neglecting the $\ddot{\delta \phi}$ term and the $ \vec{\nabla}^2 \delta \phi /a^2$ (that clearly are higher order in terms of the slow-roll parameters) and, following the same procedure used in~\cite{Linde:2012bt} by approximating $\dot{\delta \phi}$ with $H \delta \phi$, Eq.~\eqref{spectrum:perturbations_final} can be expressed as:
\begin{equation}
\begin{aligned}
	\label{spectrum:perturbations_approx}
	&-\frac{\alpha}{\Lambda} \frac{\delta_{\vec{E}\cdot \vec{B}}}{K}  = \left[ 3H^2 + \dot{\phi}_0 H \frac{\textrm{d} \ln K}{\textrm{d} \phi_0}   - \frac{\pi}{ K} \left( \frac{\alpha}{\Lambda} \right)^2  \langle \vec{E} \cdot \vec{B} \rangle \right.  \, +  \\ 
	&\qquad \qquad + \left. \frac{1}{2} \frac{\textrm{d}}{\textrm{d} \phi_0} \left(\frac{\textrm{d} \ln K}{\textrm{d} \phi_0}\right) \dot{\phi}_0^{ 2}  + \frac{\textrm{d}}{\textrm{d} \phi_0} \left( \frac{V_{E,\phi_0}}{K}\right) - \left(\frac{\alpha}{\Lambda}\right) \frac{\textrm{d} \ln K}{\textrm{d} \phi_0}   \frac{\langle \vec{E}\cdot \vec{B}\rangle}{K} \right] \delta \phi \  .
\end{aligned}
\end{equation}
At this point, we can proceed by neglecting all the other terms that are not relevant in order to study the evolution of the system in the gauge field dominated regime. In particular, it is possible to show that most of the terms appearing in Eq.~\eqref{spectrum:perturbations_approx} are not relevant for our purposes. In order to show this explicitly we have substituted the solution $\phi_0(N)$ for the background equations of motion (i.e.\ the system given by Eq.~\eqref{non_minimal:background_scalar} and Eq.~\eqref{non_minimal:background_friedmann_2}) into Eq.~\eqref{spectrum:perturbations_approx}. The plots of the absolute values of the different terms appearing in Eq.~\eqref{spectrum:perturbations_approx} is shown in Fig.~\ref{fig:checks} for two of the models presented in Sec.~\ref{sec:attractors}. It should be clear from these plots that Eq.~\eqref{spectrum:perturbations_approx} can be approximated by only keeping the terms proportional to $3H^2$ and the $\frac{\pi}{ K} \left( \frac{\alpha}{\Lambda} \right)^2  \langle \vec{E} \cdot \vec{B} \rangle$. 

Using this approximation (consistent with the discussion carried out in~\cite{Linde:2012bt}) Eq.~\eqref{spectrum:perturbations_approx} can be expressed as:
\begin{equation}
	-\frac{\alpha}{\Lambda} \delta_{\vec{E}\cdot \vec{B}} =  3H^2 K \left[ 1  - \pi \left( \frac{\alpha}{\Lambda} \right)^2  \frac{\langle \vec{E}\cdot \vec{B}\rangle}{3 H^2 K}  \right] \delta \phi \ . 
	\label{eq:phiEB}
\end{equation}
At this point we can proceed by defining:
\begin{equation}
	b \equiv  1  -  \pi \left( \frac{\alpha}{\Lambda} \right)^2  \frac{\langle \vec{E}\cdot \vec{B}\rangle}{3 H^2 K} \ ,
\end{equation}
so that in the case $K(\phi) = 1$ we recover the usual expression:
\begin{equation}
	b = 1  -  \pi \left( \frac{\alpha}{\Lambda} \right)^2 \frac{\langle \vec{E}\cdot \vec{B}\rangle}{3 H^2 } \ .
\end{equation}
In the approximation $\langle \zeta^2(x) \rangle \simeq \Delta^2_s (k)  $, the scalar power spectrum in the gauge field dominated regime can thus be expressed as:
\begin{equation}
 	\left.  \Delta^2_s (k) \right|_\text{gauge} \simeq \left(\frac{H}{\dot{\phi}_0}\right)^2 \langle \delta\phi^2 \rangle \simeq \left(\frac{H}{\dot{\phi}_0}\right)^2 \left(\frac{\alpha}{\Lambda}\right)^2 \left(\frac{1}{3 b H^2 K }\right)^2 \langle (\delta_{\vec{E}\cdot \vec{B}})^2 \rangle \ ,
 \end{equation} 
 and finally, using $\langle (\delta_{\vec{E}\cdot \vec{B}})^2 \rangle \simeq \langle \vec{E}\cdot \vec{B} \rangle^2$, we get:
 \begin{equation}
 	\label{spectrum:gauge_sourced}
 	\left.  \Delta^2_s (k) \right|_\text{gauge} \simeq  \left(\frac{\alpha \langle \vec{E}^a \cdot \vec{B}^a  \rangle / \sqrt{\mathcal{N}} }{3 b \Lambda \dot{\phi}_0 H K }\right)^2  \ ,
 \end{equation}
  where we also have generalized the definition of $b$:
  \begin{equation}
	b \equiv  1  -  \pi \left( \frac{\alpha}{\Lambda} \right)^2  \frac{\langle \vec{E}^a \cdot \vec{B}^a  \rangle }{3 H^2 K} \ ,
 \end{equation}
in order to extend our results to the case with $\mathcal{N}$ Abelian gauge fields.

\section{Uncertainties in the PBH production}
\label{app:PBH}
PBH formation has been subject to many dedicated studies over the past decades and there are several effects which we did not take into account in the analysis of the main text. Here we list some of these effects and estimate their impact on our results, see also Ref.~\cite{Carr:2016drx} for a more detailed overview.

\subsubsection*{Critical collapse}

Primordial scalar fluctuations with a power spectrum $P(\zeta)$ form PBHs upon re-entry into the horizon if  $\zeta > \zeta_c$. A reference scale for  the mass of the resulting black hole is the horizon mass $M_H$ at the time of horizon re-entry $t_N$,
\begin{equation}
M_H(N) 
 \simeq \frac{4 \pi m_p^2}{H_\text{inf}} e^{j N} \simeq 55 \, g \, \left( \frac{10^{-6} \, m_p}{H_\text{inf}}\right) \, e^{j N} \,,
 \label{eq:horizonmass}
\end{equation}
More accurately the mass spectrum of PBHs (at the time of formation) follows a critical scaling relation~\cite{Choptuik:1992jv,Koike:1995jm,Niemeyer:1999ak,Gundlach:1999cu,Gundlach:2002sx},
\begin{equation}
M(\zeta, N) = \kappa M_H(N) (\zeta - \zeta_c)^y \,,
\label{eq:critical}
\end{equation}
for $\zeta > \zeta_c$, where the dimensionless parameters $\kappa$, $\zeta_c$ and $y$ have been studied numerically for the case of spherical, Gaussian perturbations in a radiation dominated background. Following~\cite{Musco:2004ak,Musco:2008hv,Musco:2012au,Niemeyer:1997mt} we will use $\kappa = 3.3$, $\zeta_c = 0.45$ and $y = 0.36$ in the following -- subject to the caveat that these values have been obtained for the spherical collapse of Gaussian fluctuations. Employing Eq.~\eqref{eq:critical} instead of Eq.~\eqref{eq:MH}, the simple one-to-one correlation between a fluctuation sourced at a given e-fold $N$ and a PBH mass $M$ is lost, for any given $N$ there will be a tail of low-mass PBHs as $\zeta$ approaches the critical value $\zeta_c$.

We can determine the fraction of space that collapses to form black holes at a given time (i.e.\ when the fluctuations corresponding to a given $N$ re-enter the horizon) as
\begin{equation}
\beta(N) = \int_{\zeta_c}^\infty  \frac{M(\zeta, N)}{M_H(N)} P_N(\zeta) \textrm{d}\zeta =  \int_{\zeta_c}^\infty  \kappa (\zeta - \zeta_c)^y P_N(\zeta) \textrm{d}\zeta\,.
\label{eq:beta1}
\end{equation}
Here $P_N(\zeta)$ denotes the probability distribution of fluctuations sourced at $N$ e-folds before the end of inflation, which in our case is given by the non-Gaussian distribution~\eqref{eq:Pnongauss}. Using Eq.~\eqref{eq:beta1} and Eq.~\eqref{eq:critical} the fraction of space collapsing at a fixed $t_N$ in a given mass interval $[M, M + \textrm{d}M]$ is given as
\begin{equation}
\tilde \beta(N,M) \, \textrm{d}M = \frac{M}{M_H(N)} P_N(\zeta(M)) \frac{\textrm{d} \zeta(M)}{\textrm{d} M} \textrm{d}M \,,
\end{equation}
with $\zeta(M) = (M/(\kappa M_H))^{1/y} + \zeta_c$. One can easily confirm that $\int_0^\infty \tilde \beta(N,M) \textrm{d}M = \beta(N)$, i.e.\ the function $\tilde \beta(N,M)$ describes the mass distribution of PBHs formed at a given $t_N$, normalized to the total fraction of space collapsing at $t_N$. Note that $\tilde \beta(N,M)$ carries an inverse mass dimension. This function (with a somewhat different normalization) has been referred to as `initial mass function'~\cite{Niemeyer:1997mt,Carr:2016drx}. Finally integrating over $N$ we obtain the total fraction of space collapsing into PBHs in the mass range $[M, M + \textrm{d}M]$ as
\begin{equation}
\beta(M) = \int_0^{N_\text{max}} \tilde \beta(N,M) \, \frac{\textrm{d}M}{\textrm{d}N} \, \textrm{d}N \,,
\label{eq:betaMc1}
\end{equation}
with $\textrm{d}M/\textrm{d}N = j M_H(N)$ and $N_\text{max} < 60$ denotes the largest $N$ which leads to significant PBH production (within the radiation dominated regime).

\begin{figure}
\centering
\includegraphics[width=0.48\textwidth]{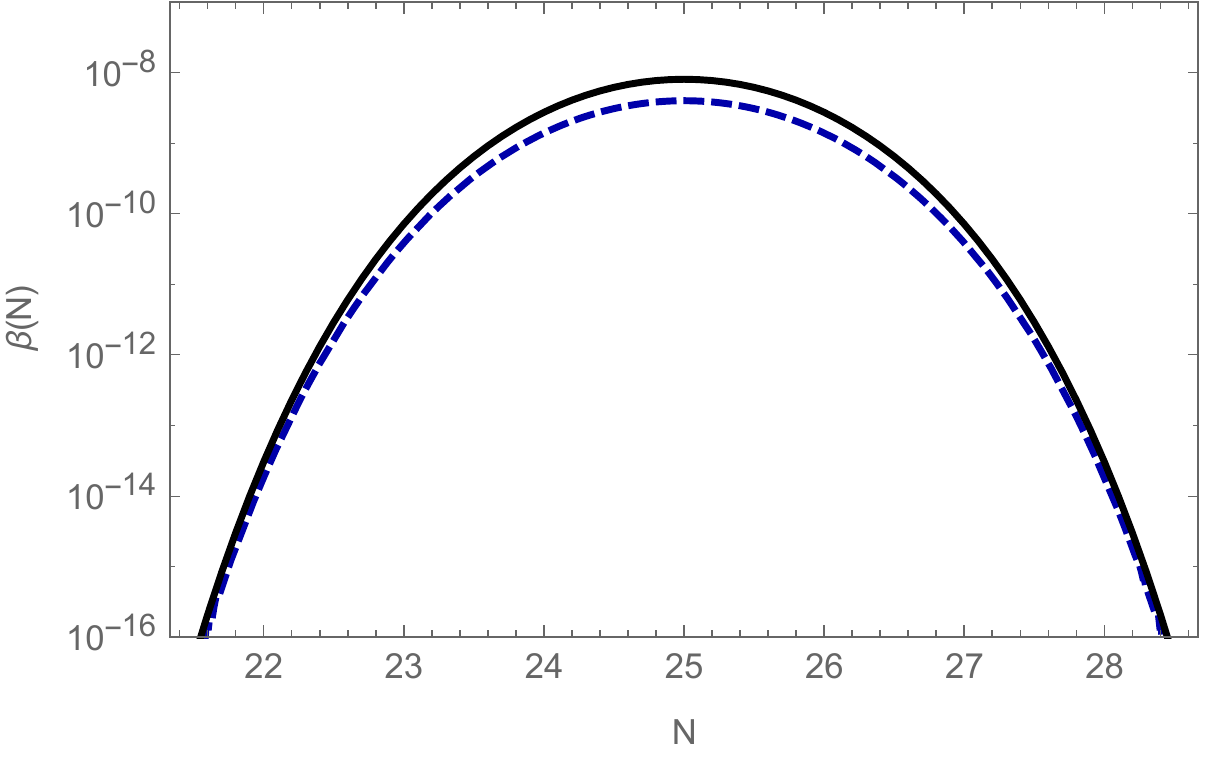}\hfill
\includegraphics[width=0.48\textwidth]{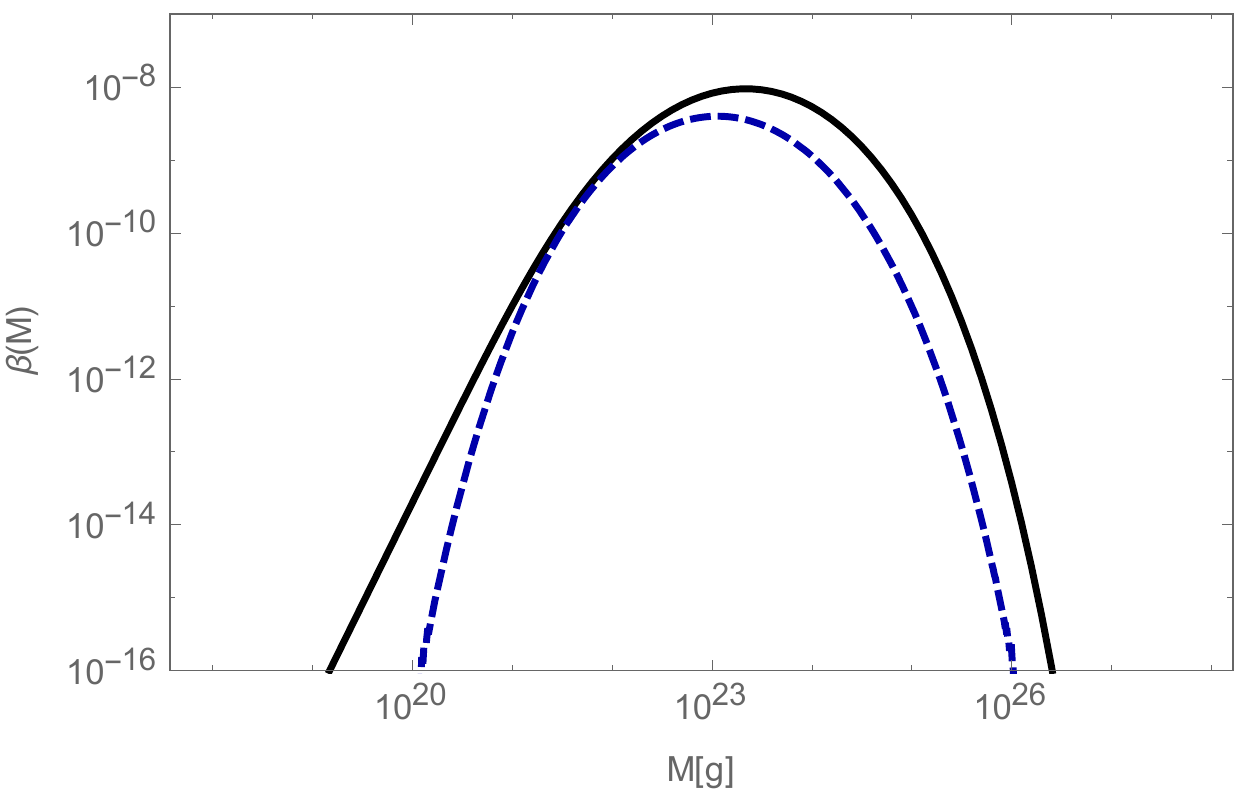}
\caption{Critical collapse versus horizon mass collapse. The dashed blue curves show the results obtained with the procedure implemented in the main text, the solid black curves are obtained with the critical collapse calculation explained in this appendix.}
\label{fig:betas}
\end{figure}

In Fig.~\ref{fig:betas} we compare the amount of black holes formed in the horizon mass scenario and taking into account critical collapse, for a toy model with a scalar spectrum $\Delta_s^2(N)$ which features a Gaussian peak around $N = 25$ with a variance of $\Delta N = 2$. We use a non-Gaussian power spectrum as in Eq.~\eqref{eq:Pnongauss}. In both panels, the dashed blue curve shows the result obtained in the horizon mass collapse setup (as implemented in the main text), whereas the solid black curve denotes the results obtained with the critical collapse formalism reviewed in this appendix. As expected, the differences are larger in the right panel, where we consider the distribution in terms of PBH masses instead of in terms of formation time $t_N$. However, in most of the parameter space the effect is only about an ${\cal O}(1)$-factor in $\beta(M)$, which turns out to be essentially irrelevant compared to the variations in $\Delta_s^2$ (to which $\beta(N)$ is exponentially sensitive) which are induced by different choices of the non-minimal coupling $\varsigma$, the number of gauge fields ${\cal N}$ and the choice of functional form of the non-minimal coupling, $h(\phi)$, see Sec.~\ref{sec:attractors}.

\subsubsection*{Non-spherical collapse}

In general, the fluctuations entering the horizon will not be exactly spherical, which can significantly alter the predictions of PBH formation. As detailed in Ref.~\cite{Kuhnel:2016exn}, the shape of the mass function is expected to remain essentially the same, but there can be an overall decrease in the amplitude. We note that in our context this effect is essentially degenerate with the number of gauge fields in the model, i.e.\ we expect that taking into account the ellipticity of the fluctuations might allow us to reproduce the mass distributions discussed in the main text with a smaller number ${\cal N}$ of gauge fields.

\subsubsection*{Mergers and accretion}

So far, we have considered primordial black holes as essentially isolated objects. While this picture is quite appropriate in the radiation dominated regime (where most of the PBH formation takes place), things become more complicated with the onset of the matter dominated regime. Due to the slower expansion rate of the universe, clustering of matter becomes efficient and black holes begin to grow due to the accretion matter and due to mergers with other PBHs. These processes stabilize the PBHs against decay through Hawking radiation (an isolated PBH will have decayed by today if $M_* \lesssim 5 \cdot 10^{14}$~g, whereas the corresponding value at matter radiation equality is $M_*^\text{eq} \simeq 3 \cdot 10^{12}$~g). It has thus been argued~\cite{Garcia-Bellido:2017mdw}, that the overall effect of the merging process may be estimated by considering all PBHs with a mass heavier than $M_*^\text{eq}$ as stable and by applying an overall shift $M \rightarrow 10^3 \, M$ to obtain the mass function today from the mass function at matter radiation equality. This moreover implies that the fraction of dark matter consisting of PBHs remains unchanged after the onset of the matter dominated regime.

In our analysis of the model determined by Eq.~\eqref{attractors:inv_definition} (which is the only model giving a significant PBH dark matter contribution), the total amount of PBH dark matter is rather insensitive to these uncertainties, since our mass distributions are only significant for $M \gg M_*$. Furthermore, we note that shifting the mass distribution even by three orders in magnitude in e.g.\ Fig.~\ref{fig:BH2}, will not significantly alter the compatibility of the spectrum with the existing bounds. This is simply due to the scale invariance of the bounds over wide ranges of the parameter space. Note however that this discussion does become relevant when comparing to the mass distribution of PBH observed e.g.\ by the LIGO/VIRGO collaboration. Given a peak-like structure in the BH distribution, extracting parameter values in the context of the scenario presented here will require a more careful study of this question.

\providecommand{\href}[2]{#2}\begingroup\raggedright


\begin{thebibliography}{10}   

\bibitem{Abbott:2016blz}
  B.~P.~Abbott {\it et al.} [LIGO Scientific and Virgo Collaborations],
  Phys.\ Rev.\ Lett.\  {\bf 116} (2016) no.6,  061102,
  [arXiv:1602.03837 [gr-qc]].


\bibitem{LISA}
  K.~Danzmann \textit{et al.}, \href{https://www.elisascience.org/files/publications/LISA_L3_20170120.pdf}{Laser Interferometer Space Antenna (LISA), L3 mission proposal}.

\bibitem{Bird:2016dcv}
  S.~Bird, I.~Cholis, J.~B.~Muñoz, Y.~Ali-Haïmoud, M.~Kamionkowski, E.~D.~Kovetz, A.~Raccanelli and A.~G.~Riess,
  Phys.\ Rev.\ Lett.\  {\bf 116} (2016) no.20,  201301,
  [arXiv:1603.00464 [astro-ph.CO]].


\bibitem{Clesse:2016vqa}
  S.~Clesse and J.~García-Bellido,
  Phys.\ Dark Univ.\  {\bf 10} (2016) 002,
  [arXiv:1603.05234 [astro-ph.CO]].


\bibitem{Sasaki:2016jop}
  M.~Sasaki, T.~Suyama, T.~Tanaka and S.~Yokoyama,
  Phys.\ Rev.\ Lett.\  {\bf 117} (2016) no.6,  061101,
  [arXiv:1603.08338 [astro-ph.CO]].


\bibitem{Carr:2016drx}
  B.~Carr, F.~Kuhnel and M.~Sandstad,
  Phys.\ Rev.\ D {\bf 94} (2016) no.8,  083504,
  [arXiv:1607.06077 [astro-ph.CO]].



\bibitem{Clesse:2016ajp}
  S.~Clesse and J.~García-Bellido,
  arXiv:1610.08479 [astro-ph.CO].

\bibitem{Blinnikov:2016bxu}
  S.~Blinnikov, A.~Dolgov, N.~K.~Porayko and K.~Postnov,
  JCAP {\bf 1611} (2016) no.11,  036
  [arXiv:1611.00541 [astro-ph.HE]].

\bibitem{Garcia-Bellido:2017fdg}
  J.~García-Bellido,
  arXiv:1702.08275 [astro-ph.CO].


\bibitem{Georg:2017mqk}
  J.~Georg and S.~Watson,
  arXiv:1703.04825 [astro-ph.CO].


\bibitem{Cotner:2016cvr}
  E.~Cotner and A.~Kusenko,
  arXiv:1612.02529 [astro-ph.CO].


\bibitem{Davoudiasl:2016mwf}
  H.~Davoudiasl and P.~P.~Giardino,
  Phys.\ Lett.\ B {\bf 768} (2017) 198,
  [arXiv:1609.00907 [gr-qc]].


\bibitem{Carr:2009jm}
  B.~J.~Carr, K.~Kohri, Y.~Sendouda and J.~Yokoyama,
  Phys.\ Rev.\ D {\bf 81} (2010) 104019,
  [arXiv:0912.5297 [astro-ph.CO]].


\bibitem{GarciaBellido:1996qt}
  J.~García-Bellido, A.~D.~Linde and D.~Wands,
  Phys.\ Rev.\ D {\bf 54} (1996) 6040,
  [astro-ph/9605094].


\bibitem{Lyth:2011kj}
  D.~H.~Lyth,
  arXiv:1107.1681 [astro-ph.CO].


\bibitem{Bugaev:2011wy}
  E.~Bugaev and P.~Klimai,
  Phys.\ Rev.\ D {\bf 85} (2012) 103504,
  [arXiv:1112.5601 [astro-ph.CO]].


\bibitem{Clesse:2015wea}
  S.~Clesse and J.~García-Bellido,
  Phys.\ Rev.\ D {\bf 92} (2015) no.2,  023524,
  [arXiv:1501.07565 [astro-ph.CO]].


\bibitem{Kohri:2012yw}
  K.~Kohri, C.~M.~Lin and T.~Matsuda,
  Phys.\ Rev.\ D {\bf 87} (2013) no.10,  103527,
  [arXiv:1211.2371 [hep-ph]].


\bibitem{Kawasaki:2012wr}
  M.~Kawasaki, N.~Kitajima and T.~T.~Yanagida,
  Phys.\ Rev.\ D {\bf 87} (2013) no.6,  063519,
  [arXiv:1207.2550 [hep-ph]].


\bibitem{Bugaev:2013vba}
  E.~V.~Bugaev and P.~A.~Klimai,
  Int.\ J.\ Mod.\ Phys.\ D {\bf 22} (2013) 1350034,
  [arXiv:1303.3146 [astro-ph.CO]].


\bibitem{Inomata:2017okj}
  K.~Inomata, M.~Kawasaki, K.~Mukaida, Y.~Tada and T.~T.~Yanagida,
  arXiv:1701.02544 [astro-ph.CO].


\bibitem{Inomata:2016rbd}
  K.~Inomata, M.~Kawasaki, K.~Mukaida, Y.~Tada and T.~T.~Yanagida,
  arXiv:1611.06130 [astro-ph.CO].


\bibitem{Kawasaki:2016pql}
  M.~Kawasaki, A.~Kusenko, Y.~Tada and T.~T.~Yanagida,
  Phys.\ Rev.\ D {\bf 94} (2016) no.8,  083523
  [arXiv:1606.07631 [astro-ph.CO]].

\bibitem{Cheng:2016qzb}
  S.~L.~Cheng, W.~Lee and K.~W.~Ng,
  JHEP {\bf 1702} (2017) 008
  [arXiv:1606.00206 [astro-ph.CO]].


\bibitem{Garcia-Bellido:2017mdw}
  J.~García-Bellido and E.~Ruiz Morales,
  arXiv:1702.03901 [astro-ph.CO].


\bibitem{Garcia-Bellido:2016dkw}
  J.~García-Bellido, M.~Peloso and C.~Unal,
  JCAP {\bf 1612} (2016) no.12,  031
  [arXiv:1610.03763 [astro-ph.CO]].


\bibitem{0605206}
  P.~Svrcek and E.~Witten,
  JHEP {\bf 0606} (2006) 051
  [hep-th/0605206].

\bibitem{Turner:1987bw}
  M.~S.~Turner and L.~M.~Widrow,
  Phys.\ Rev.\ D {\bf 37} (1988) 2743.


\bibitem{Garretson:1992vt}
  W.~D.~Garretson, G.~B.~Field and S.~M.~Carroll,
  Phys.\ Rev.\ D {\bf 46} (1992) 5346
  [hep-ph/9209238].


\bibitem{Anber:2006xt}
  M.~M.~Anber and L.~Sorbo,
  JCAP {\bf 0610} (2006) 018
  [astro-ph/0606534].


\bibitem{Barnaby:2011qe}
  N.~Barnaby, E.~Pajer and M.~Peloso,
  Phys.\ Rev.\ D {\bf 85} (2012) 023525
  [arXiv:1110.3327 [astro-ph.CO]].


\bibitem{Barnaby:2010vf}
  N.~Barnaby and M.~Peloso,
  Phys.\ Rev.\ Lett.\  {\bf 106} (2011) 181301
  [arXiv:1011.1500 [hep-ph]].


\bibitem{Linde:2012bt}
  A.~Linde, S.~Mooij and E.~Pajer,
  Phys.\ Rev.\ D {\bf 87} (2013) no.10,  103506
  [arXiv:1212.1693 [hep-th]].


\bibitem{Lin:2012gs}
  C.~M.~Lin and K.~W.~Ng,
  Phys.\ Lett.\ B {\bf 718} (2013) 1181
  [arXiv:1206.1685 [hep-ph]].


\bibitem{Meerburg:2012id}
  P.~D.~Meerburg and E.~Pajer,
  JCAP {\bf 1302} (2013) 017
  [arXiv:1203.6076 [astro-ph.CO]].


\bibitem{Bugaev:2013fya}
  E.~Bugaev and P.~Klimai,
  Phys.\ Rev.\ D {\bf 90} (2014) no.10,  103501
  [arXiv:1312.7435 [astro-ph.CO]].


\bibitem{Cook:2011hg}
  J.~L.~Cook and L.~Sorbo,
  Phys.\ Rev.\ D {\bf 85} (2012) 023534
   Erratum: [Phys.\ Rev.\ D {\bf 86} (2012) 069901]
  [arXiv:1109.0022 [astro-ph.CO]].


\bibitem{Anber:2012du}
  M.~M.~Anber and L.~Sorbo,
  Phys.\ Rev.\ D {\bf 85} (2012) 123537
  [arXiv:1203.5849 [astro-ph.CO]].


\bibitem{Bartolo:2016ami}
  N.~Bartolo {\it et al.},
  JCAP {\bf 1612} (2016) no.12,  026
  [arXiv:1610.06481 [astro-ph.CO]].


\bibitem{Domcke:2016bkh}
  V.~Domcke, M.~Pieroni and P.~Binétruy,
  JCAP {\bf 1606} (2016) 031
  [arXiv:1603.01287 [astro-ph.CO]].


\bibitem{Kallosh:2000ve}
  R.~Kallosh, L.~Kofman, A.~D.~Linde and A.~Van Proeyen,
  Class.\ Quant.\ Grav.\  {\bf 17} (2000) 4269
   Erratum: [Class.\ Quant.\ Grav.\  {\bf 21} (2004) 5017]
  [hep-th/0006179].


\bibitem{Bezrukov:2007ep}
  F.~L.~Bezrukov and M.~Shaposhnikov,
  Phys.\ Lett.\ B {\bf 659} (2008) 703
  [arXiv:0710.3755 [hep-th]].


\bibitem{Kallosh:2013hoa}
  R.~Kallosh and A.~Linde,
  JCAP {\bf 1307} (2013) 002
  [arXiv:1306.5220 [hep-th]].


\bibitem{Kallosh:2013tua}
  R.~Kallosh, A.~Linde and D.~Roest,
  Phys.\ Rev.\ Lett.\  {\bf 112} (2014) no.1,  011303
  [arXiv:1310.3950 [hep-th]].


\bibitem{Ferrara:2013rsa}
  S.~Ferrara, R.~Kallosh, A.~Linde and M.~Porrati,
  Phys.\ Rev.\ D {\bf 88} (2013) no.8,  085038
  [arXiv:1307.7696 [hep-th]].


\bibitem{Galante:2014ifa}
  M.~Galante, R.~Kallosh, A.~Linde and D.~Roest,
  Phys.\ Rev.\ Lett.\  {\bf 114} (2015) no.14,  141302
  [arXiv:1412.3797 [hep-th]].


\bibitem{Broy:2015qna}
  B.~J.~Broy, M.~Galante, D.~Roest and A.~Westphal,
  JHEP {\bf 1512} (2015) 149
  [arXiv:1507.02277 [hep-th]].


\bibitem{Das:2016kwz}
  K.~Das, V.~Domcke and K.~Dutta,
  JCAP {\bf 1703} (2017) no.03,  036
  [arXiv:1612.07075 [hep-ph]].


\bibitem{Kallosh:2013yoa}
  R.~Kallosh, A.~Linde and D.~Roest,
  JHEP {\bf 1311} (2013) 198
  [arXiv:1311.0472 [hep-th]].


\bibitem{Einhorn:2009bh}
  M.~B.~Einhorn and D.~R.~T.~Jones,
  JHEP {\bf 1003} (2010) 026
  [arXiv:0912.2718 [hep-ph]].


\bibitem{Ferrara:2010yw}
  S.~Ferrara, R.~Kallosh, A.~Linde, A.~Marrani and A.~Van Proeyen,
  Phys.\ Rev.\ D {\bf 82} (2010) 045003
  [arXiv:1004.0712 [hep-th]].


\bibitem{Buchmuller:2013zfa}
  W.~Buchmuller, V.~Domcke and K.~Kamada,
  Phys.\ Lett.\ B {\bf 726} (2013) 467
  [arXiv:1306.3471 [hep-th]].


\bibitem{Giudice:2014toa}
  G.~F.~Giudice and H.~M.~Lee,
  Phys.\ Lett.\ B {\bf 733} (2014) 58
  [arXiv:1402.2129 [hep-ph]].


\bibitem{Pallis:2013yda}
  C.~Pallis,
  JCAP {\bf 1404} (2014) 024
  [arXiv:1312.3623 [hep-ph]].


\bibitem{Pallis:2014dma}
  C.~Pallis,
  JCAP {\bf 1408} (2014) 057
  [arXiv:1403.5486 [hep-ph]].


\bibitem{Pallis:2014boa}
  C.~Pallis,
  JCAP {\bf 1410} (2014) no.10,  058
  [arXiv:1407.8522 [hep-ph]].


\bibitem{Ellis:2013xoa}
  J.~Ellis, D.~V.~Nanopoulos and K.~A.~Olive,
  Phys.\ Rev.\ Lett.\  {\bf 111} (2013) 111301
   Erratum: [Phys.\ Rev.\ Lett.\  {\bf 111} (2013) no.12,  129902]
  [arXiv:1305.1247 [hep-th]].


\bibitem{Kallosh:2013xya}
  R.~Kallosh and A.~Linde,
  JCAP {\bf 1306} (2013) 028
  [arXiv:1306.3214 [hep-th]].


\bibitem{Nakayama:2010ga}
  K.~Nakayama and F.~Takahashi,
  JCAP {\bf 1011} (2010) 039
  [arXiv:1009.3399 [hep-ph]].

\bibitem{Pieroni:2015cma}
  M.~Pieroni,
  JCAP {\bf 1602} (2016) no.02,  012
  [arXiv:1510.03691 [hep-ph]].


\bibitem{Ade:2015lrj}
  P.~A.~R.~Ade {\it et al.} [Planck Collaboration],
  Astron.\ Astrophys.\  {\bf 594} (2016) A20
  [arXiv:1502.02114 [astro-ph.CO]].


\bibitem{Anber:2009ua}
  M.~M.~Anber and L.~Sorbo,
  Phys.\ Rev.\ D {\bf 81} (2010) 043534
  [arXiv:0908.4089 [hep-th]].


\bibitem{Barnaby:2011vw}
  N.~Barnaby, R.~Namba and M.~Peloso,
  JCAP {\bf 1104} (2011) 009
  [arXiv:1102.4333 [astro-ph.CO]].


\bibitem{Freese:1990rb}
  K.~Freese, J.~A.~Frieman and A.~V.~Olinto,
  Phys.\ Rev.\ Lett.\  {\bf 65} (1990) 3233.


\bibitem{Pieroni:2016gdg}
  M.~Pieroni,
  arXiv:1611.03732 [gr-qc].


\bibitem{Ade:2015ava}
  P.~A.~R.~Ade {\it et al.} [Planck Collaboration],
  Astron.\ Astrophys.\  {\bf 594} (2016) A17
  [arXiv:1502.01592 [astro-ph.CO]].


\bibitem{Maggiore:1900zz}
  M.~Maggiore,
  ``Gravitational Waves. Vol. 1: Theory and Experiments,''
  Oxford University Press (2007).


\bibitem{Crowder:2012ik}
  S.~G.~Crowder, R.~Namba, V.~Mandic, S.~Mukohyama and M.~Peloso,
  Phys.\ Lett.\ B {\bf 726} (2013) 66
  [arXiv:1212.4165 [astro-ph.CO]].


\bibitem{Birrel:1984} 
  N.~D.~Birrell and P.~C.~W.~Davies, ``Quantum Fields in Curved Space'', Cambridge Monographs on Mathematical Physics (1984).

\bibitem{Bezrukov:2009db}
  F.~Bezrukov and M.~Shaposhnikov,
  JHEP {\bf 0907} (2009) 089
  [arXiv:0904.1537 [hep-ph]].


\bibitem{Kaiser:2013sna}
  D.~I.~Kaiser and E.~I.~Sfakianakis,
  Phys.\ Rev.\ Lett.\  {\bf 112} (2014) no.1,  011302
  [arXiv:1304.0363 [astro-ph.CO]].


\bibitem{Kallosh:2013daa}
  R.~Kallosh and A.~Linde,
  JCAP {\bf 1312} (2013) 006
  [arXiv:1309.2015 [hep-th]].


\bibitem{Kallosh:2014rga}
  R.~Kallosh, A.~Linde and D.~Roest,
  JHEP {\bf 1408} (2014) 052
  [arXiv:1405.3646 [hep-th]].


\bibitem{Kallosh:2015lwa}
  R.~Kallosh and A.~Linde,
  Phys.\ Rev.\ D {\bf 91} (2015) 083528
  [arXiv:1502.07733 [astro-ph.CO]].


\bibitem{Ade:2015xua}
  P.~A.~R.~Ade {\it et al.} [Planck Collaboration],
  Astron.\ Astrophys.\  {\bf 594} (2016) A13
  [arXiv:1502.01589 [astro-ph.CO]].


\bibitem{Mukhanov:2013tua}
  V.~Mukhanov,
  Eur.\ Phys.\ J.\ C {\bf 73} (2013) 2486
  [arXiv:1303.3925 [astro-ph.CO]].


\bibitem{Roest:2013fha}
  D.~Roest,
  JCAP {\bf 1401} (2014) 007
  [arXiv:1309.1285 [hep-th]].


\bibitem{Binetruy:2014zya}
  P.~Binetruy, E.~Kiritsis, J.~Mabillard, M.~Pieroni and C.~Rosset,
  JCAP {\bf 1504} (2015) no.04,  033
  [arXiv:1407.0820 [astro-ph.CO]].


\bibitem{Starobinsky:1980te}
  A.~A.~Starobinsky,
  Phys.\ Lett.\  {\bf 91B} (1980) 99.


\bibitem{Smith:2005mm}
  T.~L.~Smith, M.~Kamionkowski and A.~Cooray,
  Phys.\ Rev.\ D {\bf 73} (2006) 023504
  [astro-ph/0506422].


\bibitem{vanHaasteren:2011ni}
  R.~van Haasteren {\it et al.},
  Mon.\ Not.\ Roy.\ Astron.\ Soc.\  {\bf 414} (2011) no.4,  3117
   Erratum: [Mon.\ Not.\ Roy.\ Astron.\ Soc.\  {\bf 425} (2012) no.2,  1597]
  [arXiv:1103.0576 [astro-ph.CO]].


\bibitem{Kramer:2004rwa}
  M.~Kramer, 
  ``Fundamental physics with the SKA: Strong-field tests of gravity using pulsars and black holes,''
  astro-ph/0409020.


\bibitem{Petiteau}
 A.~Petiteau, in preparation.

\bibitem{Caprini:2015zlo}
  C.~Caprini {\it et al.},
  JCAP {\bf 1604} (2016) no.04,  001
  [arXiv:1512.06239 [astro-ph.CO]].


\bibitem{TheLIGOScientific:2016wyq}
  B.~P.~Abbott {\it et al.} [LIGO Scientific and Virgo Collaborations],
  Phys.\ Rev.\ Lett.\  {\bf 116} (2016) no.13,  131102
  [arXiv:1602.03847 [gr-qc]].


\bibitem{Peloso:2016gqs}
  M.~Peloso, L.~Sorbo and C.~Unal,
  JCAP {\bf 1609} (2016) no.09,  001
  [arXiv:1606.00459 [astro-ph.CO]].


\bibitem{Green:2004wb}
  A.~M.~Green, A.~R.~Liddle, K.~A.~Malik and M.~Sasaki,
  Phys.\ Rev.\ D {\bf 70} (2004) 041502
  [astro-ph/0403181].

\bibitem{0510052}
  P.~Pina Avelino,
  Phys.\ Rev.\ D {\bf 72} (2005) 124004
  doi:10.1103/PhysRevD.72.124004
  [astro-ph/0510052].

\bibitem{1201.4312}
  D.~H.~Lyth,
  JCAP {\bf 1205} (2012) 022
  doi:10.1088/1475-7516/2012/05/022
  [arXiv:1201.4312 [astro-ph.CO]].

\bibitem{Byrnes:2012yx} 
  C.~T.~Byrnes, E.~J.~Copeland and A.~M.~Green,
  Phys.\ Rev.\ D {\bf 86}, 043512 (2012)
  [arXiv:1206.4188 [astro-ph.CO]].


\bibitem{Capela:2013yf}
  F.~Capela, M.~Pshirkov and P.~Tinyakov,
  Phys.\ Rev.\ D {\bf 87} (2013) no.12,  123524
  [arXiv:1301.4984 [astro-ph.CO]].


\bibitem{Griest:2013aaa}
  K.~Griest, A.~M.~Cieplak and M.~J.~Lehner,
  Astrophys.\ J.\  {\bf 786} (2014) no.2,  158
  [arXiv:1307.5798 [astro-ph.CO]].


\bibitem{Tisserand:2006zx}
  P.~Tisserand {\it et al.} [EROS-2 Collaboration],
  Astron.\ Astrophys.\  {\bf 469} (2007) 387
  [astro-ph/0607207].


\bibitem{Novati:2013fxa}
  S.~Calchi Novati, S.~Mirzoyan, P.~Jetzer and G.~Scarpetta,
  Mon.\ Not.\ Roy.\ Astron.\ Soc.\  {\bf 435} (2013) 1582
  [arXiv:1308.4281 [astro-ph.GA]].

\bibitem{Niikura:2017zjd}
  H.~Niikura {\it et al.},
  arXiv:1701.02151 [astro-ph.CO].


\bibitem{TheLIGOScientific:2016agk}
  B.~P.~Abbott {\it et al.} [LIGO Scientific and Virgo Collaborations],
  Phys.\ Rev.\ Lett.\  {\bf 116} (2016) no.13,  131103
  [arXiv:1602.03838 [gr-qc]].


\bibitem{Misner:1974qy}
  C.~W.~Misner, K.~S.~Thorne and J.~A.~Wheeler,
  San Francisco 1973, 1279p.


\bibitem{Adshead:2015pva}
  P.~Adshead, J.~T.~Giblin, T.~R.~Scully and E.~I.~Sfakianakis,
  JCAP {\bf 1512} (2015) no.12,  034
  [arXiv:1502.06506 [astro-ph.CO]].


\bibitem{Kusenko:2014uta}
  A.~Kusenko, K.~Schmitz and T.~T.~Yanagida,
  Phys.\ Rev.\ Lett.\  {\bf 115} (2015) no.1,  011302
  [arXiv:1412.2043 [hep-ph]].


\bibitem{Anber:2015yca}
  M.~M.~Anber and E.~Sabancilar,
  Phys.\ Rev.\ D {\bf 92} (2015) no.10,  101501
  [arXiv:1507.00744 [hep-th]].


\bibitem{Adshead:2015jza}
  P.~Adshead and E.~I.~Sfakianakis,
  Phys.\ Rev.\ Lett.\  {\bf 116} (2016) no.9,  091301
  [arXiv:1508.00881 [hep-ph]].


\bibitem{Crewther:1979pi}
  R.~J.~Crewther, P.~Di Vecchia, G.~Veneziano and E.~Witten,
  Phys.\ Lett.\  {\bf 88B} (1979) 123
   Erratum: [Phys.\ Lett.\  {\bf 91B} (1980) 487].


\bibitem{Baker:2006ts}
  C.~A.~Baker {\it et al.},
  Phys.\ Rev.\ Lett.\  {\bf 97} (2006) 131801
  [hep-ex/0602020].


\bibitem{Peccei:1977hh}
  R.~D.~Peccei and H.~R.~Quinn,
  Phys.\ Rev.\ Lett.\  {\bf 38} (1977) 1440.


\bibitem{Wilczek:1977pj}
  F.~Wilczek,
  Phys.\ Rev.\ Lett.\  {\bf 40} (1978) 279.


\bibitem{Weinberg:1977ma}
  S.~Weinberg,
  Phys.\ Rev.\ Lett.\  {\bf 40} (1978) 223.


\bibitem{Vafa:1984xg}
  C.~Vafa and E.~Witten,
  Phys.\ Rev.\ Lett.\  {\bf 53} (1984) 535.


\bibitem{0605225}
  P.~Anastasopoulos, M.~Bianchi, E.~Dudas and E.~Kiritsis,
  JHEP {\bf 0611} (2006) 057
  [hep-th/0605225].

\bibitem{0507215}
  G.~Dvali,
  hep-th/0507215.

\bibitem{0811.1989}
  N.~Kaloper and L.~Sorbo,
  Phys.\ Rev.\ Lett.\  {\bf 102} (2009) 121301
  [arXiv:0811.1989 [hep-th]].


\bibitem{1101.0026}
  N.~Kaloper, A.~Lawrence and L.~Sorbo,
  JCAP {\bf 1103} (2011) 023
  [arXiv:1101.0026 [hep-th]].


\bibitem{ArkaniHamed:2006dz}
  N.~Arkani-Hamed, L.~Motl, A.~Nicolis and C.~Vafa,
  JHEP {\bf 0706} (2007) 060
  [hep-th/0601001].


\bibitem{Conlon:2012tz}
  J.~P.~Conlon,
  JCAP {\bf 1209} (2012) 019
  [arXiv:1203.5476 [hep-th]].


\bibitem{1404.3040}
  F.~Marchesano, G.~Shiu and A.~M.~Uranga,
  JHEP {\bf 1409} (2014) 184
  [arXiv:1404.3040 [hep-th]].


\bibitem{1404.3542}
  R.~Blumenhagen and E.~Plauschinn,
  Phys.\ Lett.\ B {\bf 736} (2014) 482
  [arXiv:1404.3542 [hep-th]].


\bibitem{1404.3711}
  A.~Hebecker, S.~C.~Kraus and L.~T.~Witkowski,
  Phys.\ Lett.\ B {\bf 737} (2014) 16
  [arXiv:1404.3711 [hep-th]].


\bibitem{Ferrara:2010in}
  S.~Ferrara, R.~Kallosh, A.~Linde, A.~Marrani and A.~Van Proeyen,
  Phys.\ Rev.\ D {\bf 83} (2011) 025008
  [arXiv:1008.2942 [hep-th]].


\bibitem{Buchmuller:2012ex}
  W.~Buchmüller, V.~Domcke and K.~Schmitz,
  JCAP {\bf 1304} (2013) 019
  [arXiv:1210.4105 [hep-ph]].


\bibitem{Gaillard:1993es}
  M.~K.~Gaillard and V.~Jain,
  Phys.\ Rev.\ D {\bf 49} (1994) 1951
  [hep-th/9308090].


\bibitem{Stewart:1996ey}
  E.~D.~Stewart,
  Phys.\ Lett.\ B {\bf 391} (1997) 34
  [hep-ph/9606241].


\bibitem{Stewart:1997wg}
  E.~D.~Stewart,
  Phys.\ Rev.\ D {\bf 56} (1997) 2019
  [hep-ph/9703232].


\bibitem{1404.2601}
  D.~Baumann and L.~McAllister,
  arXiv:1404.2601 [hep-th].


\bibitem{Arvanitaki:2009fg}
  A.~Arvanitaki, S.~Dimopoulos, S.~Dubovsky, N.~Kaloper and J.~March-Russell,
  Phys.\ Rev.\ D {\bf 81} (2010) 123530
  [arXiv:0905.4720 [hep-th]].


\bibitem{Cicoli:2012sz}
  M.~Cicoli, M.~Goodsell and A.~Ringwald,
  JHEP {\bf 1210} (2012) 146
  [arXiv:1206.0819 [hep-th]].


\bibitem{1409.5350}
  A.~Westphal,
  Int.\ J.\ Mod.\ Phys.\ A {\bf 30} (2015) no.09,  1530024
  [arXiv:1409.5350 [hep-th]].


\bibitem{Kim:2004rp}
  J.~E.~Kim, H.~P.~Nilles and M.~Peloso,
  JCAP {\bf 0501} (2005) 005
  [hep-ph/0409138].


\bibitem{1404.7496}
  T.~C.~Bachlechner, M.~Dias, J.~Frazer and L.~McAllister,
  Phys.\ Rev.\ D {\bf 91} (2015) no.2,  023520
  [arXiv:1404.7496 [hep-th]].


\bibitem{Shiu:2015xda}
  G.~Shiu, W.~Staessens and F.~Ye,
  JHEP {\bf 1506} (2015) 026
  [arXiv:1503.02965 [hep-th]].


\bibitem{Dimopoulos:2005ac}
  S.~Dimopoulos, S.~Kachru, J.~McGreevy and J.~G.~Wacker,
  JCAP {\bf 0808} (2008) 003
  [hep-th/0507205].


\bibitem{Cicoli:2014sva}
  M.~Cicoli, K.~Dutta and A.~Maharana,
  JCAP {\bf 1408} (2014) 012
  [arXiv:1401.2579 [hep-th]].


\bibitem{Das:2014gua}
  K.~Das and K.~Dutta,
  Phys.\ Lett.\ B {\bf 738} (2014) 457
  [arXiv:1408.6376 [hep-ph]].


\bibitem{Silverstein:2008sg}
  E.~Silverstein and A.~Westphal,
  Phys.\ Rev.\ D {\bf 78} (2008) 106003
  [arXiv:0803.3085 [hep-th]].


\bibitem{0808.0706}
  L.~McAllister, E.~Silverstein and A.~Westphal,
  Phys.\ Rev.\ D {\bf 82} (2010) 046003
  [arXiv:0808.0706 [hep-th]].


\bibitem{1503.00795}
  T.~Rudelius,
  JCAP {\bf 1509} (2015) no.09,  020
  [arXiv:1503.00795 [hep-th]].


\bibitem{1506.03447}
  B.~Heidenreich, M.~Reece and T.~Rudelius,
  JHEP {\bf 1512} (2015) 108
  [arXiv:1506.03447 [hep-th]].


\bibitem{Cicoli:2008gp}
  M.~Cicoli, C.~P.~Burgess and F.~Quevedo,
  JCAP {\bf 0903} (2009) 013
  [arXiv:0808.0691 [hep-th]].


\bibitem{Burgess:2016owb}
  C.~P.~Burgess, M.~Cicoli, S.~de Alwis and F.~Quevedo,
  JCAP {\bf 1605} (2016) no.05,  032
  [arXiv:1603.06789 [hep-th]].


\bibitem{Cicoli:2016chb}
  M.~Cicoli, D.~Ciupke, S.~de Alwis and F.~Muia,
  JHEP {\bf 1609} (2016) 026
  [arXiv:1607.01395 [hep-th]].


\bibitem{Cicoli:2016xae}
  M.~Cicoli, F.~Muia and P.~Shukla,
  JHEP {\bf 1611} (2016) 182
  [arXiv:1611.04612 [hep-th]].

\bibitem{ChiralGlobalFibre}
  M.~Cicoli, D.~Ciupke, V.~A.~Diaz, V.~Guidetti, F.~Muia and P.~Shukla,
  \textit{in preparation}.  

\bibitem{1011.4521}
  X.~Dong, B.~Horn, E.~Silverstein and A.~Westphal,
  Phys.\ Rev.\ D {\bf 84} (2011) 026011
  [arXiv:1011.4521 [hep-th]].


\bibitem{1405.3652}
  L.~McAllister, E.~Silverstein, A.~Westphal and T.~Wrase,
  JHEP {\bf 1409} (2014) 123
  [arXiv:1405.3652 [hep-th]].


\bibitem{0906.1920}
  J.~P.~Conlon and E.~Palti,
  JHEP {\bf 0909} (2009) 019
  [arXiv:0906.1920 [hep-th]].


\bibitem{Choptuik:1992jv}
  M.~W.~Choptuik,
  Phys.\ Rev.\ Lett.\  {\bf 70} (1993) 9.


\bibitem{Koike:1995jm}
  T.~Koike, T.~Hara and S.~Adachi,
  Phys.\ Rev.\ Lett.\  {\bf 74} (1995) 5170
  [gr-qc/9503007].


\bibitem{Niemeyer:1999ak}
  J.~C.~Niemeyer and K.~Jedamzik,
  Phys.\ Rev.\ D {\bf 59} (1999) 124013
  [astro-ph/9901292].


\bibitem{Gundlach:1999cu}
  C.~Gundlach,
  Living Rev.\ Rel.\  {\bf 2} (1999) 4
  [gr-qc/0001046].


\bibitem{Gundlach:2002sx}
  C.~Gundlach,
  Phys.\ Rept.\  {\bf 376} (2003) 339
  [gr-qc/0210101].


\bibitem{Musco:2004ak}
  I.~Musco, J.~C.~Miller and L.~Rezzolla,
  Class.\ Quant.\ Grav.\  {\bf 22} (2005) 1405
  [gr-qc/0412063].


\bibitem{Musco:2008hv}
  I.~Musco, J.~C.~Miller and A.~G.~Polnarev,
  Class.\ Quant.\ Grav.\  {\bf 26} (2009) 235001
  [arXiv:0811.1452 [gr-qc]].


\bibitem{Musco:2012au}
  I.~Musco and J.~C.~Miller,
  Class.\ Quant.\ Grav.\  {\bf 30} (2013) 145009
  [arXiv:1201.2379 [gr-qc]].


\bibitem{Niemeyer:1997mt}
  J.~C.~Niemeyer and K.~Jedamzik,
  Phys.\ Rev.\ Lett.\  {\bf 80} (1998) 5481
  [astro-ph/9709072].


\bibitem{Kuhnel:2016exn}
  F.~Kühnel and M.~Sandstad,
  Phys.\ Rev.\ D {\bf 94} (2016) no.6,  063514
  [arXiv:1602.04815 [astro-ph.CO]].


\end{thebibliography}
\end{document}